\documentclass[twocolumn]{aastex631}

\usepackage{amsmath}
\usepackage{cuted}
\usepackage{lipsum}
\usepackage{siunitx}
\usepackage{xcolor}
\usepackage{amsthm}

\newtheorem{teo}{Theorem}[section]

\newtheorem{lema}{Lemma}[section]

\shorttitle{Joint Photometry}
\shortauthors{Vicuña et al.}

\graphicspath{{./}{images/}}

\begin{document}

\title{Optimal photometry of point sources: Joint source flux and background determination on array detectors - from theory to practical implementation}

\author{Mario L. Vicuña}
\affiliation{Information and Decision Systems Group, Department of Electrical Engineering, Facultad de Ciencias Físicas y Matemáticas, Universidad de Chile, Beauchef 850, Santiago, Chile}

\author[0000-0002-0256-282X]{Jorge F. Silva}
\affiliation{Information and Decision Systems Group, Department of Electrical Engineering, Facultad de Ciencias Físicas y Matemáticas, Universidad de Chile, Beauchef 850, Santiago, Chile}

\author[0000-0003-1454-0596]{Rene A. Mendez}
\affiliation{Departamento de Astronomía, Facultad de Ciencias Físicas y Matemáticas, Universidad de Chile, Casilla 36-D, Santiago, Chile}

\author[0000-0003-4778-2719]{Marcos E. Orchard}
\affiliation{Information and Decision Systems Group, Department of Electrical Engineering, Facultad de Ciencias Físicas y Matemáticas, Universidad de Chile, Beauchef 850, Santiago, Chile}

\author{Sebastian Espinosa}
\affiliation{Information and Decision Systems Group, Department of Electrical Engineering, Facultad de Ciencias Físicas y Matemáticas, Universidad de Chile, Beauchef 850, Santiago, Chile}

\author[0000-0002-9024-4185]{Jeremy Tregloan-Reed}
\affiliation{Instituto de Investigación en Astronomía y Ciencias Planetarias, Universidad de Atacama, Copiapó, Atacama, Chile}


\begin{abstract}
In this paper we study the joint determination of source and background flux for point sources as observed by digital array detectors. We explicitly compute the two-dimensional Cram\'er-Rao absolute lower bound (CRLB) as well as the performance bounds for high-dimensional implicit estimators from a generalized Taylor expansion. This later approach allows us to obtain computable prescriptions for the bias and variance of the joint estimators. We compare these prescriptions with empirical results from numerical simulations in the case of the weighted least squares estimator (introducing an improved version, denoted stochastic weighted least-squares) as well as with the maximum likelihood estimator, finding excellent agreement. We demonstrate that these estimators provide quasi-unbiased joint estimations of the flux and background, with a variance that approaches the CRLB very tightly and are, hence, optimal, unlike the case of sequential estimation used commonly in astronomical photometry which is sub-optimal. We compare our predictions with numerical simulations of realistic observations, as well as with observations of a bona-fide non-variable stellar source observed with TESS, and compare it to the results from the sequential estimation of background and flux, confirming our theoretical expectations. Our practical estimators can be used as benchmarks for general photometric pipelines, or for applications that require maximum precision and accuracy in absolute photometry.
\end{abstract}

\section{Introduction}
\label{sec:intro}
Astronomical photometry, understood as the precise measurement of the brightness of celestial objects \citep{kitchin2020astrophysical}, plays a crucial role in determining the properties of these objects, and allows us to study them in detail \citep{milone2011astronomical}. Photometric techniques have a wide range of applications in astrophysics, ranging from stellar classification, studying the internal constitution of stars (asteroseismology), detection of exoplanets (transits), studies of variable stars and various transients (including supernovae), distance determination, dating open and globular clusters, studies of resolved and un-resolved stellar populations, AGN and quasars, and Cosmology, among others (see e.g., \citet{carroll2017introduction}).


As light sources in the celestial sphere are not isolated, their observation conveys additional light flux from adjacent objects, instrumental noise, and sky background, among other phenomena. Isolating the target and sky background estimation are broadly recognized as critical steps to achieve precise characterization of an object's flux. To cope with these nuisances, classic approaches rely on aperture methods based on a signal-to-noise ratio (SNR hereafter) maximization \citep{howell1989, 1995ExA.....6..163M}, which perform source detection and separation, background inference, and source photometry in three different and sequential stages (see, e.g., \citet{1987PASP...99..191S, 1993PASP..105.1342S, sextractor}). Consequently, the notion of an optimal aperture emerges, beyond which the photometric precision seems to deteriorate as the measurement area on the detector increases \citep{naylor1998}.

In relation to the inference of flux and background (that is, not taking into account the problem of identifying and separating different objects within the field of view), the Data Processing Inequality \citep{coverthomas} suggests that this sequential estimation process, in which the background estimate is taken as an input of a subsequent photometry algorithm may be a sub-optimal approach, as pixels carrying information about the source's flux also include light coming from the background. Therefore, both estimates may benefit from a simultaneous joint inference, as more background-containing pixels are incorporated in its estimation, and the precision gain for that parameter should allow better performance for the estimation of the source's flux itself.

From a practical standpoint, previous work on joint inference has been explored in the field of astronomy with promising results, such as in \citet{guglielmetti2009,guglielmettiThesis} for photometry, and \citet{gai2017} for photometry and astrometry. However, the effects and relevance of information use on the statistical precision of the estimates have, to our knowledge, not been addressed. In this paper we tackle this important issue from a theoretical perspective, including realistic simulations, as well as comparisons to real data.

On the methodological side, the present work is influenced by our previous research on the Cramér-Rao lower bound (CRLB hereafter) analysis as applied to astronomy by \citet{mendez2013, mendez2014}, and the performance characterization of practical estimators carried out by \citet{fessler1996} and \citet{espinosa2018}. Here, we extend those previous works, which allow us to thoroughly study the theoretical properties and potential gains and viability of a joint inference, through practical estimation algorithms. Our theoretical findings for the joint inference approach are compared to standard aperture methods, supporting the improved performance of the former through, both, simulations with TESS-like parameters, and actual TESS data (TESS stands for Transiting Exoplanet Survey Satellite \citep{2015JATIS...1a4003R}).

The paper is organized as follows: Section \ref{sec:preliminaries} introduces the basics of the observational model and Cramér-Rao theory. In Section \ref{sec:CRLB_analysis}, we briefly present a discussion on Cramér-Rao applied to joint inference of source flux and background. Section \ref{sec:impl_extension} presents the main methodological contribution. Sections \ref{sec:WLSE} and \ref{sec:MLE} present numerical analysis of the performance of the Weighted Least-Squares (WLS hereafter) and Maximum Likelihood (ML hereafter) estimators, respectively. These sections also provide a validation and alternative uses of our bounding methodology. In Section \ref{sec:comparison}, the WLS and ML estimators are compared against each other and with a more standard (sequential) pipeline for background and flux inference. Section \ref{sec:TESS} presents the results of joint inference on real satellite data obtained for a selected target. Finally, in Section \ref{sec:conclu} we summarize our main conclusions.

\section{Preliminaries}
\label{sec:preliminaries}
In this section, we present basic definitions and the problem setting for the joint estimation of source flux and background in the context
of nearly-isolated sources, i.e. in which the aperture size may be chosen at will, and could be 
arbitrarily large, without additional light flux from adjacent objects. Also, for completeness, some important technical results related to the parameter estimation task are also outlined here.

\subsection{The Joint Source Flux and Background Estimation Problem}
\label{sec:problem_setting}
We introduce the main inference problem of the joint estimation of the brightness and background measured by a CCD-like solid-state detector. The background noise includes open-sky photon emissions and noise from the detector itself. A point source imaged with a one-dimensional array of this kind\footnote{The extension to the two-dimensional array is presented in \citet{mendez2013}.} is parameterized by the pair $(x_{c}, \tilde{F}) \in \mathbb{X} \equiv \mathbb{R} \times \mathbb{R}^{+}$, where $x_{c}$ denotes the relative position of the point source in the array (in arcseconds) and $\tilde{F}$ corresponds to the source's brightness in photo-electrons (hereafter photo-e$^{-}$). A point source translates into a nominal brightness profile in the device, which can be expressed as:
\begin{equation}
    \tilde{F}_{x_{c},\tilde{F}}(x) = \tilde{F} \cdot \phi(x - x_{c}, \sigma) \,,
    \label{eq:nom_profile}
\end{equation}

\noindent where $\phi(x - x_{c}, \sigma)$ denotes the one-dimensional Point Spread Function (PSF), which determines the light distribution coming from the point source over the array. The PSF is parameterized by $\sigma$ (also measured in arcseconds) which is an indication of the quality of the observing site \citep{mendez2013, mendez2014}, inasmuch as it is a direct measurement of the atmospherically induced spread of the image on the detector, usually denoted as the "astronomical seeing" \citep{chromey2016measure}.

Let us consider a photon integrating device (such as a CCD or a CMOS) that measures a countable collection of independent and non-identically distributed random variables (observations or counts on the detector) represented as a vector $\mathbf{I} = (I_1,...,I_n), n \in \mathbb{N}$, where $I_{i} \sim Poisson(\lambda_{i}(x_{c}, \tilde{F})) $ corresponds to the photon count measured by the $i^{th}$ pixel in the device. If $\lambda_{i}(x_{c}, \tilde{F})$ is the \textit{total expected} flux in that pixel, then:
\begin{equation}
    \lambda_{i}(x_{c}, \tilde{F}) = \mathbb{E}\left\{ I_{i} \right\} = \underbrace{\tilde{F} \cdot g_{i}(x_{c})}_{ \equiv \tilde{F}_{i}(x_{c},\tilde{F})} + \tilde{B}_{i}, \hspace{3mm}\forall i=1,\dots,n \,,
    \label{eq:lambda_i_Bi}
\end{equation}
and
\begin{equation}
    g_{i}(x_{c}) = \int_{x_{i} - \frac{\Delta x}{2}}^{x_{i} + \frac{\Delta x}{2}} \phi(x - x_{c}, \sigma) dx \hspace{3mm}\forall i=1,\dots,n \,.
    \label{eq:g_i}
\end{equation}

$\tilde{F}_{i}(x_{c},\tilde{F})$ in Eq.~(\ref{eq:lambda_i_Bi}) comes from Eq.~(\ref{eq:nom_profile}) and denotes the fraction of (\textit{expected}) source flux on the $i^{th}$ pixel. On the other hand,  $\tilde{B}_{i}$ represents the (\textit{expected}) additive background noise coming from external sources such as photon emissions from the open-sky and the noise of the instrument itself (read-out noise and dark-current \citep{janesick2001, howell2006, janesick2007, mclean2008}).

In Equation~(\ref{eq:g_i}), $x_{i}$ denotes the relative position of the $i^{th}$ pixel's center projected into the sky $\forall i=1,\dots, n$, while $\Delta x = x_{i+1} - x_{i}$ denotes the array resolution or pixel size. A basic common assumption is that we have a good spatial coverage of the studied object \citep{lobos2015, espinosa2018}, in the sense that given its relative position $x_{c}$ we have that:
\begin{equation}
    \sum_{i=1}^{n} g_{i}(x_{c}) \approx \sum_{i \in \mathbb{N}} g_{i}(x_{c}) = \int_{-\infty}^{\infty} \phi(x - x_{c}, \sigma) = 1 \,.
    \label{eq:coverage}
\end{equation}

Supported by classical photometry estimation problems \citep{lindegren2008, gai2017}, we will assume 
 an \textit{homogeneous background}, which states that $\tilde{B}_{i} = \tilde{B}$, $\forall i=1,\dots,n$. This allows us to rewrite Equation~$(\ref{eq:lambda_i_Bi})$ as:
\begin{equation}
    \lambda_{i}(x_{c}, \tilde{F}) = \mathbb{E}\left\{ I_{i} \right\} = \tilde{F} \cdot g_{i}(x_{c}) + \tilde{B}, \hspace{3mm}\forall i=1,\dots,n \,.
    \label{eq:lambda_i}
\end{equation}
Then, the likelihood of the joint observation vector $\mathbf{I}$, given the source parameters $(x_{c}, \tilde{F})$ and background $\tilde{B}$, is expressed as:
\begin{equation}
    \begin{split}
        L(\mathbf{I}; x_{c}, \tilde{F}, \tilde{B}) 
        &\equiv p(\mathbf{I}| x_{c}, \tilde{F}, \tilde{B}) \\
        &= \prod_{i=1}^{n} \frac{e^{-\lambda_{i}(x_{c}, \tilde{F})} \cdot \lambda_{i}^{I_{i}}(x_{c}, \tilde{F})}{I_{i}!} \,.
    \end{split}
    \label{eq:likelihood}
\end{equation}

The last equality in Eq.~(\ref{eq:likelihood}) comes from the independence assumption. 

\subsection{The Inference Task}
Finally, $x_c$ is assumed to be known and the estimation task reduces to finding the parameter estimator $\hat{\boldsymbol{\theta}}~=~(\hat{\tilde{F}}, \hat{\tilde{B}})$ which is the inference of the underlying parameters $(\tilde{F}, \tilde{B})$ from $\mathbf{I}$. Formally, the inference task consists on defining a regression rule $\tau(\mathbf{I}):~\mathbb{N}^{n}~\to~\boldsymbol{\Theta}~=~\mathbb{R}^{+} \times \mathbb{R}^{+}$ such that $(\hat{\tilde{F}}, \hat{\tilde{B}}) \equiv \tau(\mathbf{I})$.

\subsection{Model Considerations}
\label{sec:considerations}
We adopt some simplifying but realistic design and model considerations to address this task. On one hand, we assume a Gaussian PSF, {\em i.e.},
\begin{equation}
    \phi(x) = \frac{1}{\sqrt{2\pi\sigma^{2}}}\exp{\left( -\frac{1}{2} \frac{x^{2}}{\sigma^{2}} \right)} \,,
    \label{eq:Gaussian_PSF}
\end{equation}
which is a reasonable choice in the considered context of a ground-based observational setting (see, e.g., \citet{mendez2010}). 

As mentioned earlier, $\sigma$ is usually related to image quality since narrower PSFs translate into less light-spreading. An alternative measure of quality is that of ``Full-Width at Half-Maximum" ($FWHM$), which is related to the Gaussian $\sigma$ through
    $FWHM = 2\sqrt{2\ln{2}}\sigma$. 

In astronomical applications, the detector noise is usually dominated by the diffuse light coming from the sky due to the long exposure times (unless the targeted source is too bright and requires very short exposures), then the background per pixel $\tilde{B}$ is dependent on the pixel size $\Delta x$. Such a background model can be formulated as follows \citep{winick1986, mendez2013}:
\begin{equation}
    \tilde{B} = G \Delta x f_{s} + D + RON^{2} \,,
    \label{eq:B_tilde}
\end{equation}
where $G$ denotes the detector's (inverse-) gain in units of photo-e$^{-}$/ADU (ADU stands for ``Analog to Digital Unit"), $f_{s}$ denotes the sky background in units of ADUs/arcsec, and $D$ and $RON$ denote dark-current and read-out-noise, respectively, in units of photo-e$^{-}$. This model allows to translate noise from the sky into individual pixels, taking into account instrumental noise too, while being able to capture the effect of exposure time through $f_{s}$ \citep{winick1986}. 

Note that the source flux and the background can be transformed into ADUs through $G$, and vice versa. Those conversions are denoted as $F$ and $B$, respectively, and relate to the quantities measured in photo-e$^{-}$ by means of
    $\tilde{F} = G \cdot F$ and 
    $\tilde{B} = G \cdot B$. 


\subsection{The Cramér-Rao Lower Bound (CRLB)}
\label{sec:CRLB_theory}
The celebrated CRLB  offers a performance bound on the variance 
of the family of unbiased estimators.\footnote{In the sense that $\mathbb{E}\{ \hat{\boldsymbol{\theta}} \} = \boldsymbol{\theta}$.} The multidimensional version of this result is:

\begin{teo} \citep{rao1945, cramer1946} 
Let $\{I_{i}\}_{i=1}^{n}$ be a collection of independent observations, whose probability density (or mass) function $L(\cdot; \boldsymbol{\theta})$ is induced by a parameter vector $\boldsymbol{\theta} = (\theta_{1},\dots,\theta_{m}) \in \boldsymbol{\Theta}, m \in \mathbb{N}$ over a parameter space $\boldsymbol{\Theta}$ (typically $\boldsymbol{\Theta} = \mathbb{R}^{m}$), such that the following ``regularity condition" is satisfied: 
    \begin{equation}
        \mathbb{E}_{\mathbf{I} \sim L(\mathbf{I};\boldsymbol{\theta})} \left\{ \frac{\partial \ln{L(\mathbf{I};\boldsymbol{\theta})}}{\partial \theta_{i}} \right\} = 0 \,, \hspace{1mm} \forall i \in \{1,\dots,m\} \,, \forall \boldsymbol{\theta} \in \boldsymbol{\Theta} \,.
        \label{eq:CRLB_reg}
    \end{equation}
    (i) Then, any unbiased estimator $\hat{\boldsymbol{\theta}}$ of $\boldsymbol{\theta}$, given by a regression rule $\tau: \mathbb{N}^{n} \to \boldsymbol{\Theta}, \hat{\boldsymbol{\theta}} \equiv \tau(\mathbf{I})$, has a covariance matrix $K_{\hat{\boldsymbol{\theta}}}\equiv\mathbb{E} \left\{ \left( \hat{\boldsymbol{\theta}} - \bar{\hat{\boldsymbol{\theta}}} \right) \cdot \left( \hat{\boldsymbol{\theta}} - \bar{\hat{\boldsymbol{\theta}}} \right)^{\dagger} \right\}$ that satisfies:
    \begin{equation}
        K_{\hat{\boldsymbol{\theta}}} - \mathcal{I}_{\boldsymbol{\theta}}^{-1} \succeq \mathbf{0}.\footnote{$\succeq \mathbf{0}$ means that the matrix on the LHS of Eq.~(\ref{eq:CRLB_psd}) is positive semi-definite: $\forall x \in \mathbb{R}^m$, $x^\dagger \cdot (K_{\hat{\boldsymbol{\theta}}} - \mathcal{I}_{\boldsymbol{\theta}}^{-1})\cdot x \geq 0$.}
        \label{eq:CRLB_psd}
    \end{equation}
    $\mathcal{I}_{\boldsymbol{\theta}} \in \mathcal{M}^{m \times m}$ denotes the Fisher Information Matrix, whose components are defined by: $\forall i,j \in \{1,\dots,m \},$
    
    \begin{equation}
        [\mathcal{I}_{\boldsymbol{\theta}}]_{(i,j)} 
        \equiv -\mathbb{E}_{\mathbf{I} \sim L(\mathbf{I};\boldsymbol{\theta})} \left\{ \frac{\partial^{2} \ln{L(\mathbf{I};\boldsymbol{\theta})}}{\partial \theta_{i} \partial \theta_{j}} \right\} \,.
        \label{eq:fisher_matrix}
    \end{equation}
    (ii) Furthermore, if there exists a function $\mathbf{h}: \mathbb{R}^n \to \boldsymbol{\Theta}$ such that: 
    \begin{equation}
        \frac{\partial \ln{L(\mathbf{I};\boldsymbol{\theta})}}{\partial \theta_{i}} = \left[ \mathcal{I}_{\boldsymbol{\theta}} (\mathbf{h}(\mathbf{I}) - \boldsymbol{\theta}) \right]_{i} \,, \hspace{3mm} \forall i \in \{1,\dots,m\} \,,
        \label{eq:CRLB_g_cond}
    \end{equation}
    then the Minimal Variance Unbiased Estimator (MVUE) is given by $\hat{\boldsymbol{\theta}} = \mathbf{h}
    (\mathbf{I})$, and its optimal covariance matrix is $\mathcal{I}_{\boldsymbol{\theta}}^{-1}$.
    
    \label{theo:CRLB}
\end{teo}

From Theorem~\ref{theo:CRLB}, any unbiased estimator $\hat{\theta}_i$ of $\theta_i$ satisfies that (from Eq.~(\ref{eq:CRLB_psd})):
\begin{equation}
    Var(\hat{\theta}_{i}) = [K_{\hat{\boldsymbol{\theta}}}]_{(i,i)} \geq [\mathcal{I}_{\boldsymbol{\theta}}^{-1}]_{(i,i)} \,, \hspace{1mm} \forall i \in \{1,\dots,m\}, 
\end{equation}
which is a lower bound for the  Mean Squared Error (MSE) of $\hat{\theta}_i$.\footnote{Remark: Theorem~\ref{theo:CRLB} does not guarantee the existence of an estimator that achieves the CRLB, however, it is still possible to find the MVUE \citep{kay}.}

\subsection{The Tensor Operator}
\label{sub_sec_tensor}
For the derivation of one of our main theoretical results in Section \ref{sec:impl_extension}, we need to introduce the following concept (see \citet{fessler1996}).  Let $i,j \in \mathbb{N}$ and a real function $f(\boldsymbol{\alpha}, \mathbf{I}), f: \boldsymbol{\Theta} \times \mathbb{N}^{n} \to \mathbb{R}$. Let us denote $\nabla^{ij}f(\cdot)$ by (a tensor operator on $f(\cdot)$)
\begin{equation}
    \nabla^{ij}f(\cdot) \equiv \frac{\partial^{i+j}}{\partial\boldsymbol{\alpha}^{i}\partial\mathbf{I}^{j}}f(\cdot),
    \label{eq:nabla_notation}
\end{equation}
where in (\ref{eq:nabla_notation}) $f(\cdot)$ is derived $i$ times with respect to the vector $\boldsymbol{\alpha}$ and $j$ times with respect to $\mathbf{I}$.\footnote{As the $\nabla^{ij}$ operator is defined as derivatives w.r.t vectors, its application on $f(\cdot)$ results in a $(i+j)$-dimensional array, whose size depends on the dimensions of $\boldsymbol{\alpha}$ and $\mathbf{I}$. For example, $\nabla^{20}f(\cdot)$ is a $p \times p$ matrix, while $\nabla^{12}$ is a $p \times n \times n$ array. For further examples and applications see Appendices \ref{app:proof_pdim_bounds},\ref{app:proof_WLS_FB} and \ref{app:proof_ML_FB}.} 

\section{CRLB for The Joint Source Flux and Background Estimation}
\label{sec:CRLB_analysis}
We begin characterizing and analyzing the CRLB for the joint source flux and background estimation. 
For that, we have the following result:

\begin{lema}\label{lm_fisher_joint}
If the astrometry $x_c \in \mathbb{R}$ is fixed and known, and we want to estimate the pair $(\tilde{F}, \tilde{B})$ from $\mathbf{I} \sim   L(\mathbf{I} ; x_{c}, \tilde{F}, \tilde{B})  $ in (\ref{eq:likelihood}), then the Fisher information matrix in Eq.~(\ref{eq:CRLB_psd}) is given by:
\begin{equation}
    \mathcal{I}_{\boldsymbol{\theta}} = \begin{bmatrix}
        \mathcal{I}_{1,1} & \mathcal{I}_{1,2} \\
        \mathcal{I}_{2,1} & \mathcal{I}_{2,2}
    \end{bmatrix} \equiv \begin{bmatrix}
          \sum_{i =1}^n \frac{g_{i}^{2}(x_{c})}{\lambda_{i}(x_{c}, \tilde{F})} &\sum_{i =1}^n  \frac{g_{i}(x_{c})}{\lambda_{i}(x_{c}, \tilde{F})} \\
          \sum_{i =1}^n  \frac{g_{i}(x_{c})}{\lambda_{i}(x_{c}, \tilde{F})} & \sum_{i =1}^n  \frac{1}{\lambda_{i}(x_{c}, \tilde{F})}
      \end{bmatrix} \,.
      \label{eq:FB_fisher}
\end{equation}  
The proof can be found in Appendix~\ref{app:proof_CRLB_FB}.
\end{lema}
Applying Theorem~\ref{theo:CRLB} and Lemma \ref{lm_fisher_joint} in this joint estimation context, the CRLBs applied to every component of $(\tilde{F}, \tilde{B})$ tells us that: 
\begin{equation}
    Var(\hat{\tilde{F}}) 
    \geq \sigma_{\tilde{F}}^{2}
    = \frac{\mathcal{I}_{2,2}}{\mathcal{I}_{1,1} \cdot \mathcal{I}_{2,2} - \mathcal{I}_{1,2}^{2}}
    = \frac{\mathcal{I}_{2,2}}{|\mathcal{I}_{\boldsymbol{\theta}}|} \,,
    \label{eq:FBound}
\end{equation}
\begin{equation}
    Var(\hat{\tilde{B}}) 
    \geq \sigma_{\tilde{B}}^{2} 
    = \frac{\mathcal{I}_{1,1}}{\mathcal{I}_{1,1} \cdot \mathcal{I}_{2,2} - \mathcal{I}_{1,2}^{2}}
    = \frac{\mathcal{I}_{1,1}}{|\mathcal{I}_{\boldsymbol{\theta}}|} \,.
    \label{eq:BBound}
\end{equation}

A key difference between the CRLB for a parameter vector $\boldsymbol{\theta} = (\theta_{1},\dots,\theta_{m})$ in (\ref{eq:CRLB_psd}) and the respective CR bound for the one-dimensional case of estimating just one of these parameters, given complete knowledge of the others, is that the former takes into account how the uncertainty on each parameter shall impact on the others' \citep{kay}. For our joint estimation problem, if we know $\tilde{B}$, the corresponding CRLB for $\tilde{F}$, denoted here as $\sigma_{\tilde{F}, 1D}^{2}$, is:
\begin{equation}
    \sigma_{\tilde{F}, 1D}^{2} = \frac{1}{\mathcal{I}_{1,1}},
    \label{eq:FBound1D}
\end{equation}
where $\mathcal{I}_{1,1}$ is just the Fisher information matrix element described in (\ref{eq:FB_fisher}).\footnote{Analogously, the bound for $\tilde{B}$ when $\tilde{F}$ and all the other parameters are assumed known corresponds to $\sigma_{\tilde{B},1D}^{2} = \frac{1}{\mathcal{I}_{2,2}}$.} The statistical interaction between components is captured 
by the off-diagonal elements of the Fisher Information matrix in (\ref{eq:FBound}) and (\ref{eq:BBound}). In the special case that $\mathcal{I}_{1,2} = 0$, the estimates become decoupled
and one could consider that the joint estimation task reduces to two isolated 1D estimation problems. 
\section{Performance Bounds for high-dimensional Implicit Estimators}
\label{sec:impl_extension}
In this section, we present a result that bounds the performance of any estimator that is the implicit solution of an optimization problem.  
Let us consider an estimator, or regression rule $\tau_J: \mathbb{N}^{n} \to \boldsymbol{\Theta}$, $\tau_{J}(\cdot)~=~[\tau_{J,1}(\cdot) ,...,\tau_{J,p}(\cdot)]$, which is defined by the solution of: 
\begin{equation}
     \tau_{J}(\mathbf{I}) \equiv \underset{\boldsymbol{\alpha} \in \boldsymbol{\Theta}}{\operatorname{argmin}} J(\boldsymbol{\alpha}, \mathbf{I}),\footnote{Note that $\operatorname{argmin}$ refers to the point(s) in domain space that minimize the objective function. More formally, it can defined as $\underset{x \in \mathcal{X}}{\operatorname{argmin}}f(x) \equiv \{\ x \in \mathcal{X} | f(x) \leq f(x'), \forall x' \in \mathcal{X} \}$, for $f: \mathcal{X} \to \mathcal{Y}$.}
     \label{eq:implicit_estimator}
\end{equation}
where $\boldsymbol{\alpha}$ represents a $p$-dimensional parameter vector, and $\mathbf{I}$ is a $n$-dimensional observation vector. We say that this estimator is implicit as no closed-form expression (of the data) is assumed for solving Eq.~(\ref{eq:implicit_estimator}) \citep{fessler1996, lobos2015,  espinosa2018}. Then, this case is difficult because not having an expression for $\tau_{J}(\cdot)$ prevents one to determine its performance.

To address this technical issue, in this work we extend the approach proposed in \citet{fessler1996} and \citet{espinosa2018} to bound both the estimator's bias and variance. In particular, we extend the theory presented in \citet{espinosa2018} 
from the scalar to the challenging multidimensional case.

Let us assume that the cost function $J(\boldsymbol{\alpha}, \mathbf{I})$ has a unique optimal value at $\boldsymbol{\alpha} = \tau_{J}(\mathbf{I})$ such that it also satisfies the first order condition given by: 
\begin{equation}
    \begin{split}
        \mathbf{0} &= \nabla J(\boldsymbol{\alpha}, \mathbf{I})\biggr\rvert_{\boldsymbol{\alpha} = \tau_{J}(\mathbf{I})} = \begin{bmatrix}
             \frac{\partial}{\partial\alpha_{1}}J(\tau_{J}(\mathbf{I}), \mathbf{I}) \\
             \vdots  \\
             \frac{\partial}{\partial\alpha_{p}}J(\tau_{J}(\mathbf{I}), \mathbf{I})
           \end{bmatrix} \,.
    \end{split}
    \label{eq:implicit_foc}
\end{equation}

Under these assumptions, the following result offers bounds on the bias and variance associated to each of the components of an intrinsic estimator vector.
\begin{teo}
\label{theorem1}
     Let $\boldsymbol{\alpha}^{\star}$ be the ground-truth parameter vector and $\tau_{J,j}(\mathbf{I})$ be the estimator solution of Eq.~(\ref{eq:implicit_estimator}). Then, for each $j \in \{1,...,p \}$ we can define three new quantities $\epsilon'_{J,j}$, $\left[ \sigma_{J}^{2} \right]_{(j,j)}$ and $\beta_{J,j}$ such that
    \begin{equation}
    \begin{split}
        |\mathbb{E}_{\mathbf{I} \sim L(\mathbf{I};\boldsymbol{\theta})}  \left\{  \tau_{J,j}(\mathbf{I}) \right\} - \alpha_{j}^{\star}| &\leq \epsilon'_{J,j} \,,
    \end{split}
    \label{eq:pdim_bias_bound}
    \end{equation}
\begin{equation}
    \underbrace{Var(\tau_{J,j}(\mathbf{I}))}_{=[K_{\tau_{J}(\mathbf{I})}]_{(j,j)}} \in \left[ \left[ \sigma_{J}^{2} \right]_{(j,j)} - \beta_{J,j}, \left[ \sigma_{J}^{2} \right]_{(j,j)} + \beta_{J,j} \right] \,.
    \label{eq:pdim_var_interval}
\end{equation}

\end{teo}

To keep the content direct and focus, the closed-form expressions that determine the bias and variance bounds in (\ref{eq:pdim_bias_bound}) and (\ref{eq:pdim_var_interval}), {\em i.e.}, $\epsilon'_{J,j}$, $\left[ \sigma_{J}^{2} \right]_{(j,j)}$ and $\beta_{J,j}$, are presented in Appendix~\ref{app:proof_pdim_bounds} in conjunction with the proof of this result.

Interpretation of Theorem \ref{theorem1}:
\begin{itemize}
\item Theorem~\ref{theorem1} provides
sufficient conditions to bound, for each $j \in \{1,...,p \}$, the bias and variance of $\tau_{J,j}(\mathbf{I})$. 
The performance bounds presented here do not depend on the exact value of $\tau_{J}(\mathbf{I})$, but on its first- and second-order derivatives (see details in Appendix~\ref{app:proof_pdim_bounds}). The calculation of these derivatives is critical to apply this result and heavily relies on the use of the operator presented in Section \ref{sub_sec_tensor}. This  calculation will be the focus of the following two sections when applying Theorem~\ref{theorem1} to our joint flux and background estimation task.  

\item For the mentioned application, it is worth presenting here the analytic form of $\left[ \sigma_{J}^{2} \right]_{(j,j)}$ given by (see Appendix~\ref{app:proof_pdim_bounds})
\begin{equation}
    \sigma_{J}^{2} \equiv \tau_{J}'(\bar{\mathbf{I}}) \cdot K_{\mathbf{I}} \cdot \left[ \tau_{J}'(\bar{\mathbf{I}}) \right]^{\dagger} \,.
    \label{eq:pdim_interval_center}
\end{equation}
$\tau_{J}'(\bar{\mathbf{I}})$ in (\ref{eq:pdim_interval_center}) is a matrix obtained as (see Section \ref{sub_sec_tensor}): 
\begin{equation}
    \tau_{J}'(\mathbf{I}) \equiv - \left[ \nabla^{20} J(\tau_{J}(\mathbf{I}), \mathbf{I}) \right]^{-1} \cdot \left[ \nabla^{11} J(\tau_{J}(\mathbf{I}), \mathbf{I}) \right] \,.
    \label{eq:pdim_tau_tilde1}
\end{equation}

\item It is worth noting that
if $\frac{\beta_{J,j}}{\left[ \sigma_{J}^{2} \right]_{(j,j)}}\ll 1$, then Theorem~\ref{theorem1} implies a tight interval, which means that 
$Var(\tau_{J,j}(\mathbf{I})) \approx \left[ \sigma_{J}^{2} \right]_{(j,j)}$. In this tight regime, we could compare $\left[ \sigma_{J}^{2} \right]_{(j,j)}$ with the CRLB to evaluate the optimality of this implicit estimator.   

\item 
Finally, for the one-dimensional case, {\em i.e.}, $p~=~1$, we recover the bounds presented in \citep{espinosa2018}. Therefore, Theorem~\ref{theorem1} offers a non-trivial multidimensional extension of \citet[Theorem 1]{espinosa2018}.
\end{itemize}

In the following two sections, we move to the application of this result to our main estimation task, Sections \ref{sec:WLSE} and \ref{sec:MLE}. Importantly, we will show that the main assumptions in Eq.~(\ref{eq:implicit_foc}) are satisfied. 

\section{WLS Estimator for Joint Source Flux and Background}
\label{sec:WLSE}
 In this section, we analyze the bias and variance of the WLS estimator in the context of joint point-source photometry and background estimation. The appeal of this estimator resides in its inherent simplicity, its wide adoption and the evidence that the Least-Squares estimator is optimal under low SNR conditions, given the knowledge of $\tilde{B}$ \citep{perryman1989}. 

Let $\{ w_{i} \}_{i=1}^{n}$ be a set of 
weights, the WLS cost function is given by:
\begin{equation}
    J_{WLS}(\boldsymbol{\alpha}, \mathbf{I}) = \sum_{i =1}^n w_{i}(I_{i} - \alpha_{1} \cdot g_{i}(x_{c}) - \alpha_{2})^{2} \,,
    \label{eq:WLS_cost}
\end{equation}
where $\boldsymbol{\alpha} \in \boldsymbol{\Theta}$. Then, the WLS estimator $\tau_{WLS}(\mathbf{I})$ is the solution of: 
\begin{equation}
    \tau_{WLS}(\mathbf{I}) \equiv \underset{\boldsymbol{\alpha} \in \boldsymbol{\Theta}}{\operatorname{argmin}} J_{WLS}(\boldsymbol{\alpha}, \mathbf{I}) \,.
    \label{eq:WLS_tau_imp}
\end{equation}

It is well known that the solution of Eq.~(\ref{eq:WLS_tau_imp}) is a linear estimator w.r.t. $\mathbf{I}$. In this linear context, Theorem~\ref{theorem1} can be applied to obtain that: 
\begin{teo}\label{Th_WLS}
Let us consider $\tau_{WLS}(\mathbf{I})$ in Eq.~(\ref{eq:WLS_tau_imp}). Its mean $\bar{\tau}_{WLS}$ and covariance matrix $K_{\tau_{WLS}(\mathbf{I})}$ can be precisely expressed as:
\begin{equation}
    \begin{split}
        \bar{\tau}_{WLS}=\mathbb{E}_{\mathbb{I} \sim L(\mathbf{I};\boldsymbol{\theta})}  \left\{  \tau_{WLS}(\mathbf{I}) \right\} = \tau_{WLS}'(\bar{\mathbf{I}}) \cdot \bar{\mathbf{I}}
    \end{split} \,,
    \label{eq:mean_WLSE}
\end{equation}
\begin{equation}
    K_{\tau_{WLS}(\mathbf{I})} = \tau_{WLS}'(\bar{\mathbf{I}}) \cdot K_\mathbf{I} \cdot \left[ \tau_{WLS}'(\bar{\mathbf{I}}) \right]^{\dagger} \,,
    \label{eq:Cov_WLSE}
\end{equation}
with
\begin{equation}
    \begin{split}
      \tau'_{WLS}(\bar{\mathbf{I}})  &= -[\nabla^{20} J_{WLS}(\tau_{WLS}(\bar{\mathbf{I}}),\bar{\mathbf{I}})]^{-1} \\
      &\hspace{3mm} \cdot \nabla^{11} J_{WLS}(\tau_{WLS}(\bar{\mathbf{I}}),\bar{\mathbf{I}})
    \end{split} \,,
    \label{eq:WLS_tau_exp}
\end{equation}
$\nabla^{20} J_{WLS}(\tau_{WLS}(\bar{\mathbf{I}}),\bar{\mathbf{I}}) =$
\begin{equation}
     2\begin{bmatrix}
            \sum_{i =1}^n w_{i} \cdot g_{i}^{2}(x_{c}) & \sum_{i =1}^n w_{i} \cdot g_{i}(x_{c}) \\
            \sum_{i =1}^n w_{i} \cdot g_{i}(x_{c}) & \sum_{i =1}^n w_{i}
        \end{bmatrix} \,,
\end{equation}
and $\nabla^{11} J_{WLS}(\tau_{WLS}(\bar{\mathbf{I}}),\bar{\mathbf{I}}) =$
\begin{equation}
     2\begin{bmatrix}
            w_{1} \cdot g_{1}(x_{c}) & \dots & w_{n} \cdot g_{n}(x_{c}) \\
            w_{1} & \dots & w_{n}
        \end{bmatrix} \,.
\end{equation}
\end{teo}

The proof of this result is presented in Appendix~\ref{app:proof_WLS_FB}. 

Given the expression in (\ref{eq:WLS_cost}), 
we can find a closed expression for $\tau_{WLS}(\mathbf{I})$. While the utilization of Theorem~\ref{theorem1} may not be strictly needed,
\footnote{However, closed forms performances can be derived from Theorem~\ref{theorem1}, which are consistent with those of Eqs.~(\ref{eq:mean_WLSE}) and~(\ref{eq:Cov_WLSE}). See Appendix~\ref{app:proof_WLS_FB} for more details.} its applicability becomes relevant for a scenario where the weights are data-driven. In the following subsections, we motivate and introduce a stochastic version of the WLS estimator, where Theorem~\ref{theorem1} proves to be a valuable tool.

\subsection{The Optimal (Oracle) Weight Selection}\label{sec:oracle_WLS}
The weight selection for Eq.~(\ref{eq:WLS_cost}) is a critical design consideration. 
From Theorem \ref{Th_WLS}, we can show that if the weights are chosen such that:
\begin{equation}
    w_{i} = \frac{K}{\lambda_{i}(x_{c}, \tilde{F})}, \hspace{1mm} \forall i \in \{1,\dots,n\} \,,
    \label{eq:opt_weights}
\end{equation}
for some $K>0$, $K_{\tau_{WLS}(\mathbf{I})}$ in Eq.~(\ref{eq:Cov_WLSE}) matches the inverse of the Fisher Information matrix in Eq.~(\ref{eq:FB_fisher}) (details are presented in Appendix~\ref{app:proof_WLS_FB}). Adding up the fact that the WLS estimator with Eq.~(\ref{eq:opt_weights}) is unbiased, we conclude that this WLS is optimal in the minimum variance sense. However, this is an oracle selection because it requires knowledge of the parameters to be estimated (see Eq.~(\ref{eq:opt_weights})), which contradicts the very essence of the inference task. Inspired by this ideal but impractical solution, an alternative data-driven design for the weights is presented next.

\subsection{The SWLS Estimator}
\label{sub_sec_SWLS}
We consider a proxy of the oracle weight set given by:~\footnote{This selection is similar to what is done in some popular photometric packages, like DAOPHOT \citep{1987PASP...99..191S}.}
\begin{equation}
    \hat{w}_{i}(I_{i}) = \frac{1}{I_{i}} \,, \hspace{3mm} \forall i = 1,\dots,n \,.
    \label{eq:SWLSE_weights}
\end{equation}
This data-driven solution is referred in this work as Stochastic WLS (SWLS). Indeed, the selection in Eq.~(\ref{eq:SWLSE_weights}) can be seen as a noisy version of Eq. (\ref{eq:opt_weights}) considering that $I_i$ is a random variable where $\mathbb{E} \left\{ I_i \right\}=\lambda_{i}(x_{c}, \tilde{F})$. 
%
By replacing $\{ \hat{w}_{i}(I_{i}) \}_{i =1}^n$ into Eq.~(\ref{eq:WLS_cost}), we have that: 
\begin{equation}
    J_{SWLS}(\boldsymbol{\alpha}, \mathbf{I}) = \sum_{i =1}^n \frac{1}{I_{i}}(I_{i} - \alpha_{1} \cdot g_{i}(x_{c}) - \alpha_{2})^{2} \,,
    \label{eq:SWLS_cost}
\end{equation}
and the SWLS estimator is defined as in Eq.~(\ref{eq:SWLS_exp}).


\begin{table*}
    \centering
    \begin{minipage}{0.75\textwidth}
    \begin{align}
        \begin{split}
        \tau_{SWLS}(\mathbf{I}) &= \underset{\boldsymbol{\alpha} \in \boldsymbol{\Theta}}{\operatorname{argmin}} J_{SWLS}(\boldsymbol{\alpha}, \mathbf{I}) \\
        &= 
        \begin{bmatrix}
            \sum_{i =1}^n \frac{1}{I_{i}} \cdot g_{i}^{2}(x_{c}) & \sum_{i =1}^n \frac{1}{I_{i}} \cdot g_{i}(x_{c}) \\
            \sum_{i =1}^n \frac{1}{I_{i}} \cdot g_{i}(x_{c}) & \sum_{i =1}^n \frac{1}{I_{i}}
        \end{bmatrix}^{-1} \cdot
        \begin{bmatrix}
            \frac{1}{I_{1}} \cdot g_{1}(x_{c}) & \dots & \frac{1}{I_{n}} \cdot g_{n}(x_{c}) \\
            \frac{1}{I_{1}} & \dots & \frac{1}{I_{n}}
        \end{bmatrix} \cdot \mathbf{I}.
    \end{split}
    \label{eq:SWLS_exp}
    \end{align}
    \end{minipage}
\vspace{1.5mm}\hrule
\end{table*}

Importantly, $\tau_{SWLS}(\cdot)$ is non-linear w.r.t. $\mathbf{I}$, which justifies the non-trivial adoption of  Theorem \ref{theorem1} to determine its performance (bias and variance bounds). 
From the expressions in Eqs.~(\ref{eq:SWLS_cost}) and~(\ref{eq:SWLS_exp}), the application of Theorem~\ref{theorem1} reduces to calculate the 
$\nabla^{ij}(\cdot)$ operators (see Section \ref{sub_sec_tensor})  on the cost function $J_{SWLS}(\cdot,\cdot)$, 
yielding the following result:

\begin{teo}
\label{th_swls}
Let us consider $\tau_{SWLS}(\mathbf{I})$ in Eq. (\ref{eq:SWLS_exp}). For any $j \in \{1,2\}$, we can determine $\epsilon'_{SWLS,j}>0$ and $\beta_{SWLS,j}$ such that:~\footnote{These expressions are elaborated in the Appendix~\ref{app:proof_WLS_FB}.}
\begin{equation}
    \begin{split}
        |\mathbb{E} \left\{  \tau_{SWLS,j}(\mathbf{I}) \right\} - \alpha_{j}^{\star}| &\leq \epsilon'_{SWLS,j}
    \end{split} \,,
    \label{eq:SWLS_bias_bound}
\end{equation}
\begin{align}
    &Var(\tau_{SWLS,j}(\mathbf{I})) \in \nonumber\\ 
    &\left[ \left[ \sigma_{SWLS}^{2} \right]_{(j,j)} - \beta_{SWLS,j}, \left[ \sigma_{SWLS}^{2} \right]_{(j,j)} + \beta_{SWLS,j} \right],
    \label{eq:SWLS_var_interval}
\end{align}
with
\begin{equation}
   \sigma_{SWLS}^{2} = \tau_{SWLS}'(\bar{\mathbf{I}}) \cdot K_{\mathbf{I}} \cdot \left[ \tau_{SWLS}'(\bar{\mathbf{I}}) \right]^{\dagger},
    \label{eq:Cov_SWLS}
\end{equation}
\begin{equation}
    \begin{split}
      \tau'_{SWLS}(\bar{\mathbf{I}})   &= -[\nabla^{20} J_{SWLS}(\tau_{SWLS}(\bar{\mathbf{I}}),\bar{\mathbf{I}})]^{-1} \\
      &\hspace{3mm} \cdot \nabla^{11} J_{SWLS}(\tau_{SWLS}(\bar{\mathbf{I}}),\bar{\mathbf{I}}) 
    \end{split},
    \label{eq:SWLS_tau_exp}
\end{equation}
where
\begin{equation}
    \begin{split}
        \nabla^{11} J_{SWLS}(\tau(\bar{\mathbf{I}}),\bar{\mathbf{I}}) &= 2\begin{bmatrix}
            \frac{-g_{1}(x_{c})}{\lambda_{1}(x_{c},\tilde{F})} & \cdots & \frac{-g_{n}(x_{c})}{\lambda_{n}(x_{c},\tilde{F})} \\
            \frac{-1}{\lambda_{1}(x_{c},\tilde{F})} & \cdots & \frac{-1}{\lambda_{n}(x_{c},\tilde{F})} 
        \end{bmatrix} \,,
    \end{split}
\end{equation}
\begin{equation}
    \begin{split}
        \nabla^{20} J_{SWLS}(\tau(\bar{\mathbf{I}}),\bar{\mathbf{I}}) &= 2\begin{bmatrix}
            \sum_{i =1}^n \frac{g_{i}^{2}(x_{c})}{\lambda_{i}(x_{c},\tilde{F})} & \sum_{i =1}^n \frac{g_{i}(x_{c})}{\lambda_{i}(x_{c},\tilde{F})} \\
            \sum_{i =1}^n \frac{g_{i}(x_{c})}{\lambda_{i}(x_{c}\tilde{F})} & \sum_{i =1}^n \frac{1}{\lambda_{i}(x_{c},\tilde{F})}
        \end{bmatrix}
    \end{split}.
\end{equation}
\end{teo}

The proof is presented in Appendix~\ref{app:proof_WLS_FB}.

From (\ref{eq:SWLS_var_interval}), we can interpret the scalar $\left[ \sigma_{SWLS}^{2} \right]_{(j,j)}~>~0$ (diagonal elements of $\sigma_{SWLS}^{2}$ in Eq.~(\ref{eq:Cov_SWLS})) as the nominal value that predicts the variance of $\tau_{SWLS,j}(\mathbf{I})$. Importantly, we have that (see the proof in Appendix~\ref{app:proof_WLS_FB})
\begin{equation}
    \begin{split}
       \sigma_{SWLS}^{2} &= \tau_{SWLS}'(\bar{\mathbf{I}}) \cdot K_{\mathbf{I}} \cdot \left [ \tau_{SWLS}'(\bar{\mathbf{I}}) \right]^{\dagger}  \\
       &= \mathcal{I}_{\boldsymbol{\alpha^{\star}}}^{-1} 
    \end{split}
    \label{eq:SWLSVar_centered}
\end{equation}
which is the CRLB of this joint estimation problem. 
In other words, $Var(\hat{\tilde{F}}_{SWLS})= Var(\tau_{SWLS,1}(\mathbf{I}))$  is centered around $\left[ \mathcal{I}_{\boldsymbol{\alpha}^{\star}}^{-1} \right]_{(1,1)}$  
 and $Var(\hat{\tilde{B}}_{SWLS})= Var(\tau_{SWLS,2}(\mathbf{I}))$  is centered around $\left[ \mathcal{I}_{\boldsymbol{\alpha}^{\star}}^{-1} \right]_{(2,2)}$, which are optimal minimum variance bounds of the problem (see Lemma~\ref{lm_fisher_joint} and Theorem~\ref{theo:CRLB}). This result is meaningful in the regime when $\epsilon'_{SWLS,j}\approx 0$ ({\em i.e.}, we have unbiased estimators) and $\frac{\beta_{SWLS,j}}{\left[ \sigma_{SWLS}^{2} \right]_{(j,j)}}\approx 0 \Rightarrow  Var(\tau_{SWLS,j}(\mathbf{I})) \approx \left[ \sigma_{SWLS}^{2} \right]_{(j,j)}$ where Theorem \ref{th_swls} predicts that the joint SWLS estimator in (\ref{eq:SWLS_exp}) is optimal as it achieves the CRLB of the problem. In our next numerical analysis, we show that this optimality condition is met for a wide range of settings.



\subsection{Numerical Analysis}
Here, we test our SWLS approach on different case studies. In particular, we focus our analysis in two points: we compare the bounds obtained for this estimator with its empirical performance (based on realistic numerical simulations), while studying how these curves evolve as the photometric aperture on the source increases.
This bifocal study tackles both the validation of our bounds for the SWLS estimator 
(in Eqs.~(\ref{eq:SWLS_bias_bound}) and~(\ref{eq:SWLS_var_interval})), and the feasibility of achieving the optimal performance (given by the CRLB in Eq.~(\ref{eq:FB_fisher})) of the joint inference discussed in Section \ref{sec:CRLB_analysis}.

The experimental setup (source, device and sky conditions) considered here and in forthcoming sections, and which we denote as ``baseline case", is summarized in Table~\ref{tab:baseline}. The $dither$ parameter denotes an offset between the $n_{x_{c}}^{th}$ pixel's center and the actual position of the point source $x_{c}$ \citep{mendez2013, mendez2014}, measured in pixels.


\begin{table}[h]
\caption{Characterization of the baseline case.}
\centering
\begin{tabular}{|c|c|c|}
    \hline
    Parameter & Value & Units \\
    \hline\hline
    $D$ & 0 & [photo-e$^{-}$] \\
    $RON$ & 5 & [photo-e$^{-}$] \\
    $G$ & 2 & [photo-e$^{-}$/ADU] \\
    $\Delta x$ & 0.2 & [arcsec] \\
    $n$ & 41 & [pixels] \\
    $f_{s}$ & 1502.5 & [ADUs/arcsec] \\
    $n_{x_{c}}$ & 21 & - \\
    $dither$ & 0 & [pixels] \\
    $FWHM$ & 1.0 & [arcsec] \\
    $\tilde{F}$ & 10002 & [photo-e$^{-}$] \\
    \hline
\end{tabular}
\label{tab:baseline}
\end{table}

 Other simulated scenarios consist of variations of this baseline case, in which only one parameter is modified at once. Such study cases are described in Table~\ref{tab:scenarios} and denoted in forthcoming figures by the modified parameter and its new value.

\begin{table}[h]
\caption{Characterization of other study cases. These derive from the baseline scenario by modifying the value of a single parameter.}
\centering
\begin{tabular}{|c|c|c|c|}
    \hline
    Scenario  & \vtop{\hbox{\strut Modified}\hbox{\strut Parameter}} & Value & Units \\
    \hline\hline
    1 & $\tilde{F}$ & 20004 & [photo-e$^{-}$]\\ 
    2 & $f_{s}$ & 3005.0 & [ADUs/arcsec] \\
    3 & $FWHM$ & 1.5 & [arcsec] \\
    4 & $dither$ & 0.125 & [pixels] \\
    5 & $\tilde{F}$ & 1000 & [photo-e$^{-}$] \\
    \hline
\end{tabular}
\label{tab:scenarios}
\end{table}

It is worth noting that, as one of the cornerstones of this work is to study how the statistical behavior of different estimation schemes is affected by aperture, the experimental configurations considered here can not be characterized by a particular SNR value, as it itself is a function of aperture. However, some insight on SNR ranges for each scenario can be found in Appendix~\ref{app:SNR}.

\subsubsection{Bias Bounds}
\label{sec:SWLS_bias}
\begin{figure}[]
    \centering
    \includegraphics[width=0.45\textwidth]{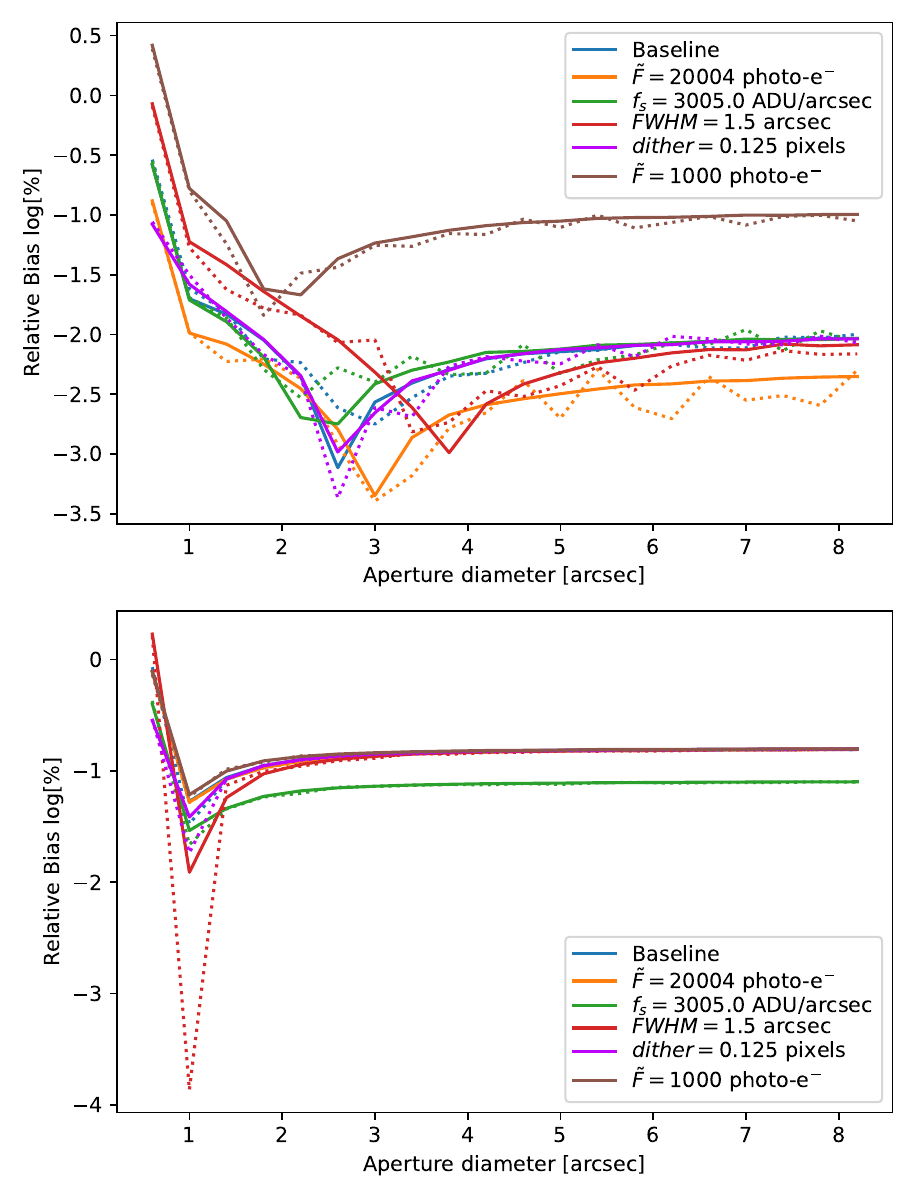}
    \caption{Relative biases for joint source flux (top) and background (bottom) estimation using the SWLS estimator, under different observational scenarios, for an aperture selection scheme. Solid lines denote the corresponding bounds as given by Equation~(\ref{eq:SWLS_bias_bound}), while dotted lines are employed for empirical bias estimates based on Eqs.~(\ref{eq:WLS_tau_imp}) and~(\ref{eq:SWLSE_weights}), and the known ground-truth. High-resolution images of this and all the figures can be found at \url{https://doi.org/10.5281/zenodo.10056229}.}
    \label{fig:SWLS_relBias_emp}
\end{figure}

Unbiasedness is a key attribute in the scope of this performance analysis. For that reason, Figure~\ref{fig:SWLS_relBias_emp} shows $\epsilon'_{SWLS,j}$ (see Eq.~(\ref{eq:SWLS_bias_bound})), displayed as relative bias, {\em i.e.} $\log{\left(100 \times \frac{\epsilon'_{J,j}}{\alpha_{j}}\right)}$ for $j \in \{1,2\}$.\footnote{Analogously, actual empirical bias expressed as relative bias is $\log{\left( 100 \times \frac{|\mathbb{E}_{\mathbf{I} \sim L(\mathbf{I};\boldsymbol{\theta})}  \left\{  \tau_{J,j}(\mathbf{I}) \right\} - \alpha_{j}^{\star}|}{\alpha_{j}^{\star}} \right)} = \log{\left( 100 \times \frac{|\bar{\tau}_{J,j} - \alpha_{j}^{\star}|}{\alpha_{j}^{\star}} \right)}$.}
As it could be expected (due to lack of information in terms of the number of pixels and PSF coverage) small apertures show relatively large values of $\epsilon'_{SWLS,j}$ (our bias bound) as well as empirical bias ranging from $0.1\%$ up to $3\%$ for $\hat{\tilde{F}}$, depending on the weakness of the source (SNR). It is interesting to note that for apertures up to $2-3$ times the FWHM both curves rapidly decrease. After that point, the bias bounds slightly increase in a smooth manner. A similar behavior is perceived for the background estimates, with the difference that its transient phase seems to be less dependent on the PSF width.

When comparing the curves for each study case, we can establish that our bounds ({\em i.e.}, $\epsilon'_{SWLS,j}$) successfully predict the estimates' measured bias, though very tightly, independently of the adopted aperture. From that, and the values attained by the curves, the SWLS solutions can be considered to be unbiased for practically any reasonable aperture regime.


\subsubsection{Variance Bounds}
\label{sec:SWLS_var}
As the SWLS estimator is shown to be unbiased, contrasting its MSE performance  against the CRLB becomes meaningful. For that matter, it is convenient to define a measure of discrepancy between the joint CRLB and the estimator's variance, as given by Eqs.~(\ref{eq:sigma_m}), (\ref{eq:mDiscrepancy_WLS}) and (\ref{eq:BDiscrepancy_WLS}).\footnote{For visualization, the variances of the  source's flux are expressed in terms of magnitudes. The corresponding standard deviation (SD) of Eq.~(\ref{eq:sigma_m}) is computed as in \citet{mendez2014}.}





\begin{table*}
\centering
\begin{minipage}{0.75\textwidth}
    \begin{equation}
        \sigma_{m} = \frac{2.5}{2} \cdot \left( \log{(\tilde{F} + \sigma_{\tilde{F}}) - \log{(\tilde{F} - \sigma_{\tilde{F}})}} \right) \,,
        \label{eq:sigma_m}
    \end{equation}

    \begin{equation}
        \Delta \sigma_{m} = \frac{\frac{2.5}{2} \cdot \left( \log{(\tilde{F} + \sqrt{Var(\tau_{J,1}(\mathbf{I}))}) - \log{(\tilde{F} - \sqrt{Var(\tau_{J,1}(\mathbf{I}))})}} \right) - \sigma_{m}}{\sigma_{m}} \times 100 \,,
        \label{eq:mDiscrepancy_WLS}
    \end{equation}

    \begin{equation}
        \Delta \sigma_{\tilde{B}} = \frac{\sqrt{Var(\tau_{J,2}(\mathbf{I}))} - \sigma_{\tilde{B}}}{\sigma_{\tilde{B}}} \times 100 \,.
        \label{eq:BDiscrepancy_WLS}
    \end{equation}
\end{minipage}
\vspace{1.5mm}\hrule
\end{table*}

\begin{figure}[!h]
    \centering
    \includegraphics[width=0.45\textwidth]{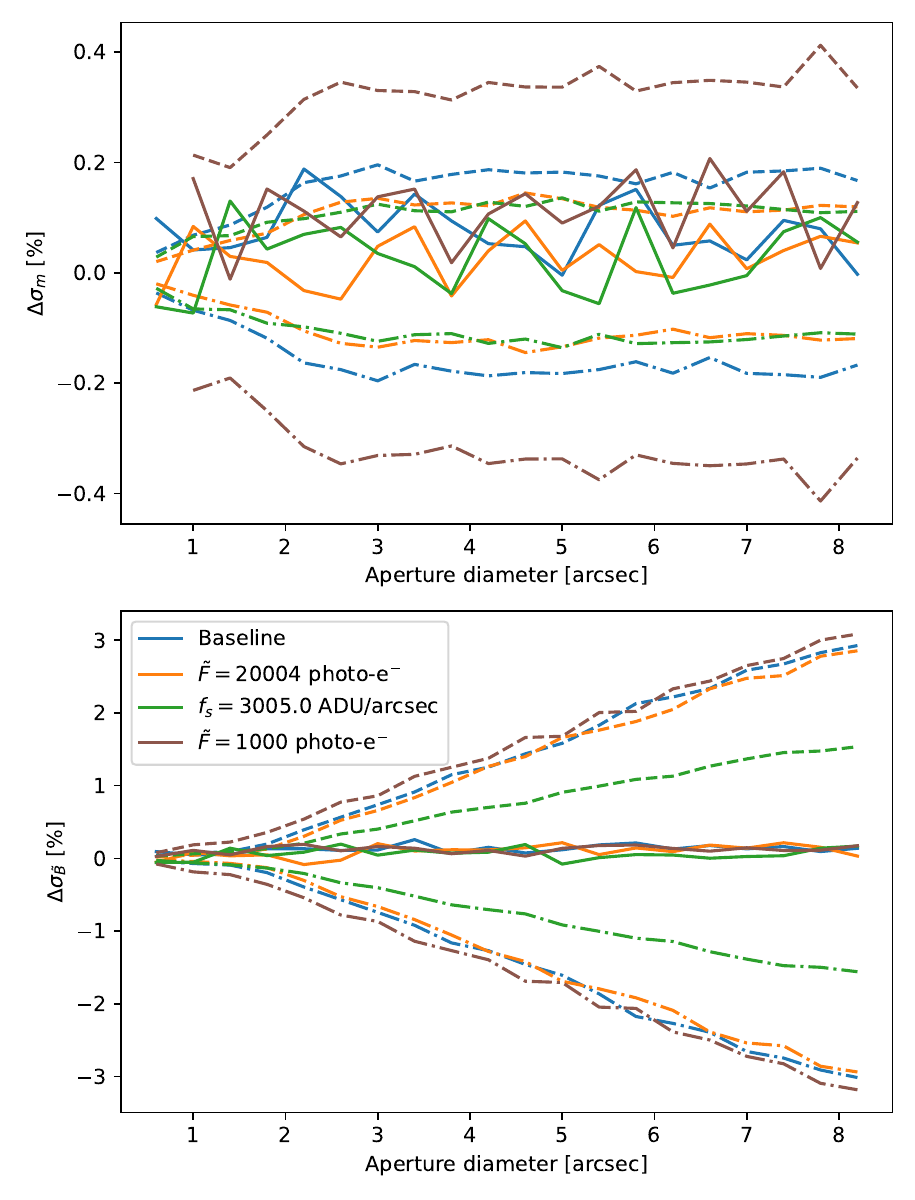}
    \caption{Performance discrepancies of the SWLS estimator for joint source flux (top) and background (bottom) estimation under different observational scenarios, for an aperture selection scheme. Dashed-dotted-, dashed- and solid lines denote the lower bound, the upper bound (based on the CRLB and Eqs.~(\ref{eq:sigma_m}) to~(\ref{eq:BDiscrepancy_WLS}), and the empirical variance estimate (based on the known ground-truth), respectively. 
    }
    \label{fig:SWLS_var_bounds_cases}
\end{figure}

As can be seen in Figure~\ref{fig:SWLS_var_bounds_cases}, the variance of the estimator lies between the upper and lower bounds predicted by Theorem~\ref{th_swls}, for both $\hat{\tilde{F}}$ and $\hat{\tilde{B}}$ and in all the cases studied here (even those not included due to space restriction).
Interestingly, the bounds corresponding to $\tau_{SWLS,1}({\bf I})$ (for the source flux) do not vary much as the aperture increases as the interval length $2\beta_{SWLS,1}$ in Eq.(\ref{eq:SWLS_var_interval}) remains relatively stable. This observation suggests that our result predicts that the performance of the SWLS is robust with respect to the choice of the aperture size. It is also worth noticing that each curve corresponds to a relative measure with respect to the ever decreasing CR bounds ($\sigma_{m}$ and $\sigma_{\tilde{B}}$). Therefore, steady discrepancy translates into higher precision as the aperture area widens.

On the other hand, the length of the interval in which $Var(\tau_{SWLS,2}({\bf I}))$ (for the background) can be found (see Eq.~(\ref{eq:SWLS_var_interval})) becomes greater and greater with aperture, almost in a linear regime. Regardless, the empirical estimates tend to be closer to zero, {\em i.e.} to the fundamental limit of the CRLB itself (see Eq.~(\ref{eq:SWLSVar_centered})). This hints that the predictive power of each bound alone diminishes with aperture, but empirical variances (solid lines in Figure~\ref{fig:SWLS_var_bounds_cases}) suggest a good, near-optimal performance in practice, which could be approximated as the mean of the bounds, or simply by $\left[ \sigma_{SWLS}^{2} \right]_{(j,j)}$.


In light of these results, our joint SWLS estimation strategy is validated (with our performance bounds and empirical evidence), showing solid evidence that joint estimation can be done efficiently (quasi-optimal, see Sections \ref{sec:comparison} and \ref{sec:TESS} for more details in realistic scenarios), while exploiting all the pixel information available for the task without the need to mask out pixels. The next section consolidates this conclusion showing another important nearly optimal joint estimation strategy for our problem. 

\section{ML Estimator for Joint Source Flux and Background}
\label{sec:MLE}
The ML estimator has been widely used for parameter estimation due to its asymptotic optimality \citep{kay}. For that reason, it has been used by the astronomical community  \citep{lindegren2008, gai2017, espinosa2018}, and many other research areas facing similar image inverse problems such as fluorescence microscopy \citep{Abraham2009, superresolution2014}.

The likelihood function of $\mathbf{I}$ given the source's relative position $x_{c}$ and its photometry $\boldsymbol{\alpha} \in \boldsymbol{\Theta}$ (see Eqs.~(\ref{eq:lambda_i}) and ~(\ref{eq:likelihood})), can be expressed as:
\begin{equation}
    L(\mathbf{I};x_{c},\boldsymbol{\alpha}) = \prod_{i =1}^n \frac{e^{-\left( \alpha_{1} \cdot g_{i}(x_{c}) + \alpha_{2} \right)} \cdot \left( \alpha_{1} \cdot g_{i}(x_{c}) + \alpha_{2} \right)^{I_{i}}}{I_{i}!} \,.
\end{equation}
Then, the ML estimator is given by:
\begin{equation}
    \tau_{ML}(\mathbf{I}) = \underset{\boldsymbol{\alpha} \in \boldsymbol{\Theta}}{\operatorname{argmin}} \underbrace{J_{ML}(\boldsymbol{\alpha}, \mathbf{I})}_{= -\ln{L(\mathbf{I};x_{c},\boldsymbol{\alpha})}}.
    \label{eq:ML_tau}
\end{equation}

$\tau_{ML}(\mathbf{I})$ is an implicitly defined estimator and, consequently,
Theorem~\ref{theorem1} is instrumental in this context. 
Interestingly, we have that $J_{ML}(\boldsymbol{\alpha}, \mathbf{I})$ is convex over the parameter space (see Appendix~\ref{app:proof_ML_FB} for the proof), then its optimization yields a unique solution. 
The adoption of Theorem~\ref{theorem1} reduces to compute 
the expressions in (\ref{eq:pdim_bias_bound}) and (\ref{eq:pdim_var_interval}) for $\tau_{ML}(\mathbf{I})$. 
The result is the following: 
\begin{teo}
\label{th_ML}
Let us consider the ML estimator $\tau_{ML}(\mathbf{I})$ in Eq.~(\ref{eq:ML_tau}). For any $j \in \{1,2\}$, we can determine $\epsilon'_{ML,j}>0$ and $\beta_{ML,j}$ such that:~\footnote{These expressions are elaborated in the Appendix~\ref{app:proof_ML_FB}.}
\begin{equation}
    \begin{split}
        |\mathbb{E} \left\{  \tau_{ML,j}(\mathbf{I}) \right\} - \alpha_{j}^{\star}| &\leq \epsilon'_{ML,j}
    \end{split} \,,
    \label{eq:ML_bias_bound}
\end{equation}
\begin{align}
    &Var(\tau_{ML,j}(\mathbf{I})) \in \nonumber\\ 
    &\left[ \left[ \sigma_{ML}^{2} \right]_{(j,j)} - \beta_{ML,j}, \left[ \sigma_{ML}^{2} \right]_{(j,j)} + \beta_{ML,j} \right],
    \label{eq:ML_var_interval}
\end{align}
with
\begin{equation}
   \sigma_{ML}^{2} \equiv \tau_{ML}'(\bar{\mathbf{I}}) \cdot K_{\mathbf{I}} \cdot \left[ \tau_{ML}'(\bar{\mathbf{I}}) \right]^{\dagger} \,,
    \label{eq:Cov_ML}
\end{equation}
\begin{equation}
    \begin{split}
      \tau'_{ML}(\bar{\mathbf{I}})   &= -[\nabla^{20} J_{ML}(\tau_{ML}(\bar{\mathbf{I}}),\bar{\mathbf{I}})]^{-1} \\
      &\hspace{3mm} \cdot \nabla^{11} J_{ML}(\tau_{ML}(\bar{\mathbf{I}}),\bar{\mathbf{I}}) 
    \end{split} \,,
    \label{eq:ML_tau_exp}
\end{equation}
where
\begin{equation}
    \begin{split}
        \nabla^{11} J_{ML}(\tau(\bar{\mathbf{I}}),\bar{\mathbf{I}}) &= \begin{bmatrix}
            \frac{-g_{1}(x_{c})}{\lambda_{1}(x_{c},\tilde{F})} & \cdots & \frac{-g_{n}(x_{c})}{\lambda_{n}(x_{c},\tilde{F})} \\
            \frac{-1}{\lambda_{1}(x_{c},\tilde{F})} & \cdots & \frac{-1}{\lambda_{n}(x_{c},\tilde{F})} 
        \end{bmatrix}
    \end{split} \,,
\end{equation}
\begin{equation}
    \begin{split}
        \nabla^{20} J_{ML}(\tau(\bar{\mathbf{I}}),\bar{\mathbf{I}}) &= \begin{bmatrix}
            \sum_{i =1}^n \frac{g_{i}^{2}(x_{c})}{\lambda_{i}(x_{c},\tilde{F})} & \sum_{i =1}^n \frac{g_{i}(x_{c})}{\lambda_{i}(x_{c},\tilde{F})} \\
            \sum_{i =1}^n \frac{g_{i}(x_{c})}{\lambda_{i}(x_{c}\tilde{F})} & \sum_{i =1}^n \frac{1}{\lambda_{i}(x_{c},\tilde{F})}
        \end{bmatrix} 
    \end{split} \,,
\end{equation}
\end{teo}

The proof of this result is presented in Appendix \ref{app:proof_ML_FB}.

Similar to the analysis made for the SWLS in Section \ref{sub_sec_SWLS}, we have in this context that: 
\begin{equation}
    \begin{split}
       \sigma_{ML}^{2} &= \tau_{ML}'(\bar{\mathbf{I}}) \cdot K_{\mathbf{I}} \cdot \left [ \tau_{ML}'(\bar{\mathbf{I}}) \right]^{\dagger}  \\
       &= \mathcal{I}_{\boldsymbol{\alpha^{\star}}}^{-1}.
    \end{split}
    \label{eq:MLVar_centered}
\end{equation}
Then again, the central (nominal) value predicted for the variance of the ML estimator in~(\ref{eq:ML_var_interval}) is precisely the CRLB.
%
This fact is very interesting because when $\epsilon'_{ML,j}\approx 0$ and 
$\frac{\beta_{ML,j} }{\left[ \sigma_{ML}^{2} \right]_{(j,j)}} \approx 0$, we can predict from Theorem~\ref{th_ML} that the performance
of the ML estimator achieves the optimal CRLB, as hinted by Eq.~(\ref{eq:MLVar_centered}). 
Remarkably, the next numerical analysis proves that this optimality condition is met for a wide range of observational settings. 

\subsection{Numerical Analysis}
In this section, we review the behavior of the performance bounds, as developed in Section \ref{sec:impl_extension}, related to the ML estimator in a more exhaustive manner, as well as its empirical exactness and precision relative to the CR bounds. The issue is addressed relying on the same study cases derived from the so called ``baseline case" described in previous sections.

\subsubsection{Bias Bounds}
\label{sec:ML_bias_theo}
\begin{figure}[!h]
    \centering
    \includegraphics[width=0.45\textwidth]{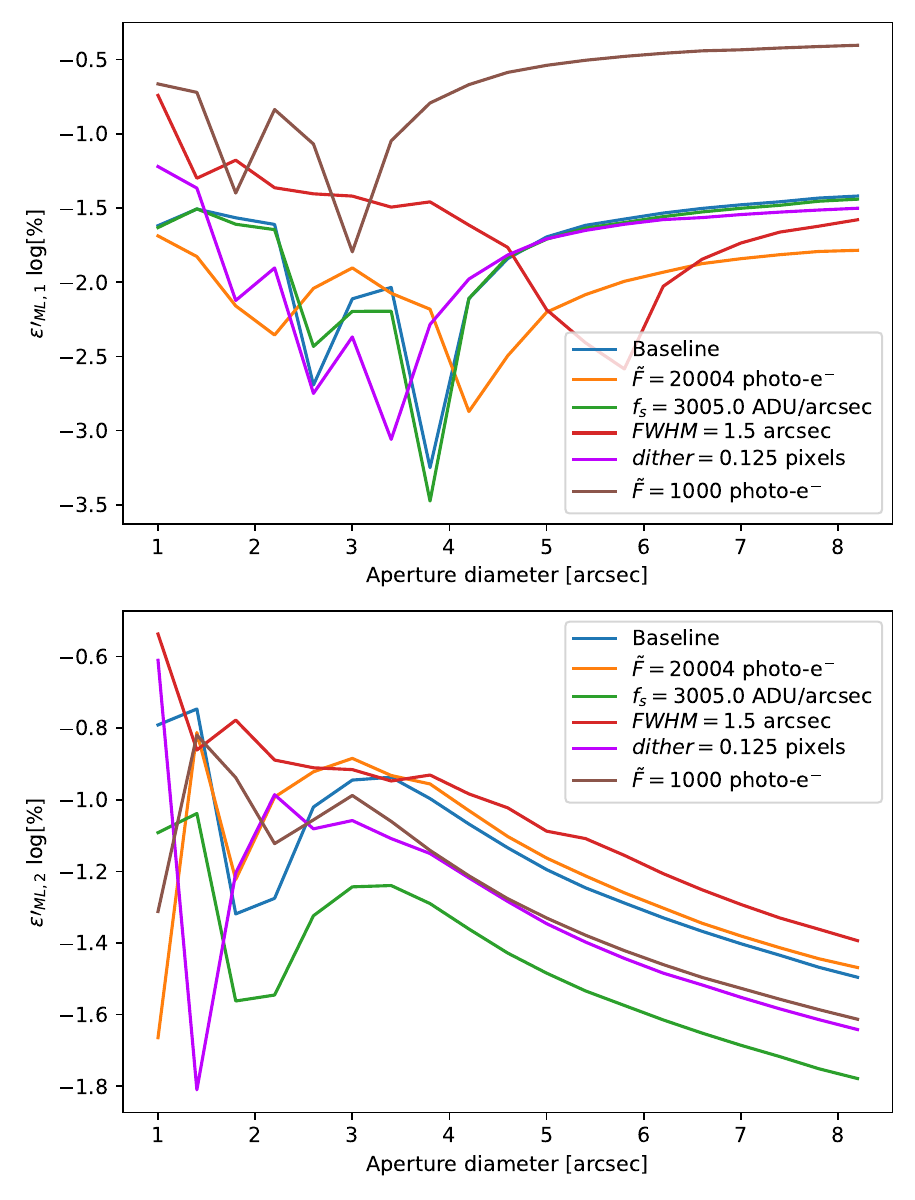}
    \caption{Bias bounds for the estimation of the joint source flux (top) and background (bottom) using the ML estimator, under different observational scenarios, for an aperture selection scheme.}
    \label{fig:ML_relbias_bounds}
\end{figure}

We begin  analysing the bounds for the bias derived from Eq.~(\ref{eq:ML_bias_bound}). Figure~\ref{fig:ML_relbias_bounds} assesses these bounds relative to the values of the true parameters, {\em i.e}., $\log{\left(100 \times \frac{\epsilon'_{ML,1}}{\tilde{F}}\right)}$ and $\log{\left(100 \times \frac{\epsilon'_{ML,2}}{\tilde{B}}\right)}$, respectively, considering different number of pixels (aperture selection). In the regime where the target's PSF is not covered completely, the bounds seem to follow an irregular oscillating behavior with a decreasing trend, which is then followed by a clear pattern. More precisely, we have the following observations: 
\begin{itemize}
    \item The curve corresponding to the object's flux increases slightly until it stabilizes. We note two exceptions, which correspond to the fainter cases where the source's light is either weaker ($F = 500$ ADU) or more extended ($FWHM = 1.5$ arcsec).The behavior seen for the fainter sources can be attributed to an insufficient aperture range displayed in the Figure.
    
    \item After slightly increasing, the curves related to background monotonically decrease.
\end{itemize}

Despite the dynamics of the bounds as aperture increases, none of them achieves values much higher than a hundredth of a percent. Therefore, we can say that  Theorem~\ref{th_ML} predicts that the joint ML estimator achieves zero bias and, consequently, measuring its precision against the CR bounds is meaningful. We also note that ML exhibits much tighter bias bounds than SWLS (please see Figures~\ref{fig:SWLS_relBias_emp} and \ref{fig:ML_relbias_bounds}), particularly for background estimates (bottom panels in both figures).

\subsubsection{Variance Bounds}
\label{sec:ML_var_theo}
As the ML estimator is predicted to be unbiased, we analyze the two elements that conform the interval depicted in Equation~(\ref{eq:ML_var_interval}). On one hand, we have its central (nominal) value, $\left[ \sigma_{ML}^{2} \right]_{(j,j)}$ in Eq.~(\ref{eq:Cov_ML}),  and, on the other hand, we have $\beta_{ML,j}$, which defines the length of the interval in Eq.~(\ref{eq:ML_var_interval}) ({\em i.e.}, $2\beta_{ML,j}$).



\begin{figure}[!h]
    \centering
    \includegraphics[width=0.45\textwidth]{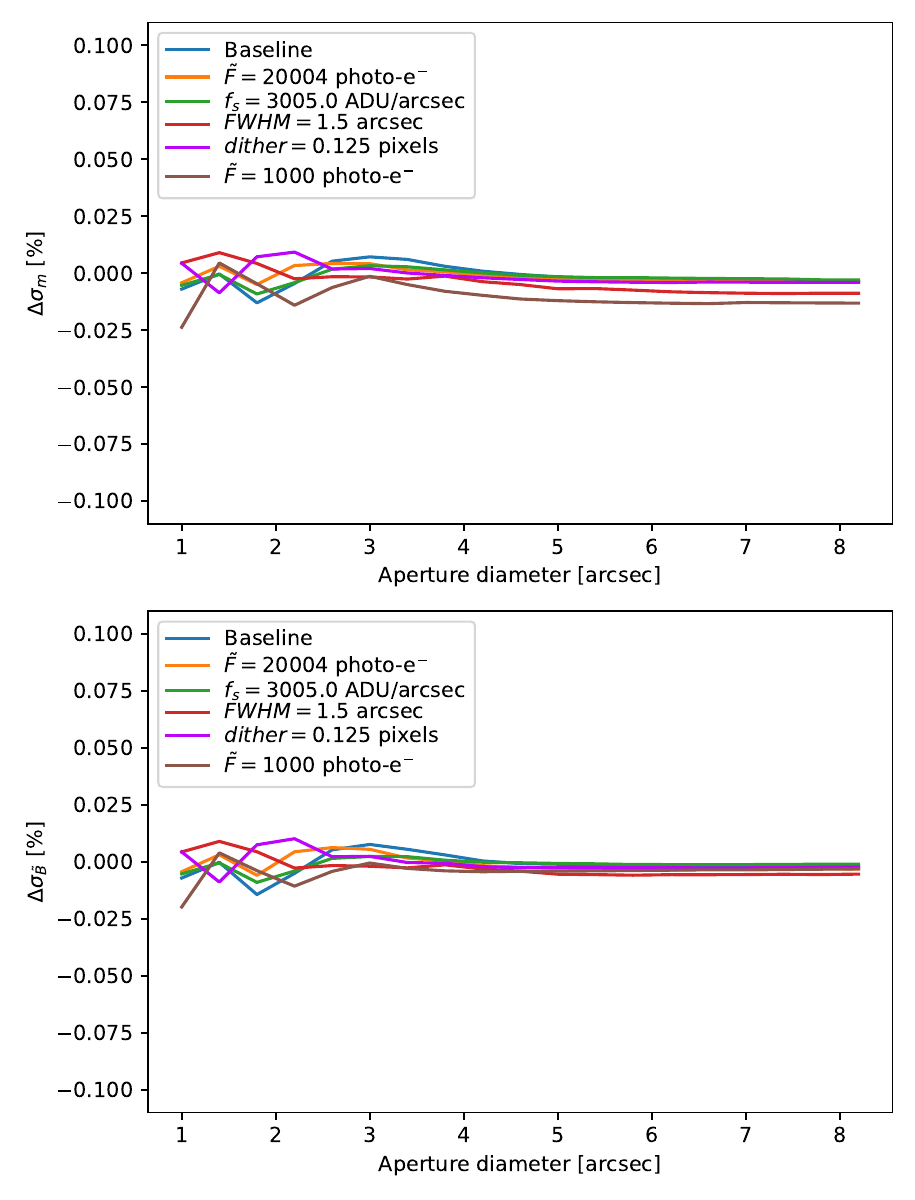}
    \caption{Performance discrepancies between $\left[ \sigma_{ML}^{2} \right]_{(1,1)}$ (top), $\left[ \sigma_{ML}^{2} \right]_{(2,2)}$ (bottom) (using the ML estimator), and their respective CRLB, under different observational scenarios, for an aperture selection scheme.}
    \label{fig:ML_centralDisc}
\end{figure}

Figure~\ref{fig:ML_centralDisc} shows the relative discrepancy between the nominal values $\left[ \sigma_{ML}^{2} \right]_{(1,1)}$ and $\left[ \sigma_{ML}^{2} \right]_{(2,2)}$ and the respective CRLBs in (\ref{eq:FBound}) and (\ref{eq:BBound}).\footnote{Note the abuse of notation in Figure~\ref{fig:ML_centralDisc}: the discrepancies in Eqs.~(\ref{eq:mDiscrepancy_WLS})~and~(\ref{eq:BDiscrepancy_WLS}) are defined as functions of~$Var(\tau_{J,j}(\mathbf{I}))~ ,~j~\in~\{1,2\}$. Here we used $\left[ \sigma_{ML}^{2} \right]_{(j,j)}$ from Eq.~(\ref{eq:Cov_ML}) instead.}
We notice some similarities between the overall tendencies of the curves in Figure~\ref{fig:ML_centralDisc} and the behavior of the bounds in Fig.~\ref{fig:ML_relbias_bounds} for the bias analysis. Two phases can be recognized: the transient stage that emerges for narrower apertures, followed by a steady smooth stage. Even more, it is shown, again, that those cases corresponding to fainter sources (either weaker or more extended across the instrument) slightly deviate from the tendency of the other scenarios. Nevertheless, the most interesting feature on the figure is that all the curves lie in a narrow area around 0\%, which indicates that our bounding strategy predicts that the ML estimator is very close to be optimal, as long as the values of $\beta_{ML,1}$ and $\beta_{ML,2}$ keep small enough.
 It is also remarkable that both $\left[ \sigma_{ML}^{2} \right]_{(1,1)}$ and $\left[ \sigma_{ML}^{2} \right]_{(2,2)}$ resemble Equation~(\ref{eq:MLVar_centered}), even if we did not assume $\tau_{ML}(\bar{\mathbf{I}}) = (\tilde{F}, \tilde{B})$.

\begin{figure}[!h]
    \centering
    \includegraphics[width=0.45\textwidth]{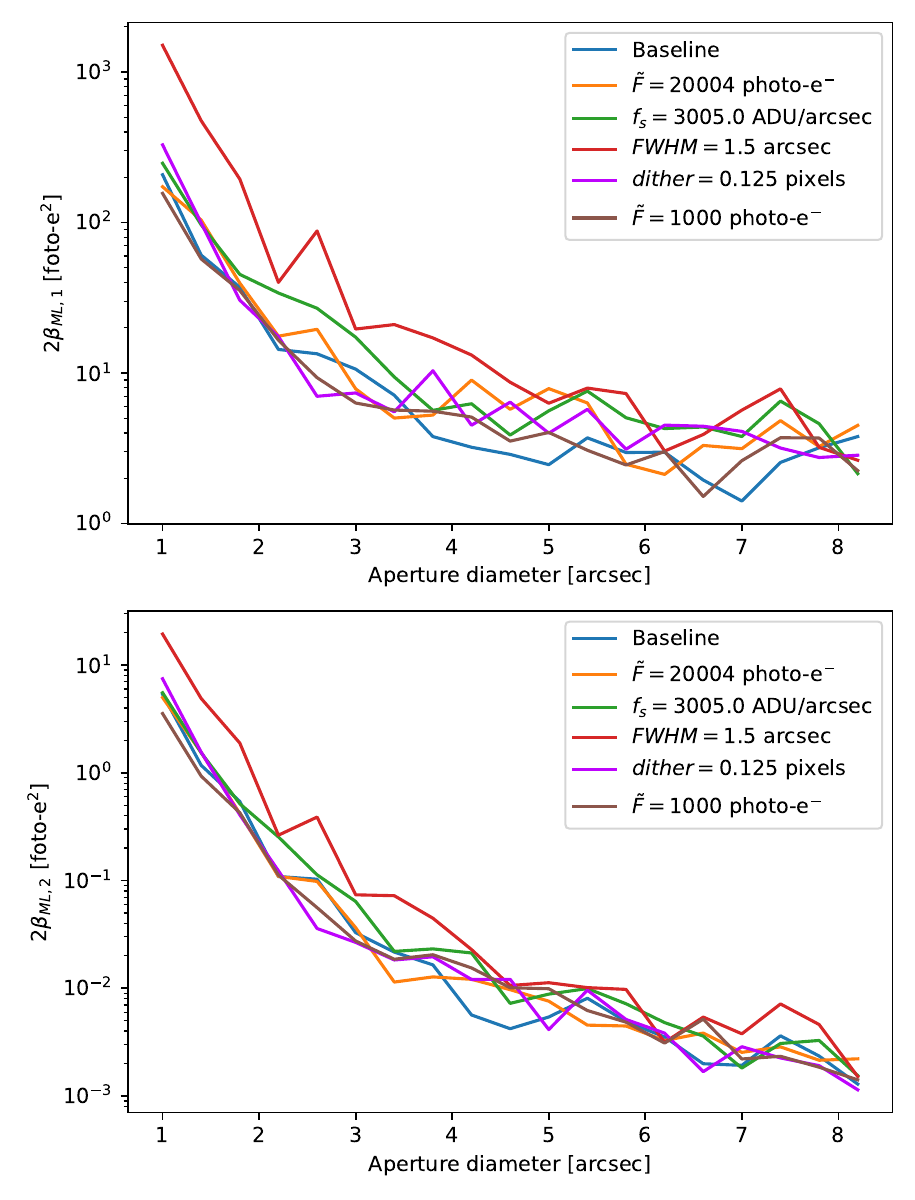}
    \caption{Length of the bounding intervals, $2\beta_{ML,j}$ for $j\in\{ 1,2 \}$ (source flux and background respectively) using the ML estimator under different observational scenarios, for an aperture selection scheme.}
    \label{fig:ML_2beta_log}
\end{figure}

The length of the bounding intervals as aperture increases are shown in Figure~\ref{fig:ML_2beta_log}. As can be observed, wider apertures and information availability allows the predicted intervals to become narrower, {\em i.e.} the predictive power of Eq.~(\ref{eq:ML_var_interval}) increases. Therefore, Theorem~\ref{th_ML} predicts that the ML estimator is capable of extracting both flux and background information from pixels farther from the object's position to allow for better estimation. If we add this to the results obtained in Figure~\ref{fig:ML_centralDisc}, we can see that the estimator's performance becomes closer to the optimal CRLB. 

It is interesting to contrast this result with the previous analysis for the SWLS estimator, which offered performance bounds that became looser with aperture (see Figure~\ref{fig:SWLS_var_bounds_cases}, particularly the lower panel), even though empirical performance was not harmed.


\subsubsection{Empirical Performance}
\label{sec:ML_emp}
\begin{figure}[!h]
    \centering
    \includegraphics[width=0.45\textwidth]{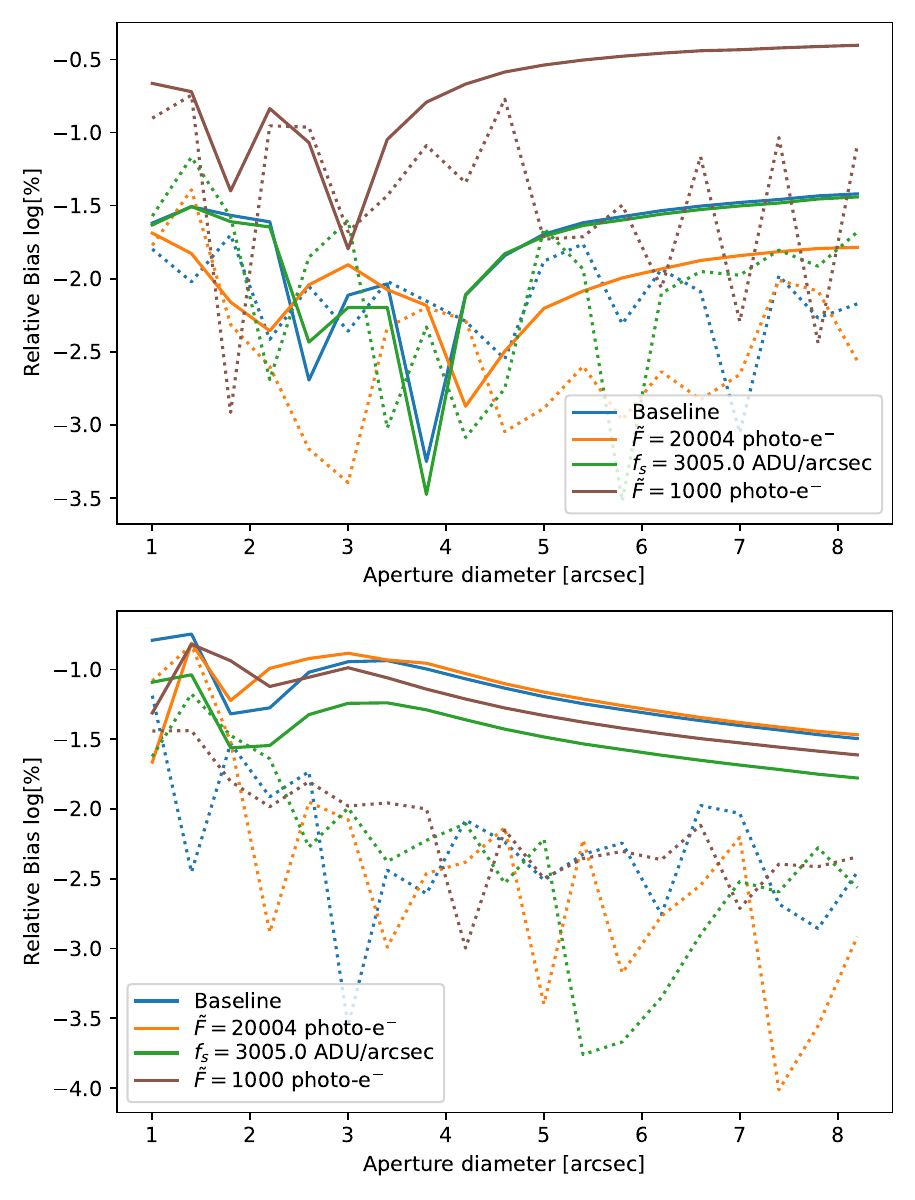}
    \caption{Relative biases for the estimation of the joint source flux (top) and background (bottom) under different observational scenarios, for an aperture selection scheme. Solid lines denote the corresponding bounds as given by Equation~(\ref{eq:pdim_bias_bound}) (also shown in Figure \ref{fig:ML_relbias_bounds}), while dotted lines are employed for empirical bias estimates.}
    \label{fig:ML_relBias_emp}
\end{figure}

Complementing the previous sections, we compare here the estimator's empirical performance in terms of its bias and variance. To begin, Figure~\ref{fig:ML_relBias_emp} compares the bounds shown in Section \ref{sec:ML_bias_theo}, for some of the designed study cases. The figure displays how the values for $\epsilon'_{ML,1}$ and $\epsilon'_{ML,2}$, as expected, bound the empirical performance of the ML estimator in terms of bias. The bounds are particularly good once good coverage of the PSF has been achieved, particularly for background estimates. 
Furthermore, the empirical bias achieved by the ML estimates are smaller than the bounds presented in Theorem \ref{th_ML}, and then we can assert that the ML estimator is, in fact, unbiased for all practical purposes.


\begin{figure}[!h]
    \centering
    \includegraphics[width=0.45\textwidth]{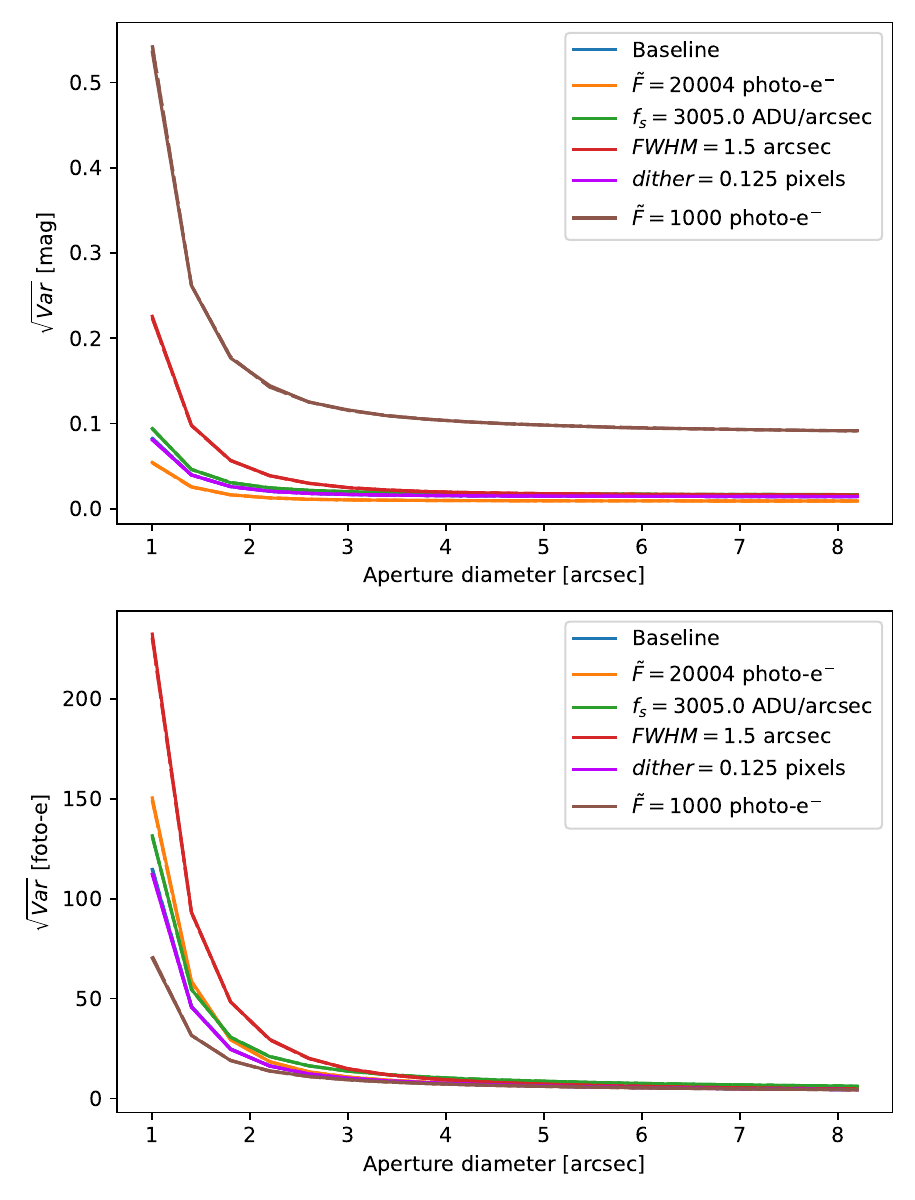}
    \caption{Variance bounds of the ML estimator for the estimation of the joint source flux (top) and background (bottom)  under different observational scenarios, for an aperture selection scheme. Dashed-dotted-, dashed- and solid lines denote the lower bound, the upper bound, and the empirical variance estimate, respectively.}
    \label{fig:ML_var_emp}
\end{figure}

Moving into the estimator's variance, Figure~\ref{fig:ML_var_emp} shows that the empirical variance curves, and the respective developed performance bounds and CRLB curves are so close to each other that the differences between them become negligible, for each of the studied scenarios. This confirms that the joint ML estimator is optimal for a wide range of scenarios and apertures. Furthermore, continuous improvement of the estimates' precision is achieved as more information is made available for inference.


\subsection{ML vs. SWLS}
Comparing the SWLS and ML estimators, we anticipate a substantial improvement of the ML estimator relative to the SWLS approach. Even though both joint estimation strategies offer near-optimal performance, {\em i.e.} variance-wise (see Figures~\ref{fig:SWLS_var_bounds_cases}~and~\ref{fig:ML_var_emp} and their analyses), the mentioned improvement is hinted when comparing Figures~\ref{fig:SWLS_relBias_emp}~and~\ref{fig:ML_relBias_emp}. 

On the details, we notice that the bounds for $\tilde{F}$ obtained from both schemes are quite similar (at least after the transient phase); though the gap between those bounds and the respective empirical bias (dotted lines) is much more noticeable in the curves for ML. On the other hand, the most clear evidence that makes ML better than SWLS relies on the differences between the curves for $\tilde{B}$.  The SWLS estimator's bounds (and measured empirical performances) become steady and apparently constant as aperture increases. In contrast, the ML bounds show a monotonic decreasing tendency, which allows to achieve lower values of bias. Furthermore, the ML empirical performance is considerably better than its bounds' predictions due to the clear gap between solid and dotted lines (see Figure~\ref{fig:ML_relBias_emp}). Therefore, from the statistical interaction between $\tilde{F}$ and $\tilde{B}$ estimates, we conclude that the improvements on the latter spread through $\hat{\tilde{F}}$, allowing it to perform better overall in the ML case, both in terms of exactness and precision. This can be verified in Table \ref{tab:comp_results} in Section \ref{sec:comparison}.


\section{Using Theorem 4.1 as an Implementation-Validation Tool}
The previous section shows the clear improvement of the ML strategy for the joint estimation of flux and background. In this section, we take advantage of the main result in Theorem \ref{theorem1} (in particular, its form presented in Theorem~\ref{th_ML}) to evaluate the appropriateness (soundness) of several optimization strategies to implement (numerically) the ML estimator in (\ref{eq:ML_tau}).  

\begin{figure}[!h]
    \centering
    \includegraphics[width=0.45\textwidth]{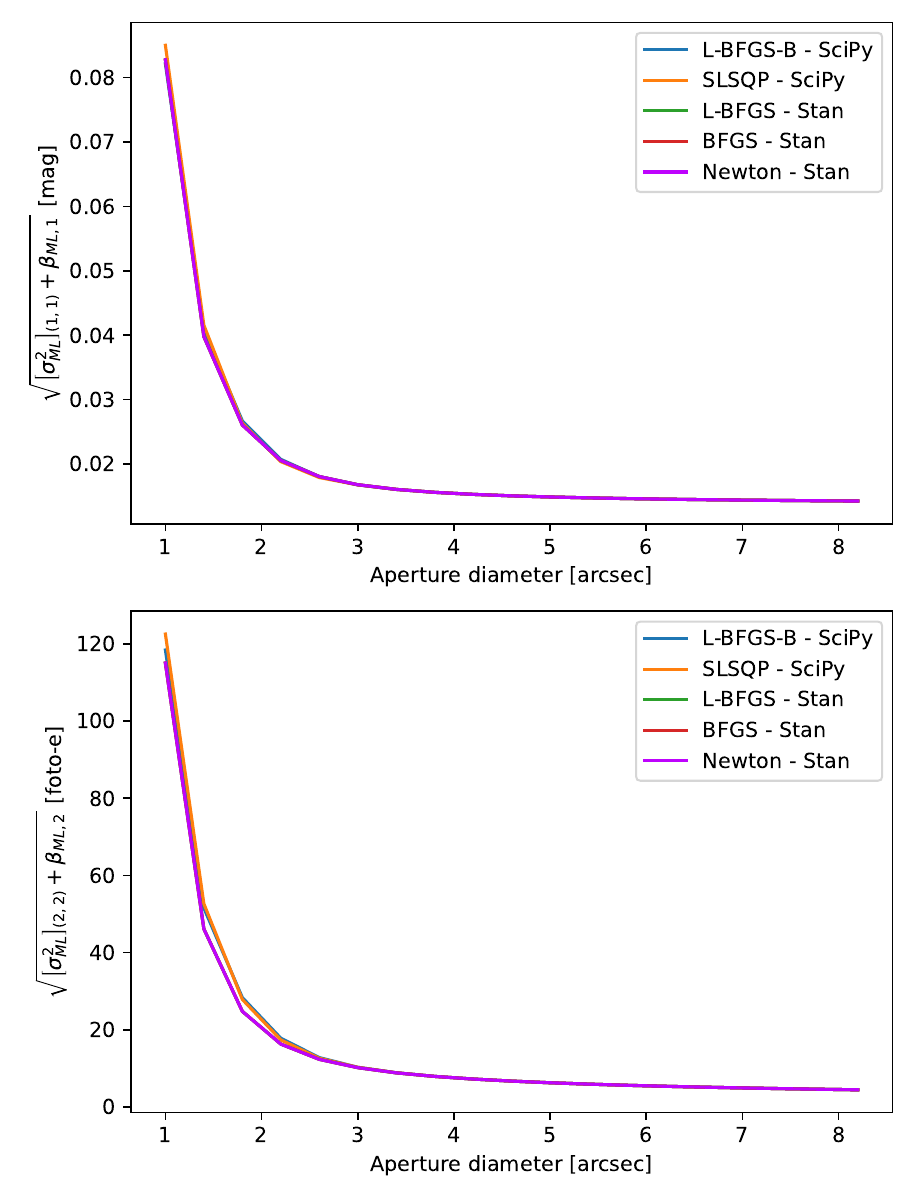}
    \caption{Upper bounds for the ML estimates' variances for different implementations of the estimation algorithm. Top and bottom panels correspond to flux and background bounds, respectively.}
    \label{fig:valid_upper_case0}
\end{figure}

The results presented so far for the ML estimator were developed with the \textit{Stan} probabilistic programming language \citep{stan_language, stan_manual}, particularly its \textit{PyStan} interface \citep{pystan}, using as optimizer the Newton method \citep{nocedal}. However, the \textit{Stan} language offers many other quasi-Newton algorithms for optimization. For further comparisons, we present in this section the results obtained with a ML estimator implemented within the \textit{SciPy} \citep{SciPy-NMeth2020} package.
%
In this context, the bounds presented in Theorem~\ref{th_ML} are instrumental tool to evaluate the expected performance behaviour of different solvers for Eq.~(\ref{eq:ML_tau}).
Evidence about the expected consistency among different numerical solvers is presented in Figure~\ref{fig:valid_upper_case0}, where the upper bounds for the variances' squared root calculated from different implementations of the ML estimator are compared. 
The key observation from this comparison is that, for a given estimator, we get almost identical ({\em i.e.} consistent with the bounds originally presented in Section~\ref{sec:MLE}) regardless of the solver used to implement the ML estimator.


In Figure~\ref{fig:valid_upper_case0}, it is worth noting that for aperture diameters greater than 3 times the $FWHM$, the difference between the curves diminishes. This trend indicates that more pixels  (information) translates into more consistent bounds among different implementations of the ML algorithm. This behavior strengthen the idea of using more pixels (ideally all the pixels) when performing joint estimation.

The fact that the bounds calculated with different implementations are very similar does not imply that the estimators behave the exact same way in practice, {\em i.e.} they present similar empirical performance. Having demonstrated that the ML estimator implemented by means of the \textit{Stan}'s Newton algorithm allows for congruent results between the empirical and predicted performances (see Section \ref{sec:ML_emp}), 
and the consistency of Theorem~\ref{th_ML} independent of the solver used, we can focus on the empirical performances of the different ML implementations.

\begin{figure}[!h]
    \centering
    \includegraphics[width=0.45\textwidth]{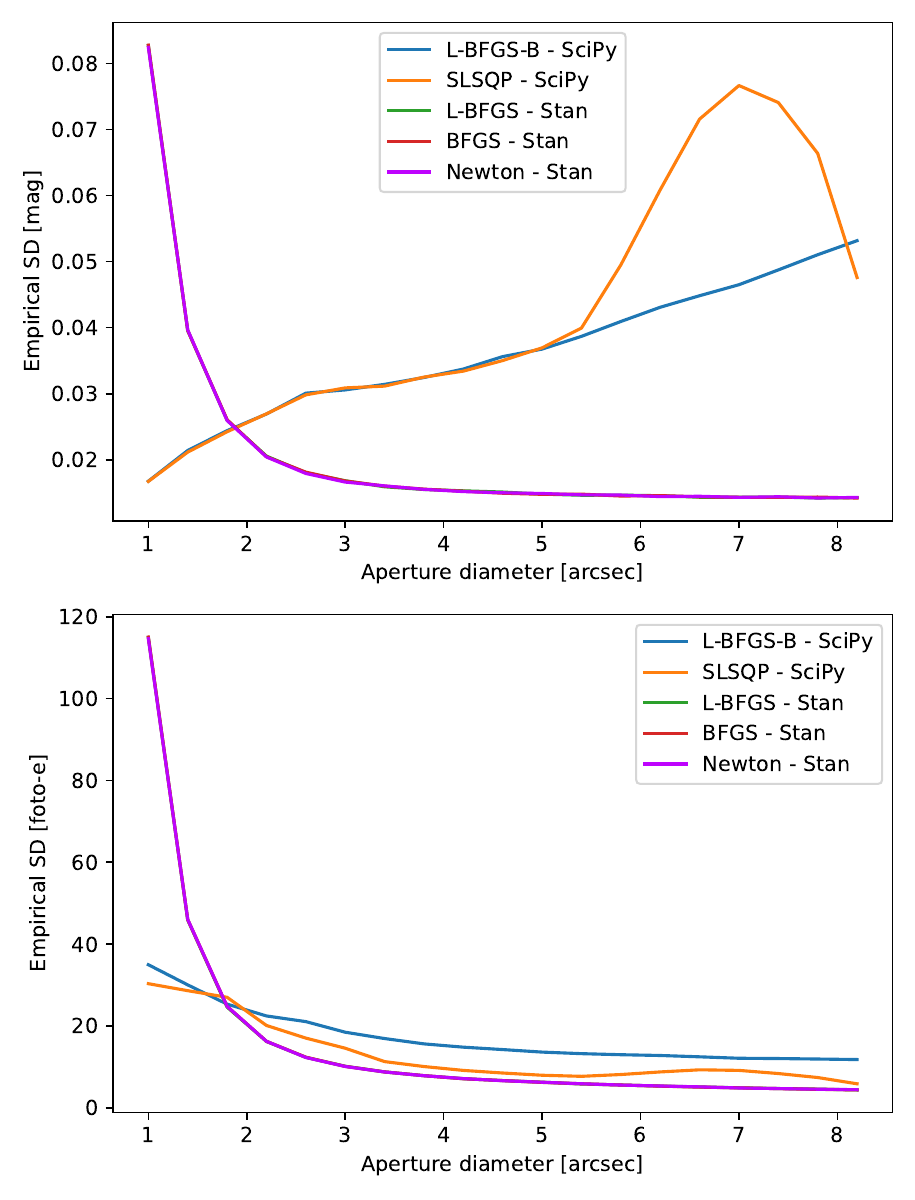}
    \caption{Empirical performances for different numerical implementations of the ML estimator. Top and bottom panels correspond to flux and background empirical performances, respectively.}
    \label{fig:valid_empVar_case0}
\end{figure}

Figure~\ref{fig:valid_empVar_case0} displays the empirical standard deviation obtained with the reviewed optimization methods for our baseline scenario. In this figure, we see how the \textit{Stan}-implemented estimators perform quite the same, attaining values similar to the ones seen in Section \ref{sec:ML_emp}. In contrast, the estimators programmed with \textit{SciPy} perform poorly\footnote{\textit{SciPy} offers plenty of alternative algorithms, which may be better than the ones reviewed here.} by showing a worsening precision on flux estimation as aperture increases, 
and background estimates that do not correspond with the curves obtained with \textit{Stan}, even though they show a decreasing tendency as aperture area increases.


Through this kind of theory-driven analysis, the developer or end user could detect the presence of implementation issues and then make informed decisions, such as inspecting the code searching for possible improvements, for example: parameter tweaking, better initial guesses for the algorithms; or, as could be the case in this scenario, to straightforwardly discard implementations that do not match the expected behavior. Therefore, the bounds developed in Section \ref{sec:impl_extension} offer a relevant framework for the development stages as part of the testing and validation processes of a photometric pipeline.

\hfill

\section{A comparison between joint inference and staged inference}
\label{sec:comparison}

In previous sections, we have demonstrated the theoretical near-optimality of the joint inference scheme. We now show some advantages of our proposal against the classical approach to the photometry problem, {\em i.e.} that of performing a sequential inference of background, followed by sky-substraction and flux estimation. To do this, we resort once more (at least for the moment) to some simplified ideal scenarios in which the relative position of the source and its PSF are known, along with a detector configuration that resembles the CCD array onboard TESS \citep{2015JATIS...1a4003R},\footnote{See also \url{https://heasarc.gsfc.nasa.gov/docs/tess/the-tess-space-telescope.html}} and the sky conditions of its observations. One more modification to the simulations presented in previous sections is that here we consider an extended 2-dimensional observational model, which allows for more realistic results.

Specifically, we consider a $11\times11$ pixel FOV that observes a centered point source. Each pixel measures $\Delta x = \Delta y = 21''$ when projected onto the sky, which is consistent with the specifications of the TESS mission \citep{TESS_mission}. Other detector parameters are also chosen to be comparable to those of the satellite's cameras such as $D=0$, $RON=8.25$ [photo-e$^{-}$], $G=5.1$ [photo-e$^{-}$/ADU] \citep{TESS_handbook}. The source is embedded into a background parameterized through $f_{s} = 0.02$ [ADU/arcsec$^{2}$] which, along with the detector's parameters, allows for a background $\tilde{B}$ of nearly $100$~[photo-e$^{-}$] per pixel. The light-spreading profile of the source is given by a Gaussian PSF whose diagonal isotropic covariance matrix is characterized by a FWHM of $1.88 \times \Delta x$ \citep{TESS_FWHM}. 

\begin{figure}[!h]
    \centering
    \includegraphics[width=0.35\textwidth]{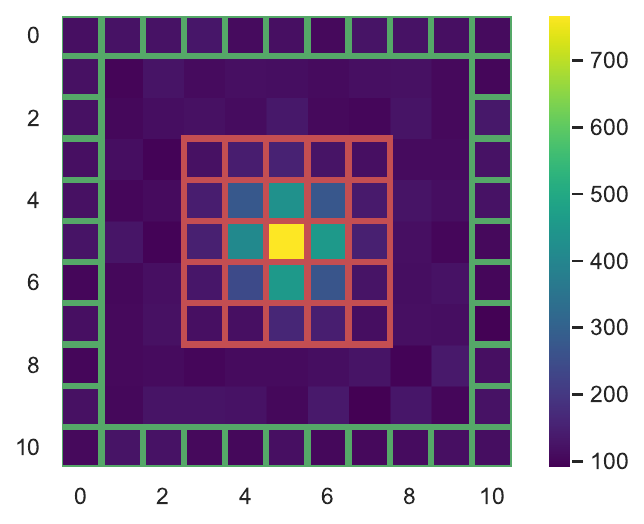}
    \caption{Simulated TESS image with a $SNR \sim 40$. Pixels used for background and flux inference are marked with green and red squares, respectively.}
    \label{fig:simulated_TESS_image}
\end{figure}

For the sequential estimation process, we design simplified squared masks to separate background pixels and source pixels. The background mask is given by the outermost ring whose width corresponds to one pixel thick, while the aperture mask consists of the $5\times5$ pixel square surrounding the source, which allows for a coverage of $99.652$\% of the source's flux, which is consistent with \citet{TESS_FWHM}. An example simulated image can be seen in Figure~\ref{fig:simulated_TESS_image}, which corresponds to a source with a $SNR \sim 40$ inside the aperture.

We test the classical approach and the joint SWLS and ML estimators in three different scenarios, each differing only in the brightness of the source. The resulting SNR  regimes (measured inside the aperture area) consider here range from high ($\sim 100$), to medium ($\sim 40$), and low ($\sim 4.5$). Naturally, the sequential estimation process considers the pixels masked in Figure~\ref{fig:simulated_TESS_image} only, while the joint inference methods consider each one of the $121$ pixels in the FOV. The results of these experiments are presented in Table~\ref{tab:comp_results}.

\begin{splitdeluxetable*}{cccccccccccBcccccBccccc}
\tabletypesize{\scriptsize}
\tablewidth{0pt} 
\tablecaption{Average performance of different estimation schemes, for high, medium and low SNR regimes. CRLB and estimates' variances for flux are parsed into units of magnitudes as in Equation~(\ref{eq:sigma_m}).}
\label{tab:comp_results}
\tablehead{
\colhead{$\tilde{B}$} & \colhead{$\tilde{F}$} & \colhead{SNR (aperture)} & \colhead{SNR (FOV)} & \colhead{$\sigma_{\tilde{B}}$} & \colhead{$\sigma_{m}$} & \multicolumn{5}{c}{Sequential Inference Scheme} & \multicolumn{5}{c}{SWLS Estimator} & \multicolumn{5}{c}{ML Estimator} \\
\colhead{[photo-e$^{-}$]} & \colhead{[photo-e$^{-}$]} & \colhead{} & \colhead{} & \colhead{[photo-e$^{-}$]} & \colhead{[mmag]} &  \colhead{Average $\hat{\tilde{B}}$} & \colhead{Average $\hat{\tilde{F}}$} & \colhead{$\sqrt{Var(\hat{\tilde{B}})}$} & \colhead{$\sqrt{Var(\hat{\tilde{F}})}$} & \colhead{\shortstack{Average Proc. \\ Time per Frame}}  & \colhead{Average $\hat{\tilde{B}}$} & \colhead{Average $\hat{\tilde{F}}$} & \colhead{$\sqrt{Var(\hat{\tilde{B}})}$} & \colhead{$\sqrt{Var(\hat{\tilde{F}})}$} & \colhead{\shortstack{Average Proc. \\ Time per Frame}}  & \colhead{Average $\hat{\tilde{B}}$} & \colhead{Average $\hat{\tilde{F}}$} & \colhead{$\sqrt{Var(\hat{\tilde{B}})}$} & \colhead{$\sqrt{Var(\hat{\tilde{F}})}$} & \colhead{\shortstack{Average Proc. \\ Time per Frame}} \\
\colhead{} & \colhead{} & \colhead{} & \colhead{} & \colhead{} & \colhead{} & \colhead{[photo-e$^{-}$]} & \colhead{[photo-e$^{-}$]} & \colhead{[photo-e$^{-}$]} & \colhead{[mmag]} & \colhead{[ms]} & \colhead{[photo-e$^{-}$]} & \colhead{[photo-e$^{-}$]} & \colhead{[photo-e$^{-}$]} & \colhead{[mmag]} & \colhead{[ms]} & \colhead{[photo-e$^{-}$]} & \colhead{[photo-e$^{-}$]} & \colhead{[photo-e$^{-}$]} & \colhead{[mmag]} & \colhead{[ms]}
}
\colnumbers
\startdata
{113.0445} & {12600} & { 101.239} & {77.727} & {1.042} & {10.494} & {113.0397} & {12557.216} & {1.686} & {11.271} & {0.00066} & {112.0446} & {12601.884} & {1.052} & {10.506} & {0.00327} & {113.0442} & {12600.863} & {1.043} & {10.498} & {15.46941} \\
{113.0445} & {3000} & {39.202} & {23.230} & {1.023} & {24.654} & {113.0438} & {2989.298} & {1.685} & {31.598} & {0.00066} & {112.0427} & {3000.912} & {1.028} & {24.725} & {0.00306} & {113.0448} & {3000.089} & {1.023} & {24.687} & {13.79289} \\
{113.0445} & {250} & {4.492} & {2.118} & {1.008} & {166.432} & {113.0408} & {248.915} & {1.679} & {309.464} & {0.00066} & {112.0336} & {251.264} & {1.018} & {167.274} & {0.00308} & {113.0361} & {250.004} & {1.009} & {166.442} & {10.67232} \\
\enddata
\end{splitdeluxetable*}

A key feature that arises from the results shown in the table is the fact that even though the classic aperture method allows for very exact background estimates (better than those from the SWLS estimator), the subsequent flux estimation is slightly biased, as it performs relative photometry as a consequence of the aperture selection, while the other two methods lead to exact estimates both of background and - more relevant - source brightness. The capability of these last inference approaches
is a natural consequence of the larger coverage granted by the increase of the imaging area, and may be critical for a wide range of astrophysical applications that require absolute photometry (\citet{2006A&A...451..375A, 2008PASP..120..328L}). It is also worth noticing that the ML joint estimator is more exact than the SWLS one, independent of the SNR regime, which is consistent with the results presented in previous sections.

Regarding the estimator's precision, both joint estimators outperform the more standard procedure for both estimated parameters. Note that the performance gain for flux may seem more remarkable as the SNR diminishes, which can  be explained by the logarithmic nature of the units used, and the Poissonian behavior of the measurements. It is also worth noting that even if the one-dimensional cases from previous sections are much simpler than the one studied here, the results shown in Table~\ref{tab:comp_results} are consistent and comparable to the bounds and empirical performances of the simpler experiments.

Another performance measure to track is the computation time that is needed to perform inference from a frame (see columns (11), (16) and (21) in Table~\ref{tab:comp_results}). We note an interesting trade-off between precision and processing time: in this simplified scenario where the masks and target's Pixel Response Function (PRF hereafter) are already defined, the sequential scheme is $\sim 5 \times$ faster than the joint SWLS, even though the latter returns both estimates simultaneously, but at the cost of falling into greater variance, farther from the joint CRLB than the concurrent method. This behavior is even more pronounced for the ML estimator, which offers better statistical performance and tightness to the fundamental bounds, but requiring execution times several orders of magnitude greater, which comes from the optimization by means of numerical methods not needed by the other methods, as they can be defined algebraically.\footnote{Note that even though processing times may be improved (particularly for methods that require numerical optimization such as ML), similar trends as those shown in Table~\ref{tab:comp_results} can be expected.}
Due to this trade-off between precision and computation time (along with the theoretic guarantees of the joint inference shown in previous sections), developers and users might be able to ponder which method is better suited for their requirements in terms of these two aspects, as well as other factors, such as the need for absolute photometry.

\section{A Real Case Study: Joint Estimation with TESS Data}
\label{sec:TESS}

\begin{figure}
    \centering
    \includegraphics[width=0.5\textwidth]{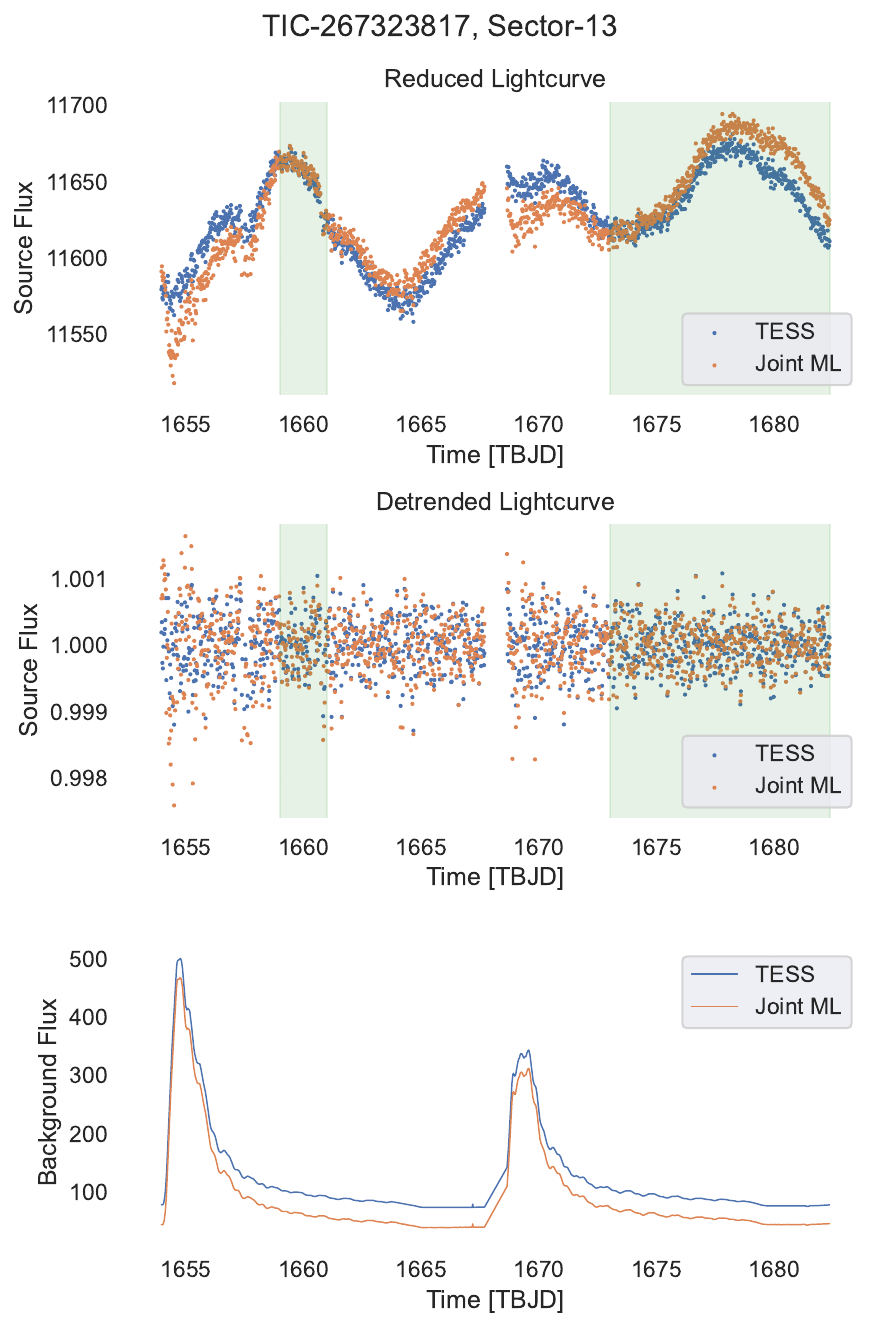}
    \caption{Reduced (top), detrended (middle) source flux lightcurves and estimated background (bottom). Highlighted areas on the flux graphs correspond to the time windows depicted in Table~\ref{tab:TESS_comp}.}
    \label{fig:TIC-267323817_S13}
\end{figure}

The joint inference methods studied throughout this work (particularly the ML estimator) show very promising results that make them appealing to apply in real data. Given that, we designed the experiments of Section~\ref{sec:comparison} to resemble to some extent the properties of the TESS equipment (which was designed to deliver exquisite precision photometry), it is only natural to retrieve for this section data obtained by this satellite, and compare our estimates with theirs (which basically use the classical two-stage sequential approach). For this matter, we retrieved measurements from the target TIC-267323817, Sector 13, as a suitable bona-fide non-variable test candidate. The main reasons behind this selection are the clean FOV around the source, the relative stability of the target's reported reduced lightcurve (see Figure~\ref{fig:TIC-267323817_S13}) and the seemingly good fit of the PRF obtained by means of interpolation of the available PRF templates.\footnote{Note that the PRF implicitly determines the target's astrometry when fitted, not only the light-spreading on the image.}\footnote{FITS files made public at \url{https://archive.stsci.edu/missions/tess/models/prf_fitsfiles/}.} Each frame consists of a $30$ minutes cadence of a $11\times11$ pixel area with the target placed at the central pixel. 

After preprocessing the data according to the TESS pipeline\footnote{See \url{https://outerspace.stsci.edu/display/TESS/2.1+Levels+of+data+processing}.} and feeding it to our implementation of the ML estimator, we were able to obtain simultaneous estimates of the target's flux and the homogeneous background surrounding it. As our algorithm uses the whole frame, and therefore performs ``absolute" photometry if the source is isolated, our lightcurves must be scaled down in order to compare it with the (TESS) reference. The results of this comparison (after scaling our light curve) are shown in Figure~\ref{fig:TIC-267323817_S13}.

Even though the actual source and background fluxes are not constant, the ML estimates show very similar behavior in terms of peaks and tendencies, even overlapping with the reference TESS lightcurve. Regarding the statistical differences between TESS and the joint ML, Table~\ref{tab:TESS_comp} presents the scatter of the time windows highlighted in Figure~\ref{fig:TIC-267323817_S13}. The window between 1659 and 1661 TBJD shows a very good correspondence between both reduced curves, yet the dispersion of the joint ML estimates is slightly smaller. However, once the lightcurves are further processed and detrended, we note how this apparent precision gain may be an artifact of the non-steady underlying flux. 

Something similar happens with the subsequent window, spanning from day 1661 to 1673, but the difference between the scatters of the curves is greater, being that of our joint approach considerably lower than TESS'. This difference is likely explained by a combination of actual flux variations and the apparent flux drop detected by the ML algorithm around day 1668 (right after the gap in the data). This notorious decline in brightness obtained with the Joint ML estimator may draw the scattering estimates to appear far more precise than the reference value (18.913 against 27.766), but once the major tendencies and fluctuations are removed, we notice how this gain doesn't hold.

\begin{deluxetable*}{c|cc|cc}[ht]
\tabletypesize{\scriptsize}
\tablewidth{0pt} 
\tablecaption{Scatter comparison between TESS' flux estimates and our ML ones for different time windows, calculated as the sample standard deviation ($\sigma$).}
\label{tab:TESS_comp}
\tablehead{
\colhead{Time Window} & \multicolumn{2}{c}{Reduced Lightcurve Scatter ($\sigma$)} & 
\multicolumn{2}{c}{Detrended Lightcurve Scatter ($\sigma$)} \\
\colhead{[TBJD]} & \colhead{[photo-e$^{-}$/s]} & \colhead{[photo-e$^{-}$/s]} & \colhead{} & \colhead{} \\
\colhead{} & \colhead{TESS} & \colhead{Joint ML} & \colhead{TESS} & \colhead{Joint ML}
}
\startdata
{[1654 - 1659]} & {26.458} & {33.267} & {\num{4.391e-04}} & {\num{6.795e-04}} \\ 
{[1659 - 1661]} & {12.683} & 
{{12.577}} & {\num{3.686e-04}} & {\num{4.157e-04}} \\
{[1661 - 1673]} & {27.766} & {18.913} & {\num{3.742e-04}} & {\num{4.228e-04}} \\
{[1673 - 1682.3]} & {20.533} & {24.767} & 
{\num{3.202e-04}} & {\num{3.185e-04}} \\
\enddata
\end{deluxetable*}

In the last highlighted window, starting from day 1673, both reduced curves show very similar flux values at the beginning of the time span, but then the one obtained with joint inference shows that the target is slightly brighter than what the TESS pipeline predicts (in contrast to the preceding window). This flux discrepancy produces a scatter $\sim$25\% higher than TESS' (see last row, second and third columns in Table \ref{tab:TESS_comp}). The cornerstone result of the analysis of this last time window is that despite the higher flux reported by the joint method, once the curve is detrended, the standard deviation is slightly lower than the reference (see last row, fourth and fifth columns in Table \ref{tab:TESS_comp}). This can be interpreted by saying that our estimates during that time period are less scattered around the global flux's tendency, {\em i.e.} more precise by about 0.5\%. It is also worth noticing that the time window corresponds to a relatively long period of $>$9 days, accounting for $\sim$450 valid measurements.

Despite the fact that our estimation methods show evidence of outperforming the classical approach regarding flux inference, it is important to keep in mind that, since we have no access to a ground-truth value, both schemes can bear with the bias-variance trade-off. Therefore, some biases may be incurred by both approaches. However, as was shown previously (Sections \ref{sec:MLE} and \ref{sec:comparison}), there are several performance guarantees of the joint ML estimator, along with other well-known asymptotic properties of the ML estimation \citep{kay}, as long as accurate astrometry and PRF estimation is provided (which however may not be  necessarily true for real data). On the other side, it is known that, when performing sequential inference, poor background estimation may significantly hinder and bias the flux estimates \citep{guglielmetti2009}, which is consistent with the results shown on Table~\ref{tab:comp_results}.

The key conclusion of this section is that, for certain circumstances, joint inference may pose comparable and competitive results with traditional estimation methods, without the need of introducing masks to segment and filter information offering a cleaner data processing.

\section{Summary, Conclusions and Future Work}
\label{sec:conclu}
In this paper, we develop expressions for the CRLB 
for the case of the joint flux and background estimation using a photon-integrating device (see Section~\ref{sec:CRLB_analysis}). We also develop the theory to compute the bias and variance of implicit estimators and we apply this result
to the context of a joint (multidimensional) estimation of flux and background of well-isolated point sources. We demonstrate that our theory (see Section~\ref{sec:impl_extension}) provides meaningful effective bounds for the bias and the variance (see Eqs.~(\ref{eq:pdim_bias_bound}) and~(\ref{eq:pdim_var_interval})) of implicit estimators: SWLS in Section \ref{sub_sec_SWLS} and ML in Section \ref{sec:MLE}. Through numerical simulations, we show that the WLS, with a judicious choice of weights (which leads to what we call the SWLS), as well as the ML estimators provide unbiased estimations of the flux and background, and that their variance closely approaches, above an adequate aperture size, the fundamental bounds for this problem given by the CRLB. Then, we demonstrate that the task of joint estimation of flux and background is feasible and present two practical strategies that achieve optimal performance results. 

We apply the SWLS and ML to simulated TESS-like images under realistic conditions, and compare it to the classical two-stage photometric approach. Remarkably, we demonstrate in this controlled setting that our joint estimation framework offers better performance than the convectional approach. To conclude, we analyze our approach applied on real observations of a non-variable source observed with TESS, comparing it with the results obtained from the TESS pipeline (which uses the classical approach). These results are less conclusive, however they show that our joint estimation approach with the ML principle is a competitive strategy offering comparable results with the TESS pipeline. This last comparison is challenging for several reasons: first, the ground-truth is unknown, there are slight flux differences between the two methods given that the TESS pipeline uses an aperture mask (a two-stage approach), and there are cyclic trends in the photometric signal (possibly due to instrumental effects) that have to be corrected and compensated (de-trended).

Please note that high-resolution images of all the figures in this article can be found at \url{https://doi.org/10.5281/zenodo.10056229}.

\hfill

\subsection{Future Work}
Future extensions of our work include incorporating a joint estimate of photometry and astrometry, which implies simultaneously determining at least six parameters. In this case the high-dimensionality of the problem is not prone to analytical analysis, but the methodology outlined in Section~\ref{sec:impl_extension} is directly applicable.

Finally, it would be interesting to apply our methodology on a routine base to pipelines such as those on TESS, JWST, Euclid, or the future Vera-Rubin survey, specially for isolated objects in which the total source flux as well as good control of systematic effects are desirable. One might also consider implementing this approach for a  future reprocessing of Gaia observations.

\section{Acknowledgements}
\label{sec:acknowledge}

\begin{acknowledgments}
M.V. acknowledges Michael Fausnaugh, MIT Research Scientist/TESS Pipeline Scientist, for his help in processing TESS data. M.V. ackowledges partial support from FONDECYT/ANID grant Nro. 1210315, and ``Proyecto enlace ENL02/23". The work of J.S. and M.O. is supported by FONDECYT/ANID grants Nro. 1210315 and 1210031 and from the Advanced Center for Electrical and Electronic Engineering, AC3E, Basal Project FB0008.
R.A.M. acknowledges support from the ``Vicerrectoría de Investigación y Desarrollo (VID) de la Universidad de Chile, project number: ENL02/23". We also acknowledge an anonymous referee, who provided a very thorough review of the paper, its content and presentation, and which helped to clarify and improve the readability of our work.
\end{acknowledgments}


\appendix

\section{Proof: Cramér-Rao Lower Bounds for the Joint Estimation Problem of Flux and Background}
\label{app:proof_CRLB_FB}

As stated in Theorem \ref{theo:CRLB},  the first necessary step to characterize the fundamental precision limits of the joint estimation problem is to check the regularity condition in Equation~(\ref{eq:CRLB_reg}) for each of the target parameters. This is verified as follows:

\begin{equation}
    \begin{split}
        \mathbb{E}_{\mathbf{I} \sim L(\mathbf{I};x_{c}, \tilde{F},\tilde{B})} \left\{ \frac{\partial}{\partial\tilde{F}} \ln{L(\mathbf{I};x_{c}, \tilde{F},\tilde{B})} \right\} &= \mathbb{E}_{\mathbf{I} \sim L(\mathbf{I};x_{c}, \tilde{F},\tilde{B})} \left\{ \frac{\partial}{\partial\tilde{F}} \sum_{i =1}^nI_{i}\ln{\lambda_{i}(x_{c}, \tilde{F})} - \lambda_{i}(x_{c}, \tilde{F}) - \ln{I_{i}!} \right\} \\
        &= \mathbb{E}_{\mathbf{I} \sim L(\mathbf{I};x_{c}, \tilde{F},\tilde{B})} \left\{ \sum_{i =1}^n I_{i} \frac{\partial}{\partial\tilde{F}}\ln{\lambda_{i}(x_{c}, \tilde{F})} - \frac{\partial}{\partial\tilde{F}} \lambda_{i}(x_{c}, \tilde{F}) \right\} \\
        &= \mathbb{E}_{\mathbf{I} \sim L(\mathbf{I};x_{c}, \tilde{F},\tilde{B})} \left\{ \sum_{i =1}^n I_{i} \frac{g_{i}(x_{c})}{\lambda_{i}(x_{c}, \tilde{F})} - g_{i}(x_{c}) \right\} \\
        &= \sum_{i =1}^n \mathbb{E}_{\mathbf{I} \sim L(\mathbf{I};x_{c}, \tilde{F},\tilde{B})} \left\{ I_{i} \right\} \frac{g_{i}(x_{c})}{\lambda_{i}(x_{c}, \tilde{F})} - g_{i}(x_{c}) \\
        &= \sum_{i =1}^n \lambda_{i}(x_{c}, \tilde{F}) \frac{g_{i}(x_{c})}{\lambda_{i}(x_{c}, \tilde{F})} - g_{i}(x_{c}) \\
        &= 0,
    \end{split}
\end{equation}

\begin{equation}
    \begin{split}
        \mathbb{E}_{\mathbf{I} \sim L(\mathbf{I};x_{c}, \tilde{F},\tilde{B})} \left\{ \frac{\partial}{\partial\tilde{B}} \ln{L(\mathbf{I};x_{c}, \tilde{F},\tilde{B})} \right\} &=  \mathbb{E}_{\mathbf{I} \sim L(\mathbf{I};x_{c}, \tilde{F},\tilde{B})} \left\{ \sum_{i =1}^n I_{i} \frac{\partial}{\partial\tilde{B}}\ln{\lambda_{i}(x_{c}, \tilde{F})} - \frac{\partial}{\partial\tilde{B}} \lambda_{i}(x_{c}, \tilde{F}) \right\} \\
        &= \mathbb{E}_{\mathbf{I} \sim L(\mathbf{I};x_{c}, \tilde{F},\tilde{B})} \left\{ \sum_{i =1}^n I_{i} \frac{1}{\lambda_{i}(x_{c}, \tilde{F})} - 1 \right\} \\
        &= \sum_{i =1}^n \mathbb{E}_{\mathbf{I} \sim L(\mathbf{I};x_{c}; \tilde{F},\tilde{B})} \left\{ I_{i} \right\} \frac{1}{\lambda_{i}(x_{c}, \tilde{F})} - 1 \\
        &= \sum_{i =1}^n \lambda_{i}(x_{c}, \tilde{F}) \frac{1}{\lambda_{i}(x_{c}, \tilde{F})} - 1 \\
        &= 0.
    \end{split}
\end{equation}

Therefore, the necessary conditions stated in Theorem \ref{theo:CRLB} are satisfied. The second step is to find the Fisher Information Matrix (FIM) by taking the negative of the expectation on the second derivatives, as in Equation~(\ref{eq:fisher_matrix}). In what follows, the different components of the FIM are calculated:

\begin{equation}
    \begin{split}
        \mathcal{I}_{1,1} &\equiv \left[ \mathcal{I}_{\boldsymbol{\theta}} \right]_{(1,1)} = - \mathbb{E}_{\mathbf{I} \sim L(\mathbf{I};x_{c}, \tilde{F},\tilde{B})} \left\{ \frac{\partial^{2}}{\partial\tilde{F}^{2}} \ln{L(\mathbf{I};x_{c}, \tilde{F},\tilde{B})} \right\} \\
        &= - \mathbb{E}_{\mathbf{I} \sim L(\mathbf{I};x_{c}, \tilde{F},\tilde{B})} \left\{ \frac{\partial}{\partial\tilde{F}} \left( \sum_{i =1}^n I_{i} \frac{g_{i}(x_{c})}{\lambda_{i}(x_{c}, \tilde{F})} - g_{i}(x_{c}) \right) \right\} = - \mathbb{E}_{\mathbf{I} \sim L(\mathbf{I};x_{c}, \tilde{F},\tilde{B})} \left\{ \sum_{i =1}^n I_{i}g_{i}(x_{c})  \frac{\partial}{\partial\tilde{F}}\frac{1}{\lambda_{i}(x_{c}, \tilde{F})} \right\} \\
        &= - \mathbb{E}_{\mathbf{I} \sim L(\mathbf{I};x_{c}, \tilde{F},\tilde{B})} \left\{ \sum_{i =1}^n I_{i}g_{i}(x_{c}) \cdot \frac{-g_{i}(x_{c})}{\lambda_{i}^{2}(x_{c}, \tilde{F})} \right\} = \sum_{i =1}^n \mathbb{E}_{\mathbf{I} \sim L(\mathbf{I};x_{c}, \tilde{F},\tilde{B})} \left\{ I_{i} \right\} \frac{g_{i}^{2}(x_{c})}{\lambda_{i}^{2}(x_{c}, \tilde{F})} \\
        &= \sum_{i =1}^n \frac{g_{i}^{2}(x_{c})}{\lambda_{i}(x_{c}, \tilde{F})},
    \end{split}
\end{equation}

\begin{equation}
    \begin{split}
        \mathcal{I}_{2,1} &\equiv \left[ \mathcal{I}_{\boldsymbol{\theta}} \right]_{(2,1)} = - \mathbb{E}_{\mathbf{I} \sim L(\mathbf{I};x_{c}, \tilde{F},\tilde{B})} \left\{ \frac{\partial^{2}}{\partial\tilde{B}\partial\tilde{F}} \ln{L(\mathbf{I};x_{c}, \tilde{F},\tilde{B})} \right\} \\
        &= - \mathbb{E}_{\mathbf{I} \sim L(\mathbf{I};x_{c}, \tilde{F},\tilde{B})} \left\{ \frac{\partial}{\partial\tilde{B}} \left( \sum_{i =1}^n I_{i} \frac{g_{i}(x_{c})}{\lambda_{i}(x_{c}, \tilde{F})} - g_{i}(x_{c}) \right) \right\} = - \mathbb{E}_{\mathbf{I} \sim L(\mathbf{I};x_{c}, \tilde{F},\tilde{B})} \left\{ \sum_{i =1}^n I_{i}g_{i}(x_{c})  \frac{\partial}{\partial\tilde{B}}\frac{1}{\lambda_{i}(x_{c}, \tilde{F})} \right\} \\
        &= - \mathbb{E}_{\mathbf{I} \sim L(\mathbf{I};x_{c}, \tilde{F},\tilde{B})} \left\{ \sum_{i =1}^n I_{i}g_{i}(x_{c}) \cdot \frac{-1}{\lambda_{i}^{2}(x_{c}, \tilde{F})} \right\} = \sum_{i =1}^n \mathbb{E}_{\mathbf{I} \sim L(\mathbf{I};x_{c}, \tilde{F},\tilde{B})} \left\{ I_{i} \right\} \frac{g_{i}(x_{c})}{\lambda_{i}^{2}(x_{c}, \tilde{F})} \\
        &= \sum_{i =1}^n \frac{g_{i}(x_{c})}{\lambda_{i}(x_{c}, \tilde{F})},
    \end{split}
\end{equation}

\begin{equation}
    \begin{split}
        \mathcal{I}_{1,2} &\equiv \left[ \mathcal{I}_{\boldsymbol{\theta}} \right]_{(1,2)} = - \mathbb{E}_{\mathbf{I} \sim L(\mathbf{I};x_{c}, \tilde{F},\tilde{B})} \left\{ \frac{\partial^{2}}{\partial\tilde{F}\partial\tilde{B}} \ln{L(\mathbf{I};x_{c}, \tilde{F},\tilde{B})} \right\} \\
        &=- \mathbb{E}_{\mathbf{I} \sim L(\mathbf{I};x_{c}, \tilde{F},\tilde{B})} \left\{ \frac{\partial}{\partial\tilde{F}} \left( \sum_{i =1}^n I_{i} \frac{1}{\lambda_{i}(x_{c}, \tilde{F})} - 1 \right) \right\} = - \mathbb{E}_{\mathbf{I} \sim L(\mathbf{I};x_{c}, \tilde{F},\tilde{B})} \left\{ \sum_{i =1}^n I_{i}  \frac{\partial}{\partial\tilde{F}}\frac{1}{\lambda_{i}(x_{c}, \tilde{F})} \right\} \\
        &= - \mathbb{E}_{\mathbf{I} \sim L(\mathbf{I};x_{c}, \tilde{F},\tilde{B})} \left\{ \sum_{i =1}^n I_{i} \cdot \frac{-g_{i}(x_{c})}{\lambda_{i}^{2}(x_{c}, \tilde{F})} \right\} = \sum_{i =1}^n \mathbb{E}_{\mathbf{I} \sim L(\mathbf{I};x_{c}, \tilde{F},\tilde{B})} \left\{ I_{i} \right\} \frac{g_{i}(x_{c})}{\lambda_{i}^{2}(x_{c}, \tilde{F})} \\
        &= \sum_{i =1}^n \frac{g_{i}(x_{c})}{\lambda_{i}(x_{c}, \tilde{F})},
    \end{split}
\end{equation}

\begin{equation}
    \begin{split}
        \mathcal{I}_{2,2} &\equiv \left[ \mathcal{I}_{\boldsymbol{\theta}} \right]_{(2,2)} = - \mathbb{E}_{\mathbf{I} \sim L(\mathbf{I};x_{c}, \tilde{F},\tilde{B})} \left\{ \frac{\partial^{2}}{\partial\tilde{B}^{2}} \ln{L(\mathbf{I};x_{c}, \tilde{F},\tilde{B})} \right\} \\
        &= - \mathbb{E}_{\mathbf{I} \sim L(\mathbf{I};x_{c}, \tilde{F},\tilde{B})} \left\{ \frac{\partial}{\partial\tilde{B}} \left( \sum_{i =1}^n I_{i} \frac{1}{\lambda_{i}(x_{c}, \tilde{F})} - 1 \right) \right\} = - \mathbb{E}_{\mathbf{I} \sim L(\mathbf{I};x_{c}, \tilde{F},\tilde{B})} \left\{ \sum_{i =1}^n I_{i}  \frac{\partial}{\partial\tilde{B}}\frac{1}{\lambda_{i}(x_{c}, \tilde{F})} \right\} \\
        &= - \mathbb{E}_{\mathbf{I} \sim L(\mathbf{I};x_{c}, \tilde{F},\tilde{B})} \left\{ \sum_{i =1}^n I_{i} \cdot \frac{-1}{\lambda_{i}^{2}(x_{c}, \tilde{F})} \right\} = \sum_{i =1}^n \mathbb{E}_{\mathbf{I} \sim L(\mathbf{I};x_{c}, \tilde{F},\tilde{B})} \left\{ I_{i} \right\} \frac{1}{\lambda_{i}^{2}(x_{c}, \tilde{F})} \\
        &= \sum_{i =1}^n \frac{1}{\lambda_{i}(x_{c}, \tilde{F})}.
    \end{split}
\end{equation}

These equations recover the Fisher Matrix described in Equation~(\ref{eq:FB_fisher}) and, as a consequence of Theorem \ref{theo:CRLB}, also imply the bounds of Equations~(\ref{eq:FBound}) and~(\ref{eq:BBound}).

It is also worth noting, that the identity $\mathcal{I}_{1,2} = \mathcal{I}_{2,1}$ is verified, {\em i.e.} $\mathcal{I}_{\boldsymbol{\theta}}$ is symmetric.

\section{Proof: Bounding the Performance of a high-dimensional Implicit Estimator}
\label{app:proof_pdim_bounds}

Let $\boldsymbol{\Theta} \subseteq \mathbb{R}^{p}$ be some parameter space with $\boldsymbol{\alpha} \in \boldsymbol{\Theta}$ a $p$-dimensional parameter vector to be estimated from a $n$-dimensional observation vector $\mathbf{I}$. Let $J: \boldsymbol{\Theta} \times \mathbb{R}^{n} \to \mathbb{R}$ be the objective function of some optimization problem which defines a decision rule $\tau_{J}(\cdot)$ as the solution of
\begin{equation}
    \tau_{J}(\mathbf{I}) \equiv \underset{\boldsymbol{\alpha} \in \boldsymbol{\Theta}}{\operatorname{argmin}} J(\boldsymbol{\alpha}, \mathbf{I}).
    \label{eq:proof_implicit_estimator}
\end{equation}

Let's also make the assumption that the cost function $J(\boldsymbol{\alpha},  \mathbf{I})$ has a unique optimal value at $\boldsymbol{\alpha} = \tau_{J}(\mathbf{I})$ which also satisfies the following first order condition: 

\begin{equation}
    \begin{split}
        \mathbf{0} &= \nabla J(\boldsymbol{\alpha}, \mathbf{I})\biggr\rvert_{\boldsymbol{\alpha} = \tau_{J}(\mathbf{I})} =  \left[  J'(\boldsymbol{\alpha}, \mathbf{I})  \right]^{\intercal} \biggr\rvert_{\boldsymbol{\alpha} = \tau_{J}(\mathbf{I})}\\
        &= \begin{bmatrix}
             \frac{\partial}{\partial\alpha_{1}}J(\tau_{J}(\mathbf{I}), \mathbf{I}) \\
             \vdots  \\
             \frac{\partial}{\partial\alpha_{p}}J(\tau_{J}(\mathbf{I}), \mathbf{I})
           \end{bmatrix}.
    \end{split}
    \label{eq:proof_implicit_foc}
\end{equation}

Assuming that the functional $J(\boldsymbol{\alpha}, \mathbf{I})$ is (at least) twice differentiable, the estimator's Taylor expansion around the mean vector $\bar{\mathbf{I}}$ is as follows:

\begin{equation}
    \tau_{J}(\mathbf{I}) = \tau_{J}(\bar{\mathbf{I}}) + \tau_{J}'(\bar{\mathbf{I}}) \cdot (\mathbf{I} - \bar{\mathbf{I}}) + e(\bar{\mathbf{I}}, \mathbf{I} - \bar{\mathbf{I}}),
    \label{eq:proof_pdim_taylor}
\end{equation}

\noindent where 

\begin{equation}
    \begin{split}
        e (\bar{\mathbf{I}}, \mathbf{I} - \bar{\mathbf{I}}) &= \frac{1}{2} \sum_{i=1}^{n} \sum_{h=1}^{n} \frac{\partial^{2}}{\partial I_{i} \partial I_{h}} \tau_{J}(\bar{\mathbf{I}} + t(\mathbf{I} - \bar{\mathbf{I}})) (I_{i} - \bar{I}_{i}) (I_{h} - \bar{I}_{h}) \\
        &= \frac{1}{2} \sum_{i=1}^{n} \sum_{h=1}^{n} \begin{bmatrix}
                                                                     \frac{\partial^{2}}{\partial I_{i} \partial I_{h}} \tau_{J,1}(\bar{\mathbf{I}} + t(\mathbf{I} - \bar{\mathbf{I}})) \\
                                                                     \vdots \\
                                                                     \frac{\partial^{2}}{\partial I_{i} \partial I_{h}} \tau_{J,p}(\bar{\mathbf{I}} + t(\mathbf{I} - \bar{\mathbf{I}}))
                                                                 \end{bmatrix} (I_{i} - \bar{I}_{i}) (I_{h} - \bar{I}_{h}),
    \end{split}
    \label{eq:proof_pdim_e}
\end{equation}

\noindent for some fixed but unknown value $t \in [0,1]$.

By known properties of the covariance matrix, the Taylor expansion in Equation~(\ref{eq:proof_pdim_taylor}) can also be used to approximate the covariance matrix of the estimator $\tau_{J}$:~\footnote{As a shorthand, we denote the covariance matrix of two random vectors $X, Y$ as $Cov(X, Y)~=~ K_{X,Y}~=~\mathbb{E} \left\{ (X - \bar{X}) \cdot (Y - \bar{Y})^{\dagger} \right\}$.}

\begin{equation}
    \begin{split}
        Var(\tau_{J}(\mathbf{I})) =& Var(\tau_{J}(\bar{\mathbf{I}}) + \tau_{J}'(\bar{\mathbf{I}}) \cdot (\mathbf{I} - \bar{\mathbf{I}}) + e(\bar{\mathbf{I}}, \mathbf{I} - \bar{\mathbf{I}})) \\
        =& Var(\tau_{J}'(\bar{\mathbf{I}}) \cdot (\mathbf{I} - \bar{\mathbf{I}}) + e(\bar{\mathbf{I}}, \mathbf{I} - \bar{\mathbf{I}})) \\
        =& Var(\tau_{J}'(\bar{\mathbf{I}}) \cdot (\mathbf{I} - \bar{\mathbf{I}})) + Cov(\tau_{J}'(\bar{\mathbf{I}}) \cdot (\mathbf{I} - \bar{\mathbf{I}}), e(\bar{\mathbf{I}}, \mathbf{I} - \bar{\mathbf{I}})) \\
        &+ Cov(e(\bar{\mathbf{I}}, \mathbf{I} - \bar{\mathbf{I}}), \tau_{J}'(\bar{\mathbf{I}}) \cdot (\mathbf{I} - \bar{\mathbf{I}})) + Var(e(\bar{\mathbf{I}}, \mathbf{I} - \bar{\mathbf{I}})).
    \end{split}
    \label{eq:proof_pdim_taylor_var}
\end{equation}

Noting that

\begin{equation}
    \begin{split}
        \mathbb{E}_{\mathbf{I} \sim L(\mathbf{I};\boldsymbol{\alpha}^{\star})} \left\{ \tau_{J}'(\bar{\mathbf{I}}) \cdot (\mathbf{I} - \bar{\mathbf{I}}) \right\} &= \tau_{J}'(\bar{\mathbf{I}}) \cdot \mathbb{E}_{\mathbf{I} \sim L(\mathbf{I};\boldsymbol{\alpha}^{\star})} \left\{ \mathbf{I} - \bar{\mathbf{I}} \right\} \\
        &= \mathbf{0},
    \end{split}
    \label{eq:proof_linerity_espinosa}
\end{equation}

\noindent the first element on the right-hand side of Equation~(\ref{eq:proof_pdim_taylor_var}) can be expressed as
\begin{equation}
    Var(\tau_{J}'(\bar{\mathbf{I}}) \cdot (\mathbf{I} - \bar{\mathbf{I}})) = \tau_{J}'(\bar{\mathbf{I}}) \cdot K_{\mathbf{I}} \cdot \left [ \tau_{J}'(\bar{\mathbf{I}}) \right]^{\dagger}.
\end{equation}

The identity in Equation~(\ref{eq:proof_linerity_espinosa}) also allows to simplify the $p \times p$ cross-covariance between $\tau_{J}'(\bar{\mathbf{I}}) \cdot (\mathbf{I} - \bar{\mathbf{I}})$ and $e(\bar{\mathbf{I}}, \mathbf{I} - \bar{\mathbf{I}})$ by:
\begin{equation}
    \begin{split}
        Cov(\tau_{J}'(\bar{\mathbf{I}}) \cdot (\mathbf{I} - \bar{\mathbf{I}}), e(\bar{\mathbf{I}}, \mathbf{I} - \bar{\mathbf{I}})) =& \mathbb{E}_{\mathbf{I} \sim L(\mathbf{I};\boldsymbol{\alpha}^{\star})} \left\{ \left[ \tau_{J}'(\bar{\mathbf{I}}) \cdot (\mathbf{I} - \bar{\mathbf{I}}) \right] \cdot \left[ e(\bar{\mathbf{I}}, \mathbf{I} - \bar{\mathbf{I}}) - \mathbb{E}_{\mathbf{I} \sim L(\mathbf{I};\boldsymbol{\alpha}^{\star})} \left\{ e(\bar{\mathbf{I}}, \mathbf{I} - \bar{\mathbf{I}}) \right\} \right]^{\dagger} \right\} \\
        =& \mathbb{E}_{\mathbf{I} \sim L(\mathbf{I};\boldsymbol{\alpha}^{\star})} \left\{ \left[ \tau_{J}'(\bar{\mathbf{I}}) \cdot (\mathbf{I} - \bar{\mathbf{I}}) \right] \cdot e^{\dagger}(\bar{\mathbf{I}}, \mathbf{I} - \bar{\mathbf{I}}) \right\} \\
        &- \mathbb{E}_{\mathbf{I} \sim L(\mathbf{I};\boldsymbol{\alpha}^{\star})} \left\{ \tau_{J}'(\bar{\mathbf{I}}) \cdot (\mathbf{I} - \bar{\mathbf{I}}) \right\} \cdot \mathbb{E}_{\mathbf{I} \sim L(\mathbf{I};\boldsymbol{\alpha}^{\star})} \left\{ e^{\dagger}(\bar{\mathbf{I}}, \mathbf{I} - \bar{\mathbf{I}}) \right\} \\
        =& \mathbb{E}_{\mathbf{I} \sim L(\mathbf{I};\boldsymbol{\alpha}^{\star})} \left\{ \left[ \tau_{J}'(\bar{\mathbf{I}}) \cdot (\mathbf{I} - \bar{\mathbf{I}}) \right] \cdot e^{\dagger}(\bar{\mathbf{I}}, \mathbf{I} - \bar{\mathbf{I}}) \right\},
    \end{split}
    \label{eq:proof_pdim_cov_tau_e}
\end{equation}

\begin{equation}
    Cov(e(\bar{\mathbf{I}}, \mathbf{I} - \bar{\mathbf{I}}), \tau_{J}'(\bar{\mathbf{I}}) \cdot (\mathbf{I} - \bar{\mathbf{I}})) = \left[ Cov(\tau_{J}'(\bar{\mathbf{I}}) \cdot (\mathbf{I} - \bar{\mathbf{I}}), e(\bar{\mathbf{I}}, \mathbf{I} - \bar{\mathbf{I}})) \right]^{\intercal}.
    \label{eq:proof_pdim_cov_transpose}
\end{equation}

Before continuing with the proof, it is worth noting that the the focus of this work's section is to address the bias and variance of each individual component of the implicit estimator given by $\tau_{J}$. Then, we are not particularly interested in every component of the matrix $K_{\tau_{J}(\mathbf{I})}$, but in its diagonal or, in other words, scalar forms of Equation~(\ref{eq:proof_pdim_taylor_var}), which result from the projection of $\tau_{J}(\mathbf{I})$ onto the $p$-dimensional canonical vectors, {\em i.e.} $\tau_{J,j}(\mathbf{I}) = \langle \hat{\imath}_{j}, \tau_{J}(\mathbf{I}) \rangle$ with $j \in \{1,\dots,p\}$, where $\hat{\imath}_{j}$ denotes the canonical vector whose only non-zero component equals $1$ at the $j$-th coordinate. Therefore, for $j \in \{1,\dots,p\}$:
\begin{equation}
    \begin{split}
        \left[ Var(\tau_{J}(\mathbf{I})) \right]_{(j,j)} =& \bm{\hat{\imath}}_{j}^{\intercal} Var(\tau_{J}(\mathbf{I})) \bm{\hat{\imath}}_{j} \\
        =& Var(\langle\bm{\hat{\imath}}_{j}, \tau_{J}(\mathbf{I}) \rangle) \\
        =& Var(\langle\bm{\hat{\imath}}_{j}, \tau_{J}(\bar{\mathbf{I}}) + \tau_{J}'(\bar{\mathbf{I}}) \cdot (\mathbf{I} - \bar{\mathbf{I}}) + e(\bar{\mathbf{I}}, \mathbf{I} - \bar{\mathbf{I}})\rangle) \\
        =& Var\left( \tau_{J,j}(\bar{\mathbf{I}}) + \left[ \tau_{J}'(\bar{\mathbf{I}}) \right]_{(j,\cdot)} \cdot (\mathbf{I} - \bar{\mathbf{I}}) + e_{j}(\bar{\mathbf{I}}, \mathbf{I} - \bar{\mathbf{I}}) \right) \\
        =& Var\left( \left[ \tau_{J}'(\bar{\mathbf{I}}) \right]_{(j,\cdot)} \cdot (\mathbf{I} - \bar{\mathbf{I}}) + e_{j}(\bar{\mathbf{I}}, \mathbf{I} - \bar{\mathbf{I}}) \right) \\
        =& \left [ \tau_{J}'(\bar{\mathbf{I}}) \right]_{(j,\cdot)} \cdot K_{\mathbf{I}} \cdot \left [ \tau_{J}'(\bar{\mathbf{I}}) \right]_{(j,\cdot)}^{\dagger} + Cov\left( \left[ \tau_{J}'(\bar{\mathbf{I}}) \right]_{(j,\cdot)} \cdot (\mathbf{I} - \bar{\mathbf{I}}), e_{j}(\bar{\mathbf{I}}, \mathbf{I} - \bar{\mathbf{I}}) \right) \\
        &+ Cov\left( e_{j}(\bar{\mathbf{I}}, \mathbf{I} - \bar{\mathbf{I}}), \left[ \tau_{J}'(\bar{\mathbf{I}}) \right]_{(j,\cdot)} \cdot (\mathbf{I} - \bar{\mathbf{I}}) \right) + Var(e_{j}(\bar{\mathbf{I}}, \mathbf{I} - \bar{\mathbf{I}})) \\
        =& \left [ \tau_{J}'(\bar{\mathbf{I}}) \right]_{(j,\cdot)} \cdot K_{\mathbf{I}} \cdot \left [ \tau_{J}'(\bar{\mathbf{I}}) \right]_{(j,\cdot)}^{\dagger} + Var(e_{j}(\bar{\mathbf{I}}, \mathbf{I} - \bar{\mathbf{I}})) \\
        &+ 2Cov\left( \left[ \tau_{J}'(\bar{\mathbf{I}}) \right]_{(j,\cdot)} \cdot (\mathbf{I} - \bar{\mathbf{I}}), e_{j}(\bar{\mathbf{I}}, \mathbf{I} - \bar{\mathbf{I}}) \right).
    \end{split}
    \label{eq:proof_pdim_taylor_var_j}
\end{equation}

Note that the last line in Equation~(\ref{eq:proof_pdim_taylor_var_j}) follows from Equation~(\ref{eq:proof_pdim_cov_transpose}), or simply by properties of the scalar covariance operator.

\subsection{Bias Bounds}

From the projection of the estimator's Taylor expansion onto the direction of the canonical vector $\bm{\hat{\imath}}_{j}$ depicted briefly in Equation~(\ref{eq:proof_pdim_taylor_var_j}), it is possible to bound the magnitude of the bias with respect to the corresponding parameter $\alpha_{j}^{\star} = \langle\bm{\hat{\imath}}_{j}, \boldsymbol{\alpha}^{\star}\rangle$ by applying the properties of the expectation operator and the triangle inequality:

\begin{equation}
    \begin{split}
        \left| \mathbb{E}_{\mathbf{I} \sim L(\mathbf{I};\boldsymbol{\alpha}^{\star})} \left\{ \tau_{J,j}(\mathbf{I}) \right\} - \alpha_{j}^{\star} \right| &= \left| \mathbb{E}_{\mathbf{I} \sim L(\mathbf{I};\boldsymbol{\alpha}^{\star})} \left\{ \tau_{J,j}(\bar{\mathbf{I}}) + \left[ \tau_{J}'(\bar{\mathbf{I}}) \right]_{(j,\cdot)} \cdot (\mathbf{I} - \bar{\mathbf{I}}) + e_{j}(\bar{\mathbf{I}}, \mathbf{I} - \bar{\mathbf{I}}) \right\} - \alpha_{j}^{\star} \right| \\
        &= \left| \tau_{J,j}(\bar{\mathbf{I}}) + \left[ \tau_{J}'(\bar{\mathbf{I}}) \right]_{(j,\cdot)} \cdot \mathbb{E}_{\mathbf{I} \sim L(\mathbf{I};\boldsymbol{\alpha}^{\star})} \left\{ (\mathbf{I} - \bar{\mathbf{I}}) \right\} + \mathbb{E}_{\mathbf{I} \sim L(\mathbf{I};\boldsymbol{\alpha}^{\star})} \left\{ e_{j}(\bar{\mathbf{I}}, \mathbf{I} - \bar{\mathbf{I}}) \right\} - \alpha_{j}^{\star} \right| \\
        &= \left| \tau_{J,j}(\bar{\mathbf{I}}) + \mathbb{E}_{\mathbf{I} \sim L(\mathbf{I};\boldsymbol{\alpha}^{\star})} \left\{ e_{j}(\bar{\mathbf{I}}, \mathbf{I} - \bar{\mathbf{I}}) \right\} - \alpha_{j}^{\star} \right| \\
        &\leq \left| \tau_{J,j}(\bar{\mathbf{I}}) - \alpha_{j}^{\star} \right| + \left| \mathbb{E}_{\mathbf{I} \sim L(\mathbf{I};\boldsymbol{\alpha}^{\star})} \left\{ e_{j}(\bar{\mathbf{I}}, \mathbf{I} - \bar{\mathbf{I}}) \right\} \right| \\
        &\leq \left| \tau_{J,j}(\bar{\mathbf{I}}) - \alpha_{j}^{\star} \right| + \underset{t \in [0,1]}{\operatorname{max}} \left| \mathbb{E}_{\mathbf{I} \sim L(\mathbf{I};\boldsymbol{\alpha}^{\star})} \left\{ e_{j}(\bar{\mathbf{I}}, \mathbf{I} - \bar{\mathbf{I}}) \right\} \right| \\
        &\equiv \left| \tau_{J,j}(\bar{\mathbf{I}}) - \alpha_{j}^{\star} \right| + \epsilon_{J,j}, \\
        &\equiv \epsilon'_{J,j},
    \end{split}
    \label{eq:proof_eps_def}
\end{equation}

\noindent which recovers the relation of Equation~(\ref{eq:pdim_bias_bound}).

\subsection{Variance Bounds}
The expression of Equation~(\ref{eq:proof_pdim_taylor_var_j}) can be decomposed into two main elements:

\begin{equation}
    \begin{split}
        Var(\tau_{J,j}(\mathbf{I})) =& \left [ \tau_{J}'(\bar{\mathbf{I}}) \right]_{(j,\cdot)} \cdot K_{\mathbf{I}} \cdot \left [ \tau_{J}'(\bar{\mathbf{I}}) \right]_{(j,\cdot)}^{\dagger} + Var(e_{j}(\bar{\mathbf{I}}, \mathbf{I} - \bar{\mathbf{I}})) \\
        &+ 2Cov\left( \left[ \tau_{J}'(\bar{\mathbf{I}}) \right]_{(j,\cdot)} \cdot (\mathbf{I} - \bar{\mathbf{I}}), e_{j}(\bar{\mathbf{I}}, \mathbf{I} - \bar{\mathbf{I}}) \right) \\
        \equiv& \left [ \tau_{J}'(\bar{\mathbf{I}}) \right]_{(j,\cdot)} \cdot K_{\mathbf{I}} \cdot \left [ \tau_{J}'(\bar{\mathbf{I}}) \right]_{(j,\cdot)}^{\dagger} + \gamma_{J,j},
    \end{split}
\end{equation}

\noindent being $\gamma_{J,j}$ the critical term that shall allow us to find the interval depicted in Equation~(\ref{eq:pdim_var_interval}). From this last equation and the reverse triangle inequality it follows that:
\begin{equation}
    \begin{split}
        Var(\tau_{J,j}(\mathbf{I})) &= \left [ \tau_{J}'(\bar{\mathbf{I}}) \right]_{(j,\cdot)} \cdot K_{\mathbf{I}} \cdot \left [ \tau_{J}'(\bar{\mathbf{I}}) \right]_{(j,\cdot)}^{\dagger} + \gamma_{J,j} \\
        \Rightarrow \left| \gamma_{J,j} \right| &= \left| Var(\tau_{J,j}(\mathbf{I})) - \left [ \tau_{J}'(\bar{\mathbf{I}}) \right]_{(j,\cdot)} \cdot K_{\mathbf{I}} \cdot \left [ \tau_{J}'(\bar{\mathbf{I}}) \right]_{(j,\cdot)}^{\dagger} \right| \\
        &\geq \left| \left| Var(\tau_{J,j}(\mathbf{I})) \right| - \left| \left [ \tau_{J}'(\bar{\mathbf{I}}) \right]_{(j,\cdot)} \cdot K_{\mathbf{I}} \cdot \left [ \tau_{J}'(\bar{\mathbf{I}}) \right]_{(j,\cdot)}^{\dagger} \right| \right| \\
        &= \left| Var(\tau_{J,j}(\mathbf{I})) - \left [ \tau_{J}'(\bar{\mathbf{I}}) \right]_{(j,\cdot)} \cdot K_{\mathbf{I}} \cdot \left [ \tau_{J}'(\bar{\mathbf{I}}) \right]_{(j,\cdot)}^{\dagger} \right| \\
        \Rightarrow Var(\tau_{J,j}(\mathbf{I})) &\in \left[ \left [ \tau_{J}'(\bar{\mathbf{I}}) \right]_{(j,\cdot)} \cdot K_{\mathbf{I}} \cdot \left [ \tau_{J}'(\bar{\mathbf{I}}) \right]_{(j,\cdot)}^{\dagger} - \left| \gamma_{J,j} \right|, \left [ \tau_{J}'(\bar{\mathbf{I}}) \right]_{(j,\cdot)} \cdot K_{\mathbf{I}} \cdot \left [ \tau_{J}'(\bar{\mathbf{I}}) \right]_{(j,\cdot)}^{\dagger} + \left| \gamma_{J,j} \right| \right].
    \end{split}
    \label{eq:proof_pdim_var_interval_partial}
\end{equation}

Note that the line after the triangle inequality application comes from the non-negativity of $Var(\tau_{J,j}(\mathbf{I}))$ and $K_{\mathbf{I}}$ being a positive semidefinite matrix.

The interval shown in the last expression of Equation~(\ref{eq:proof_pdim_var_interval_partial}) resembles that of Equation~(\ref{eq:pdim_var_interval}), but with one key difference: in Section \ref{sec:impl_extension} the bounding interval was stated in terms of $\beta_{J,j}$ (yet to be defined; see Equation (\ref{eq:proof_beta_loose})) instead of $\gamma_{J,j}$. In what follows it is proven that:
\begin{equation}
    \left| \gamma_{J,j} \right| \leq \beta_{J,j}.
\end{equation}
Note that this inequality conveys looser bounds for $Var(\tau_{J,j}(\mathbf{I}))$, which comes from $\gamma_{J,j}$ being itself a function of the unknown variable $t \in [0,1]$ and by following similar steps of those proposed by \citet{espinosa2018}.

For that matter, let's consider each one of the terms that define $\gamma_{J,j}$ independently. Firstly, we have that:
\begin{equation}
    \begin{split}
        Var(e_{j}(\bar{\mathbf{I}}, \mathbf{I} - \bar{\mathbf{I}})) &= \mathbb{E}_{\mathbf{I} \sim L(\mathbf{I};\boldsymbol{\alpha}^{\star})} \left\{ \left( e_{j}(\bar{\mathbf{I}}, \mathbf{I} - \bar{\mathbf{I}}) - \mathbb{E}_{\mathbf{I} \sim L(\mathbf{I};\boldsymbol{\alpha}^{\star})} \left\{ e_{j}(\bar{\mathbf{I}}, \mathbf{I} - \bar{\mathbf{I}}) \right\} \right)^{2} \right\} \\
        &\leq \mathbb{E}_{\mathbf{I} \sim L(\mathbf{I};\boldsymbol{\alpha}^{\star})} \left\{ \left(  e_{j}(\bar{\mathbf{I}}, \mathbf{I} - \bar{\mathbf{I}}) \right)^{2} \right\} \\
        &\leq \underset{t \in [0,1]}{\operatorname{max}} \mathbb{E}_{\mathbf{I} \sim L(\mathbf{I};\boldsymbol{\alpha}^{\star})} \left\{ \left(  e_{j}(\bar{\mathbf{I}}, \mathbf{I} - \bar{\mathbf{I}}) \right)^{2} \right\}.
    \end{split}
    \label{eq:proof_var_e_bound}
\end{equation}

On the other hand, by taking absolute value of the second term and considering again Equations~(\ref{eq:proof_linerity_espinosa}) and~(\ref{eq:proof_pdim_cov_tau_e}), we obtain that:
\begin{equation}
    \begin{split}
        \left| Cov\left( \left[ \tau_{J}'(\bar{\mathbf{I}}) \right]_{(j,\cdot)} \cdot (\mathbf{I} - \bar{\mathbf{I}}), e_{j}(\bar{\mathbf{I}}, \mathbf{I} - \bar{\mathbf{I}}) \right) \right| &= \left| \mathbb{E}_{\mathbf{I} \sim L(\mathbf{I};\boldsymbol{\alpha}^{\star})} \left\{ \left[ \left[ \tau_{J}'(\bar{\mathbf{I}}) \right]_{(j,\cdot)} \cdot (\mathbf{I} - \bar{\mathbf{I}}) \right] \cdot e_{j}(\bar{\mathbf{I}}, \mathbf{I} - \bar{\mathbf{I}}) \right\} \right| \\
        &\leq \underset{t \in [0,1]}{\operatorname{max}}\left| \mathbb{E}_{\mathbf{I} \sim L(\mathbf{I};\boldsymbol{\alpha}^{\star})} \left\{ \left[ \left[ \tau_{J}'(\bar{\mathbf{I}}) \right]_{(j,\cdot)} \cdot (\mathbf{I} - \bar{\mathbf{I}}) \right] \cdot e_{j}(\bar{\mathbf{I}}, \mathbf{I} - \bar{\mathbf{I}}) \right\} \right|.
    \end{split}
    \label{eq:proof_cov_e_bound}
\end{equation}
Finally, combining Equations~(\ref{eq:proof_var_e_bound}) and~(\ref{eq:proof_cov_e_bound}) both the desired bound and the definition for $\beta_{J,j}$ are recovered, concluding the demonstration.

\begin{equation}
    \begin{split}
        \left| \gamma_{J,j} \right| &= \left| Var(e_{j}(\bar{\mathbf{I}}, \mathbf{I} - \bar{\mathbf{I}})) + 2Cov\left( \left[ \tau_{J}'(\bar{\mathbf{I}}) \right]_{(j,\cdot)} \cdot (\mathbf{I} - \bar{\mathbf{I}}), e_{j}(\bar{\mathbf{I}}, \mathbf{I} - \bar{\mathbf{I}}) \right) \right| \\
        &\leq \left| Var(e_{j}(\bar{\mathbf{I}}, \mathbf{I} - \bar{\mathbf{I}})) \right| + 2\left| Cov\left( \left[ \tau_{J}'(\bar{\mathbf{I}}) \right]_{(j,\cdot)} \cdot (\mathbf{I} - \bar{\mathbf{I}}), e_{j}(\bar{\mathbf{I}}, \mathbf{I} - \bar{\mathbf{I}}) \right) \right| \\
        &= Var(e_{j}(\bar{\mathbf{I}}, \mathbf{I} - \bar{\mathbf{I}})) + 2\left| Cov\left( \left[ \tau_{J}'(\bar{\mathbf{I}}) \right]_{(j,\cdot)} \cdot (\mathbf{I} - \bar{\mathbf{I}}), e_{j}(\bar{\mathbf{I}}, \mathbf{I} - \bar{\mathbf{I}}) \right) \right| \\
        &\leq \underset{t \in [0,1]}{\operatorname{max}} \mathbb{E}_{\mathbf{I} \sim L(\mathbf{I};\boldsymbol{\alpha}^{\star})} \left\{ \left(  e_{j}(\bar{\mathbf{I}}, \mathbf{I} - \bar{\mathbf{I}}) \right)^{2} \right\} + 2\underset{t \in [0,1]}{\operatorname{max}}\left| \mathbb{E}_{\mathbf{I} \sim L(\mathbf{I};\boldsymbol{\alpha}^{\star})} \left\{ \left[ \left[ \tau_{J}'(\bar{\mathbf{I}}) \right]_{(j,\cdot)} \cdot (\mathbf{I} - \bar{\mathbf{I}}) \right] \cdot e_{j}(\bar{\mathbf{I}}, \mathbf{I} - \bar{\mathbf{I}}) \right\} \right| \\
        &\equiv \beta_{J,j} \\
        \Rightarrow Var(\tau_{J,j}(\mathbf{I})) &\in \left[ \left [ \tau_{J}'(\bar{\mathbf{I}}) \right]_{(j,\cdot)} \cdot K_{\mathbf{I}} \cdot \left [ \tau_{J}'(\bar{\mathbf{I}}) \right]_{(j,\cdot)}^{\dagger} - \left| \beta_{J,j} \right|, \left [ \tau_{J}'(\bar{\mathbf{I}}) \right]_{(j,\cdot)} \cdot K_{\mathbf{I}} \cdot \left [ \tau_{J}'(\bar{\mathbf{I}}) \right]_{(j,\cdot)}^{\dagger} + \left| \beta_{J,j} \right| \right].
    \end{split}
    \label{eq:proof_beta_loose}
\end{equation}

\subsection{First- and Second-Order Derivatives for $\tau_{J}$}

A key task in the calculation of the bias- and variance- intervals as defined in previous steps is the formulation of the first- and second-order derivatives of $\tau_{J}$ w.r.t $\mathbf{I}$. In particular, the first order derivative $\tau_{J}'$ allows (among others) to find the central value of the variance interval (see Eqs.~(\ref{eq:pdim_var_interval}) and~(\ref{eq:pdim_interval_center})), and the $n\times n$ elements of the second-order derivative matrix, denoted as $\frac{\partial^{2}}{\partial I_i \partial I_h} \tau_{J}, i,h\in\{1,\dots,n\}$, are required in the calculation of the error term in Equation~(\ref{eq:proof_pdim_e}) and consequently the expectations in Eqs.~(\ref{eq:proof_eps_def}) and~(\ref{eq:proof_beta_loose}).

The condition in Equation~(\ref{eq:proof_implicit_foc}) can be represented as a function $\Psi: \mathbb{R}^{n} \to \mathbb{R}^{p}$

\begin{equation}
    \Psi(\mathbf{I}) = \begin{bmatrix}
             \frac{\partial}{\partial\alpha_{1}}J(\tau_{J}(\mathbf{I}), \mathbf{I}) \\
             \vdots  \\
             \frac{\partial}{\partial\alpha_{p}}J(\tau_{J}(\mathbf{I}), \mathbf{I})
           \end{bmatrix},
    \label{eq:proof_psi_function}
\end{equation}

\noindent whose zero is $\boldsymbol{\alpha} = \tau_{J}(\mathbf{I})$.

In order to find the Taylor expansion depicted in Equations~(\ref{eq:proof_pdim_taylor}) and~(\ref{eq:proof_pdim_e}) one can draw upon the chain rule to take the partial derivative $\frac{\partial}{\partial I_{i}}$ on both sides of Equation~(\ref{eq:proof_implicit_foc}), resulting in one of the $n$ sets of $p$ equations on $p$ unknowns of the form:
\begin{equation}
    \begin{split}
        \mathbf{0} &= \frac{\partial}{\partial I_{i}} \Psi(\mathbf{I}) \\
        &= \begin{bmatrix}
            \sum_{k=1}^{p} \frac{\partial^{2}}{\partial\alpha_{1} \partial\alpha_{k}} J(\tau_{J}(\mathbf{I}),\mathbf{I}) \cdot \frac{\partial}{\partial I_{i}} \tau_{J,k}(\mathbf{I}) + \frac{\partial^{2}}{\partial\alpha_{1} \partial I_{i}} J(\tau_{J}(\mathbf{I}),\mathbf{I}) \\
            \vdots \\
            \sum_{k=1}^{p} \frac{\partial^{2}}{\partial\alpha_{p} \partial\alpha_{k}} J(\tau_{J}(\mathbf{I}),\mathbf{I}) \cdot \frac{\partial}{\partial I_{i}} \tau_{J,k}(\mathbf{I}) + \frac{\partial^{2}}{\partial\alpha_{p} \partial I_{i}} J(\tau_{J}(\mathbf{I}),\mathbf{I})
        \end{bmatrix}.
    \end{split}
    \label{eq:proof_chain1_i}
\end{equation}

For ease of notation, this last expression can be rewritten in matrix form as follows
\begin{equation}
    \begin{split}
        \mathbf{0} &= \begin{bmatrix}
        \frac{\partial^{2}}{\partial\alpha_{1} \partial\alpha_{1}} J(\tau_{J}(\mathbf{I}), \mathbf{I}) & \dots & \frac{\partial^{2}}{\partial\alpha_{1} \partial\alpha_{p}} J(\tau_{J}(\mathbf{I}), \mathbf{I}) \\
        \vdots & \ddots & \vdots \\
        \frac{\partial^{2}}{\partial\alpha_{p} \partial\alpha_{1}} J(\tau_{J}(\mathbf{I}), \mathbf{I}) & \dots & \frac{\partial^{2}}{\partial\alpha_{p} \partial\alpha_{p}} J(\tau_{J}(\mathbf{I}), \mathbf{I})
        \end{bmatrix} \cdot \begin{bmatrix}
            \frac{\partial}{\partial I_{i}} \tau_{J,1}(\mathbf{I}) \\
            \vdots \\
            \frac{\partial}{\partial I_{i}} \tau_{J,p}(\mathbf{I})
        \end{bmatrix} + \begin{bmatrix}
            \frac{\partial^{2}}{\partial\alpha_{1} \partial I_{i}} J(\tau_{J}(\mathbf{I}), \mathbf{I}) \\
            \vdots \\
            \frac{\partial^{2}}{\partial\alpha_{p} \partial I_{i}} J(\tau_{J}(\mathbf{I}), \mathbf{I})
        \end{bmatrix} \\
        &= \left[ \nabla^{20} J(\tau_{J}(\mathbf{I}), \mathbf{I}) \right] \cdot \begin{bmatrix}
            \frac{\partial}{\partial I_{i}} \tau_{J,1}(\mathbf{I}) \\
            \vdots \\
            \frac{\partial}{\partial I_{i}} \tau_{J,p}(\mathbf{I})
        \end{bmatrix} + \left[ \nabla^{11} J(\tau_{J}(\mathbf{I}), \mathbf{I}) \right]_{(\cdot, i)}.
    \end{split}
    \label{eq:proof_chain1_i_matrix}
\end{equation}

Noting that the total derivatives of $\Psi$ and $\tau_{J}$ can be written, respectively, as
\begin{equation}
    \Psi'(\mathbf{I}) = \begin{bmatrix}
        \frac{\partial}{\partial I_{1}} \Psi(\mathbf{I}) & \dots & \frac{\partial}{\partial I_{n}} \Psi(\mathbf{I})
    \end{bmatrix} \in \mathbb{R}^{p \times n},
    \label{eq:proof_psi_prima}
\end{equation}

\begin{equation}
    \tau_{J}'(\mathbf{I}) = \begin{bmatrix}
        \frac{\partial}{\partial I_{1}} \tau_{J,1}(\mathbf{I}) & \dots & \frac{\partial}{\partial I_{n}} \tau_{J,1}(\mathbf{I}) \\
        \vdots & \ddots & \vdots \\
        \frac{\partial}{\partial I_{1}} \tau_{J,p}(\mathbf{I}) & \dots & \frac{\partial}{\partial I_{n}} \tau_{J,p}(\mathbf{I})
    \end{bmatrix} \in \mathbb{R}^{p \times n},
    \label{eq:proof_tau_prima}
\end{equation}

\noindent the $n$ sets of linear systems depicted in Equation~(\ref{eq:proof_chain1_i_matrix}) can be merged into\footnote{Note the abuse of notation, where the left-hand side of Equation~(\ref{eq:proof_chain1_full_matrix}) represents a $p \times n$ null matrix.}:
\begin{equation}
    \begin{split}
        \mathbf{0} &= \Psi'(\mathbf{I}) \\
        &= \left[ \nabla^{20} J(\tau_{J}(\mathbf{I}), \mathbf{I}) \right] \cdot \begin{bmatrix}
        \frac{\partial}{\partial I_{1}} \tau_{J,1}(\mathbf{I}) & \dots & \frac{\partial}{\partial I_{n}} \tau_{J,1}(\mathbf{I}) \\
        \vdots & \ddots & \vdots \\
        \frac{\partial}{\partial I_{1}} \tau_{J,p}(\mathbf{I}) & \dots & \frac{\partial}{\partial I_{n}} \tau_{J,p}(\mathbf{I})
    \end{bmatrix} + \left[ \nabla^{11} J(\tau_{J}(\mathbf{I}), \mathbf{I}) \right] \\
    &= \left[ \nabla^{20} J(\tau_{J}(\mathbf{I}), \mathbf{I}) \right] \cdot \tau_{J}'(\mathbf{I}) + \left[ \nabla^{11} J(\tau_{J}(\mathbf{I}), \mathbf{I}) \right].
    \end{split} 
    \label{eq:proof_chain1_full_matrix}
\end{equation}

Finally, Equation~(\ref{eq:pdim_tau_tilde1}) is recovered, conditional on $\nabla^{20} J(\tau_{J}(\mathbf{I}), \mathbf{I})$ being an invertible matrix.
\begin{equation}
    \tau_{J}'(\mathbf{I}) = - \left[ \nabla^{20} J(\tau_{J}(\mathbf{I}), \mathbf{I}) \right]^{-1} \cdot \left[ \nabla^{11} J(\tau_{J}(\mathbf{I}), \mathbf{I}) \right].
\end{equation}

Employing the chain rule once again to take the partial derivative $\frac{\partial}{\partial I_{h}}$ on both sides of Equation~(\ref{eq:proof_chain1_i}) (more precisely on each one of the components of said equation), Fessler's Equation~(15) \citep{fessler1996} is recovered for each $j \in \{ 1,\dots,p \}$:
\begin{equation}
    \begin{split}
        0 = &\sum_{k=1}^{p} \left[ \left( \sum_{l=1}^{p} \left[ \frac{\partial^{3}}{\partial\alpha_{j} \partial\alpha_{k} \partial\alpha_{l}} J(\tau_{J}(\mathbf{I}),\mathbf{I}) \cdot \frac{\partial}{\partial I_{h}} \tau_{J,l}(\mathbf{I}) \right] + \frac{\partial^{3}}{\partial\alpha_{j} \partial\alpha_{k} \partial I_{h}} J(\tau_{J}(\mathbf{I}),\mathbf{I}) \right) \cdot \frac{\partial}{\partial I_{i}} \tau_{J,k}(\mathbf{I}) \right] \\
        &+ \sum_{k=1}^{p} \left[ \frac{\partial^{2}}{\partial\alpha_{j} \partial\alpha_{k}} J(\tau_{J}(\mathbf{I}),\mathbf{I}) \cdot \frac{\partial^{2}}{\partial I_{h} \partial I_{i}} \tau_{J,k}(\mathbf{I}) \right] \\
        &+ \sum_{k=1}^{p} \left[ \frac{\partial^{3}}{\partial\alpha_{j} \partial\alpha_{k} \partial I_{i}} J(\tau_{J}(\mathbf{I}),\mathbf{I}) \cdot \frac{\partial}{\partial I_{h}} \tau_{J,k}(\mathbf{I}) \right] + \frac{\partial^{3}}{\partial\alpha_{j} \partial I_{h} \partial I_{i}} J(\tau_{J}(\mathbf{I}),\mathbf{I}).
    \end{split}
    \label{eq:proof_fesslers15}
\end{equation}

Let us then denote the following:
\begin{equation}
    P_{J,j}^{i,h}(\mathbf{I}) = \sum_{k=1}^{p} \left[ \left( \sum_{l=1}^{p} \left[ \frac{\partial^{3}}{\partial\alpha_{j} \partial\alpha_{k} \partial\alpha_{l}} J(\tau_{J}(\mathbf{I}),\mathbf{I}) \cdot \frac{\partial}{\partial I_{h}} \tau_{J,l}(\mathbf{I}) \right] + \frac{\partial^{3}}{\partial\alpha_{j} \partial\alpha_{k} \partial I_{h}} J(\tau_{J}(\mathbf{I}),\mathbf{I}) \right) \cdot \frac{\partial}{\partial I_{i}} \tau_{J,k}(\mathbf{I}) \right],
    \label{eq:proof_p_j}
\end{equation}

\begin{equation}
    Q_{J,j}^{i,h}(\mathbf{I}) = \sum_{k=1}^{p} \left[ \frac{\partial^{3}}{\partial\alpha_{j} \partial\alpha_{k} \partial I_{i}} J(\tau_{J}(\mathbf{I}),\mathbf{I}) \cdot \frac{\partial}{\partial I_{h}} \tau_{J,k}(\mathbf{I}) \right] + \frac{\partial^{3}}{\partial\alpha_{j} \partial I_{h} \partial I_{i}} J(\tau_{J}(\mathbf{I}),\mathbf{I}).
    \label{eq:proof_q_j}
\end{equation}

Therefore, Equation~(\ref{eq:proof_fesslers15}) can be rewritten as
\begin{equation}
    \begin{split}
        0 &= P_{J,j}^{i,h}(\mathbf{I}) + \sum_{k=1}^{p} \left[ \frac{\partial^{2}}{\partial\alpha_{j} \partial\alpha_{k}} J(\tau_{J}(\mathbf{I}),\mathbf{I}) \cdot \frac{\partial^{2}}{\partial I_{h} \partial I_{i}} \tau_{J,k}(\mathbf{I}) \right] + Q_{J,j}^{i,h}(\mathbf{I}) \\
        &= P_{J,j}^{i,h}(\mathbf{I}) + \left[ \nabla^{20}J(\tau_{J}(\mathbf{I}),\mathbf{I}) \right]_{(j,\cdot)} \cdot \begin{bmatrix}
        \frac{\partial^{2}}{\partial I_{i} \partial I_{h}} \tau_{J,1}(\mathbf{I}) & \dots & \frac{\partial^{2}}{\partial I_{i} \partial I_{h}} \tau_{J,p}(\mathbf{I})
    \end{bmatrix} ^{\intercal} + Q_{J,j}^{i,h}(\mathbf{I}) \\
        &= P_{J,j}^{i,h}(\mathbf{I}) + \left[ \nabla^{20}J(\tau_{J}(\mathbf{I}),\mathbf{I}) \right]_{(j,\cdot)} \cdot \frac{\partial^{2}}{\partial I_{i} \partial I_{h}} \tau_{J}(\mathbf{I}) + Q_{J,j}^{i,h}(\mathbf{I}).
    \end{split}
\end{equation}

Written in that form, we can stack the $p$ equations and get the matrix form that allows to find the second derivative of $\tau_{J}(\cdot)$ as follows:
\begin{equation}
    \begin{split}
        \mathbf{0} &= \begin{bmatrix}
            P_{J,1}^{i,h}(\mathbf{I}) \\
            \vdots \\
            P_{J,p}^{i,h}(\mathbf{I})
        \end{bmatrix} + \begin{bmatrix}
            \left[ \nabla^{20}J(\tau_{J}(\mathbf{I}),\mathbf{I}) \right]_{(1,\cdot)} \\
            \vdots \\
            \left[ \nabla^{20}J(\tau_{J}(\mathbf{I}),\mathbf{I}) \right]_{(p,\cdot)}
        \end{bmatrix} \cdot \frac{\partial^{2}}{\partial I_{i} \partial I_{h}} \tau_{J}(\mathbf{I}) + \begin{bmatrix}
            Q_{J,1}^{i,h}(\mathbf{I}) \\
            \vdots \\
            Q_{J,p}^{i,h}(\mathbf{I})
        \end{bmatrix} \\
        &\equiv P_{J}^{i,h}(\mathbf{I}) + \nabla^{20}J(\tau_{J}(\mathbf{I}),\mathbf{I}) \cdot \frac{\partial^{2}}{\partial I_{i} \partial I_{h}} \tau_{J}(\mathbf{I}) + Q_{J}^{i,h}(\mathbf{I}) \\
        \Rightarrow \frac{\partial^{2}}{\partial I_{i} \partial I_{h}} \tau_{J}(\mathbf{I}) &= -\left[ \nabla^{20}J(\tau_{J}(\mathbf{I}),\mathbf{I}) \right]^{-1} \cdot \left[ P_{J}^{i,h}(\mathbf{I}) + Q_{J}^{i,h}(\mathbf{I}) \right].
    \end{split}
    \label{eq:proof_pdim_dtau2}
\end{equation}

In order to complete the demonstration in a more succint manner, it suffices to describe $P^{i,h}(\mathbf{I})$ and $Q^{i,h}(\mathbf{I})$ in a tensor form. For that matter, it serves to treat said vectors' individual components, depicted in Equations~(\ref{eq:proof_p_j}) and~(\ref{eq:proof_q_j}). Let's take then the latter, which can be easily rewritten as:
\begin{equation}
    \begin{split}
        Q_{J,j}^{i,h}(\mathbf{I}) =& \sum_{k=1}^{p} \left[ \frac{\partial^{3}}{\partial\alpha_{j} \partial\alpha_{k} \partial I_{i}} J(\tau_{J}(\mathbf{I}),\mathbf{I}) \cdot \frac{\partial}{\partial I_{h}} \tau_{J,k}(\mathbf{I}) \right] + \frac{\partial^{3}}{\partial\alpha_{j} \partial I_{h} \partial I_{i}} J(\tau_{J}(\mathbf{I}),\mathbf{I}) \\
        =& \begin{bmatrix}
            \frac{\partial^{3}}{\partial \alpha_{j} \partial \alpha_{1} \partial I_{i}} J(\tau_{J}(\mathbf{I}),\mathbf{I}) & \dots & \frac{\partial^{3}}{\partial \alpha_{j} \partial \alpha_{p} \partial I_{i}} J(\tau_{J}(\mathbf{I}),\mathbf{I})
        \end{bmatrix} \cdot \begin{bmatrix}
            \frac{\partial}{\partial I_{h}} \tau_{J,1}(\mathbf{I}) \\
            \vdots \\
            \frac{\partial}{\partial I_{h}} \tau_{J,p}(\mathbf{I})
        \end{bmatrix} + \left[ \nabla^{12} J(\tau_{J}(\mathbf{I}),\mathbf{I}) \right]_{(j,i,h)} \\
        =& \left[ \nabla^{21} J(\tau_{J}(\mathbf{I}),\mathbf{I}) \right]_{(j,\cdot,i)} \cdot \left[ \tau_{J}'(\mathbf{I}) \right]_{(\cdot,h)} + \left[ \nabla^{12} J(\tau_{J}(\mathbf{I}),\mathbf{I}) \right]_{(j,i,h)}.
    \end{split}
\end{equation}

Following similar steps, {\em i.e.} identifying matrix forms, operations and considering the notation introduced by Equation~(\ref{eq:nabla_notation}), the following can be found:
\begin{equation}
    \begin{split}
        P_{J,j}^{i,h}(\mathbf{I}) =& \sum_{k=1}^{p} \left[ \left( \sum_{l=1}^{p} \left[ \frac{\partial^{3}}{\partial\alpha_{j} \partial\alpha_{k} \partial\alpha_{l}} J(\tau_{J}(\mathbf{I}),\mathbf{I}) \cdot \frac{\partial}{\partial I_{h}} \tau_{J,l}(\mathbf{I}) \right] + \frac{\partial^{3}}{\partial\alpha_{j} \partial\alpha_{k} \partial I_{h}} J(\tau_{J}(\mathbf{I}),\mathbf{I}) \right) \cdot \frac{\partial}{\partial I_{i}} \tau_{J,k}(\mathbf{I}) \right] \\
        =& \sum_{k=1}^{p}\left[ \sum_{l=1}^{p} \left[ \frac{\partial^{3}}{\partial\alpha_{j} \partial\alpha_{k} \partial\alpha_{l}} J(\tau_{J}(\mathbf{I}),\mathbf{I}) \cdot \frac{\partial}{\partial I_{h}} \tau_{J,l}(\mathbf{I}) \cdot \frac{\partial}{\partial I_{i}} \tau_{J,k}(\mathbf{I}) \right] \right] \\
        &+ \sum_{k=1}^{p}\left[ \frac{\partial^{3}}{\partial\alpha_{j} \partial\alpha_{k} \partial I_{h}} J(\tau_{J}(\mathbf{I}),\mathbf{I}) \cdot \frac{\partial}{\partial I_{i}} \tau_{J,k}(\mathbf{I}) \right] \\
        =& \begin{bmatrix}
            \frac{\partial}{\partial I_{i}} \tau_{J,1}(\mathbf{I}) & \dots & \frac{\partial}{\partial I_{i}} \tau_{J,p}(\mathbf{I})
        \end{bmatrix} \cdot \begin{bmatrix}
            \frac{\partial^{3}}{\partial\alpha_{j} \partial\alpha_{1} \partial\alpha_{1}} J(\tau_{J}(\mathbf{I}),\mathbf{I}) & \cdots & \frac{\partial^{3}}{\partial\alpha_{j} \partial\alpha_{1} \partial\alpha_{p}} J(\tau_{J}(\mathbf{I}),\mathbf{I}) \\
            \vdots & \ddots & \vdots \\
            \frac{\partial^{3}}{\partial\alpha_{j} \partial\alpha_{p} \partial\alpha_{1}} J(\tau_{J}(\mathbf{I}),\mathbf{I}) & \cdots & \frac{\partial^{3}}{\partial\alpha_{j} \partial\alpha_{p} \partial\alpha_{p}} J(\tau_{J}(\mathbf{I}),\mathbf{I})
        \end{bmatrix} \cdot \begin{bmatrix}
            \frac{\partial}{\partial I_{h}} \tau_{J,1}(\mathbf{I}) \\
            \vdots \\
            \frac{\partial}{\partial I_{h}} \tau_{J,p}(\mathbf{I})
        \end{bmatrix} \\
        &+ \begin{bmatrix}
            \frac{\partial^{3}}{\partial \alpha_{j} \partial \alpha_{1} \partial I_{h}} J(\tau_{J}(\mathbf{I}),\mathbf{I}) & \dots & \frac{\partial^{3}}{\partial \alpha_{j} \partial \alpha_{p} \partial I_{h}} J(\tau_{J}(\mathbf{I}),\mathbf{I})
        \end{bmatrix} \cdot \begin{bmatrix}
            \frac{\partial}{\partial I_{i}} \tau_{J,1}(\mathbf{I}) \\
            \vdots \\
            \frac{\partial}{\partial I_{i}} \tau_{J,p}(\mathbf{I})
        \end{bmatrix} \\
        =&\left[ \tau_{J}'(\mathbf{I}) \right]_{(\cdot,i)}^{\intercal} \cdot \left[ \nabla^{30} J(\tau_{J}(\mathbf{I}),\mathbf{I}) \right]_{(j,\cdot,\cdot)} \cdot \left[ \tau_{J}'(\mathbf{I}) \right]_{(\cdot,h)} + \left[ \nabla^{21} J(\tau_{J}(\mathbf{I}),\mathbf{I}) \right]_{(j,\cdot,h)} \cdot \left[ \tau_{J}'(\mathbf{I}) \right]_{(\cdot,i)},
    \end{split}
\end{equation}

Then, by stacking the these expressions into the corresponding vectors $P_{J}^{i,h}(\mathbf{I})$ and $Q_{J}^{i,h}(\mathbf{I})$ used as in Equation (\ref{eq:proof_pdim_dtau2}), we have the following:
\begin{equation}
    \begin{split}
        P_{J}^{i,h}(\mathbf{I}) &\equiv \left[ \left[ \tau_{J}'(\mathbf{I}) \right]_{(\cdot,i)}^{T} \cdot \left[ \nabla^{30} J(\tau_{J}(\mathbf{I}), \mathbf{I}) \right]_{(j,\cdot,\cdot)} \cdot \left[ \tau_{J}'(\mathbf{I}) \right]_{(\cdot,h)} + \left[ \nabla^{21} J(\tau_{J}(\mathbf{I}), \mathbf{I}) \right]_{(j,\cdot,h)} \cdot\left[ \tau_{J}'(\mathbf{I}) \right]_{(\cdot,i)} \right]_{j=1}^{p} \in \mathbb{R}^{p \times 1},
    \end{split}
    \label{eq:pdim_Pih}
\end{equation}
\begin{equation}
    \begin{split}
        Q_{J}^{i,h}(\mathbf{I}) &\equiv \left[ \left[ \nabla^{21} J(\tau_{J}(\mathbf{I}), \mathbf{I}) \right]_{(j,\cdot,i)} \cdot \left[ \tau_{J}'(\mathbf{I}) \right]_{(\cdot,h)} + \left[ \nabla^{12} J(\tau_{J}(\mathbf{I}), \mathbf{I}) \right]_{(j,i,h)} \right]_{j=1}^{p} \in \mathbb{R}^{p \times 1}.
    \end{split}
    \label{eq:pdim_Qih}
\end{equation}


By plugging these expressions in Equation~(\ref{eq:proof_pdim_dtau2}), the second-order derivatives $\frac{\partial^{2}}{\partial I_{i} \partial I_{h}} \tau_{J}(\mathbf{I})$ can be calculated, along with $\mathbf{e}(\bar{\mathbf{I}}, \mathbf{I} - \bar{\mathbf{I}})$ in (\ref{eq:proof_pdim_e}) and the performance bounds formulated here.

\section{Proof: WLS Estimator for the Joint Estimation of Flux and Background}
\label{app:proof_WLS_FB}

We start our demonstration by showing that the WLS functional is convex over $\boldsymbol{\Theta}$, provided a fixed set of weights $\{w_{i}\}_{i=1}^{n}$ and observation vector $\mathbf{I}$.

For that matter, let's consider $\boldsymbol{\alpha}_{A} = (\alpha_{1}, \alpha_{2}), \boldsymbol{\alpha}_{B} = (\alpha_{1}', \alpha_{2}') \in \boldsymbol{\Theta}$ be two vectors on the parameter space and some real $t  \in [0,1]$. Then,
\begin{equation}
    \begin{split}
        J_{WLS}(t\boldsymbol{\alpha}_{A} + (1-t)\boldsymbol{\alpha}_{B}, \mathbf{I}) &= \sum_{i =1}^n w_{i}\left( I_{i} - (t\alpha_{1} + (1-t)\alpha_{1}') \cdot g_{i}(x_{c}) - (t\alpha_{2} + (1-t)\alpha_{2}') \right)^{2} \\
        &= \sum_{i =1}^n w_{i} \left( I_{i} - t(\alpha_{1} \cdot g_{i(x_{c})} + \alpha_{2}) - (1-t)(\alpha_{1}' \cdot g_{i(x_{c})} + \alpha_{2}') \right)^{2} \\
        &= \sum_{i =1}^n w_{i} \left( t(I_{i} - \alpha_{1} \cdot g_{i(x_{c})} - \alpha_{2}) + (1-t)(I_{i} - \alpha_{1}' \cdot g_{i(x_{c})} - \alpha_{2}') \right)^{2} \\
        &\leq \sum_{i =1}^n w_{i} \left( t\left( I_{i} - \alpha_{1} \cdot g_{i(x_{c})} - \alpha_{2} \right)^{2} + (1-t)\left( I_{i} - \alpha_{1}' \cdot g_{i(x_{c})} - \alpha_{2}' \right)^{2} \right) \\
        &= t\sum_{i =1}^n w_{i}\left( I_{i} - \alpha_{1} \cdot g_{i(x_{c})} - \alpha_{2} \right)^{2} + (1-t)\sum_{i =1}^n w_{i} \left( I_{i} - \alpha_{1}' \cdot g_{i(x_{c})} - \alpha_{2}' \right)^{2} \\
        &= tJ_{WLS}(\boldsymbol{\alpha}_{A},\mathbf{I}) + (1-t)J_{WLS}(\boldsymbol{\alpha}_{B},\mathbf{I}) 
    \end{split}
    \label{proof:WLS_convexity}
\end{equation}\qed

\subsection{Linear WLS Solution}

As $J_{WLS}$ is a convex function over $\boldsymbol{\Theta}$, the posed optimization problem of Equation~(\ref{eq:WLS_tau_imp}) has a unique solution. This allows us to apply the methodology proposed in Section \ref{sec:impl_extension}, but also (in this particular case) to find a closed expression for $\tau_{WLS}$. To do that, we resort to the first order condition of Equation~(\ref{eq:implicit_foc}):
\begin{equation}
    \begin{split}
        \begin{bmatrix}
            \frac{\partial}{\partial\alpha_{1}}J_{WLS}(\boldsymbol{\alpha}, \mathbf{I}) \\
            \frac{\partial}{\partial\alpha_{2}}J_{WLS}(\boldsymbol{\alpha}, \mathbf{I})
        \end{bmatrix} \biggr\rvert_{\boldsymbol{\alpha} = \tau_{WLS}(\mathbf{I})} &= \mathbf{0} \\
        \Rightarrow \begin{bmatrix}
            \sum_{i =1}^n w_{i}g_{i}(x_{c}) \left( \alpha_{1}g_{i}(x_{c}) + \alpha_{2} \right) \\
            \sum_{i =1}^n w_{i} \left( \alpha_{1}g_{i}(x_{c}) + \alpha_{2} \right)
        \end{bmatrix}\biggr\rvert_{\boldsymbol{\alpha} = \tau_{WLS}(\mathbf{I})} &= \begin{bmatrix}
            \sum_{i =1}^n w_{i}g_{i}(x_{c})I_{i} \\
            \sum_{i =1}^n w_{i}I_{i}
        \end{bmatrix} \\
        \Rightarrow \begin{bmatrix}
            \sum_{i =1}^n w_{i}g_{i}^{2}(x_{c}) & \sum_{i =1}^n w_{i}g_{i}(x_{c}) \\
            \sum_{i =1}^n w_{i}g_{i}(x_{c}) & \sum_{i =1}^n w_{i}
        \end{bmatrix} \cdot \begin{bmatrix}
            \alpha_{1} \\
            \alpha_{2}
        \end{bmatrix}\biggr\rvert_{\boldsymbol{\alpha} = \tau_{WLS}(\mathbf{I})} &= \begin{bmatrix}
            w_{1}g_{1}(x_{c}) & \dots & w_{n}g_{n}(x_{c}) \\
            w_{1} & \dots & w_{n}
        \end{bmatrix} \cdot \begin{bmatrix}
            I_{1} \\
            \vdots \\
            I_{n}
        \end{bmatrix} \\
        \Rightarrow \frac{1}{2} \nabla^{20} J_{WLS}(\tau_{WLS}(\mathbf{I}),\mathbf{I}) \cdot \tau_{WLS}(\mathbf{I}) &= -\frac{1}{2} \nabla^{11} J_{WLS}(\tau_{WLS}(\mathbf{I}),\mathbf{I}) \cdot \mathbf{I}.
    \end{split}
\end{equation}

Therefore, as $\nabla^{20} J_{WLS}(\tau_{WLS}(\mathbf{I}),\mathbf{I})$ is invertible as a result of $J_{WLS}$ being convex over $\boldsymbol{\Theta}$, we have a linear estimator given by
\begin{equation}
    \begin{split}
        \tau_{WLS}(\mathbf{I}) &= -[\nabla^{20} J_{WLS}(\tau_{WLS}(\bar{\mathbf{I}}),\bar{\mathbf{I}})]^{-1} \cdot \nabla^{11} J_{WLS}(\tau_{WLS}(\bar{\mathbf{I}}),\bar{\mathbf{I}}) \cdot \mathbf{I} \\
        &= \tau'_{WLS}(\bar{\mathbf{I}}) \cdot \mathbf{I} \,
    \end{split}
\end{equation}
from which we have $\bar{\tau}_{WLS} = \tau_{WLS}'(\bar{\mathbf{I}}) \cdot \bar{\mathbf{I}}$ and $K_{\tau_{WLS}(\mathbf{I})} = \tau_{WLS}'(\bar{\mathbf{I}}) \cdot K_{\mathbf{I}} \cdot [\tau_{WLS}'(\bar{\mathbf{I}})]^{\dagger}$.

It may be interesting to note that the same form for $\tau_{WLS}(\cdot)$ can be found by means of the methodology of Section~\ref{sec:impl_extension}. Briefly put, it can be verified that the third-order derivatives that comprise $\nabla^{12}J_{WLS}(\boldsymbol{\alpha},\mathbf{I})$, $\nabla^{21}J_{WLS}(\boldsymbol{\alpha},\mathbf{I})$ and $\nabla^{30}J_{WLS}(\boldsymbol{\alpha},\mathbf{I})$ are equal to zero, {\em i.e.} $\forall j,k,l \in \{1,2\}, \forall i,h\in\{1,\dots,n\}$
\begin{equation}
    \frac{\partial^{3}}{\partial\alpha_{j}\partial I_{i} \partial I_{h}} J_{WLS}(\boldsymbol{\alpha},\mathbf{I}) = 0, \hspace{3mm} \frac{\partial^{3}}{\partial\alpha_{j}\partial\alpha_{k}\partial I_{h}} J_{WLS}(\boldsymbol{\alpha},\mathbf{I}) = 0, \hspace{3mm} \frac{\partial^{3}}{\partial\alpha_{j}\partial\alpha_{k}\partial\alpha_{l}} J_{WLS}(\boldsymbol{\alpha},\mathbf{I}) = 0,
\end{equation}


\noindent which implies, from Equations~(\ref{eq:proof_pdim_dtau2}),~(\ref{eq:pdim_Pih}) and~(\ref{eq:pdim_Qih}) that $\tau_{WLS}''(\mathbf{I}) = \mathbf{0}$ and, consequently, $\mathbf{e}(\bar{\mathbf{I}}, \mathbf{I} - \bar{\mathbf{I}}) = \mathbf{0}$. On that account, the corresponding Taylor approximation around $\bar{\mathbf{I}}$ reduces to 
\begin{equation}
    \begin{split}
        \tau_{WLS}(\bar{\mathbf{I}}) + \tau_{WLS}'(\bar{\mathbf{I}}) \cdot (\mathbf{I} - \bar{\mathbf{I}}) &= \tau_{WLS}'(\bar{\mathbf{I}}) \cdot \mathbf{I} + \tau_{WLS}(\bar{\mathbf{I}}) - \tau_{WLS}'(\bar{\mathbf{I}}) \cdot \bar{\mathbf{I}} \\
        &= \tau_{WLS}'(\bar{\mathbf{I}}) \cdot \mathbf{I} + \tau_{WLS}'(\bar{\mathbf{I}}) \cdot \bar{\mathbf{I}} - \tau_{WLS}'(\bar{\mathbf{I}}) \cdot \bar{\mathbf{I}} \\
        &= \tau_{WLS}'(\bar{\mathbf{I}}) \cdot \mathbf{I} \\
        &= \tau_{WLS}(\mathbf{I}).
    \end{split}
\end{equation}

This equivalence between $\tau_{WLS}$ and its Taylor expansion is indeed evident from the linear nature of $\tau_{WLS}$, but the importance of this analysis is to show some level of consistency between this particular case and the general estimator studied in Section \ref{sec:impl_extension}.

\subsection{Optimal Weight Set}

For the sake of completeness, recalling that $\tau_{WLS}'(\mathbf{I}) = -[\nabla^{20} J_{WLS}(\tau_{WLS}(\mathbf{I}),\mathbf{I})]^{-1} \cdot \nabla^{11} J_{WLS}(\tau_{WLS}(\mathbf{I}),\mathbf{I})$ is a constant matrix given a set of weights $\{ w_{i} \}_{i =1}^n$, we now show that the weight selection given by Equation~(\ref{eq:opt_weights}), {\em i.e.}:
\begin{equation}
    w_{i} = \frac{K}{\lambda_{i}(x_{c}, \tilde{F})}, \hspace{3mm} \forall i \in \{1,\dots,n\}, \hspace{3mm} K>0,
\end{equation}

\noindent allows the WLS estimator to achieve optimality in the Cramér-Rao sense.

For that matter, it suffices to evaluate said weight set in $\tau_{WLS}'(\mathbf{I})$:
\begin{equation}
    \begin{split}
        \tau_{WLS}'(\mathbf{I}) &= \frac{1}{K} \begin{bmatrix}
            \sum_{i =1}^n \frac{g_{i}^{2}(x_{c})}{\lambda_{i}(x_{c},\tilde{F})} & \sum_{i =1}^n \frac{g_{i}(x_{c})}{\lambda_{i}(x_{c},\tilde{F})} \\
            \sum_{i =1}^n \frac{g_{i}(x_{c})}{\lambda_{i}(x_{c},\tilde{F})} & \sum_{i =1}^n \frac{1}{\lambda_{i}(x_{c},\tilde{F})}
        \end{bmatrix} ^{-1} \cdot K \begin{bmatrix}
            \frac{g_{1}(x_{c})}{\lambda_{1}(x_{c},\tilde{F})} & \dots & \frac{g_{n}(x_{c})}{\lambda_{n}(x_{c},\tilde{F})} \\
            \frac{1}{\lambda_{1}(x_{c},\tilde{F})} & \dots & \frac{1}{\lambda_{n}(x_{c},\tilde{F})}
        \end{bmatrix} \\
        &= \mathcal{I}_{\boldsymbol{\alpha}^{\star}}^{-1} \cdot \begin{bmatrix}
            \frac{g_{1}(x_{c})}{\lambda_{1}(x_{c},\tilde{F})} & \dots & \frac{g_{n}(x_{c})}{\lambda_{n}(x_{c},\tilde{F})} \\
            \frac{1}{\lambda_{1}(x_{c},\tilde{F})} & \dots & \frac{1}{\lambda_{n}(x_{c},\tilde{F})}
        \end{bmatrix},
    \end{split}
\end{equation}

\noindent where the structure of the estimation problem's Fisher matrix can be identified. Therefore,
\begin{equation}
    \begin{split}
        K_{\tau_{WLS}(\mathbf{I})} &= \tau_{WLS}'(\mathbf{I}) \cdot K_{\mathbf{I}} \cdot \tau_{WLS}'(\mathbf{I})^{\intercal} \\
        &= \mathcal{I}_{\boldsymbol{\alpha}^{\star}}^{-1} \cdot \begin{bmatrix}
            \frac{g_{1}(x_{c})}{\lambda_{1}(x_{c},\tilde{F})} & \dots & \frac{g_{n}(x_{c})}{\lambda_{n}(x_{c},\tilde{F})} \\
            \frac{1}{\lambda_{1}(x_{c},\tilde{F})} & \dots & \frac{1}{\lambda_{n}(x_{c},\tilde{F})}
        \end{bmatrix} \cdot K_{\mathbf{I}} \cdot \begin{bmatrix}
            \frac{g_{1}(x_{c})}{\lambda_{1}(x_{c},\tilde{F})} & \frac{1}{\lambda_{1}(x_{c},\tilde{F})} \\
            \vdots & \vdots \\
            \frac{g_{n}(x_{c})}{\lambda_{n}(x_{c},\tilde{F})} & \frac{1}{\lambda_{n}(x_{c},\tilde{F})}
        \end{bmatrix} \cdot \mathcal{I}_{\boldsymbol{\alpha}^{\star}}^{-1} \\
        &= \mathcal{I}_{\boldsymbol{\alpha}^{\star}}^{-1} \cdot \begin{bmatrix}
            \frac{g_{1}(x_{c})}{\lambda_{1}(x_{c},\tilde{F})} & \dots & \frac{g_{n}(x_{c})}{\lambda_{n}(x_{c},\tilde{F})} \\
            \frac{1}{\lambda_{1}(x_{c},\tilde{F})} & \dots & \frac{1}{\lambda_{n}(x_{c},\tilde{F})}
        \end{bmatrix} \cdot \begin{bmatrix}
            g_{1}(x_{c}) & 1 \\
            \vdots & \vdots \\
            g_{n}(x_{c}) & 1
        \end{bmatrix} \cdot \mathcal{I}_{\boldsymbol{\alpha}^{\star}}^{-1} \\
        &= \mathcal{I}_{\boldsymbol{\alpha}^{\star}}^{-1} \cdot \begin{bmatrix}
            \sum_{i =1}^n \frac{g_{i}^{2}(x_{c})}{\lambda_{i}(x_{c},\tilde{F})} & \sum_{i =1}^n \frac{g_{i}(x_{c})}{\lambda_{i}(x_{c},\tilde{F})} \\
            \sum_{i =1}^n \frac{g_{i}(x_{c})}{\lambda_{i}(x_{c},\tilde{F})} & \sum_{i =1}^n \frac{1}{\lambda_{i}(x_{c},\tilde{F})}
        \end{bmatrix} \cdot \mathcal{I}_{\boldsymbol{\alpha}^{\star}}^{-1} \\
        &= \mathcal{I}_{\boldsymbol{\alpha}^{\star}}^{-1} \cdot \mathcal{I}_{\boldsymbol{\alpha}^{\star}} \cdot \mathcal{I}_{\boldsymbol{\alpha}^{\star}}^{-1} \\
        &= \mathcal{I}_{\boldsymbol{\alpha}^{\star}}^{-1},
    \end{split}
\end{equation}

\noindent which concludes the result. However, it is worth noting that this estimator's configuration requires to know $\boldsymbol{\alpha}^{\star}$ beforehand (the reader may recall Equation~(\ref{eq:opt_weights}) in Section \ref{sec:oracle_WLS} and note that this knowledge is required to compute $\lambda_{i}(x_{c},\tilde{F})$), which contradicts the aim and very essence of the posed problem.

\subsection{SWLS as a non-linear Estimator}
\label{sec:proof_SWLSE_bounds}

Just as the general WLS estimator, the optimization problem resulting from minimization of Equation~(\ref{eq:SWLS_cost}) has a closed-form, though non-linear on $\mathbf{I}$, solution; which can be readily\footnote{Note that this new optimization problem is also convex on $\boldsymbol{\Theta}$, which can be easily demonstrated by simply replacing $w_{i}$ with $\frac{1}{I_{i}}$ into Equation~(\ref{proof:WLS_convexity}). Such evaluation can be carried further into the derivation of the closed-form expression for the SWLS estimator.} formulated as
\begin{equation}
    \tau_{SWLS}(\mathbf{I}) = \begin{bmatrix}
        \sum_{i =1}^n \frac{1}{I_{i}} \cdot g_{i}^{2}(x_{c}) & \sum_{i =1}^n \frac{1}{I_{i}} \cdot g_{i}(x_{c}) \\
        \sum_{i =1}^n \frac{1}{I_{i}} \cdot g_{i}(x_{c}) & \sum_{i =1}^n \frac{1}{I_{i}}
    \end{bmatrix}^{-1} \cdot
    \begin{bmatrix}
        \frac{1}{I_{1}} \cdot g_{1}(x_{c}) & \dots & \frac{1}{I_{n}} \cdot g_{n}(x_{c}) \\
        \frac{1}{I_{1}} & \dots & \frac{1}{I_{n}}
    \end{bmatrix} \cdot \mathbf{I}.
    \label{eq:proof_SWLS_tau_exp}
\end{equation}

In order to evaluate the performance bounds developed in Section \ref{sec:impl_extension} for this particular case, let's consider firstly the derivatives of $J_{SWLS}(\boldsymbol{\alpha},\mathbf{I})$ with respect to $\boldsymbol{\alpha}$, which define the first-order condition of Equation~(\ref{eq:implicit_foc}):
\begin{equation}
    \frac{\partial}{\partial\alpha_{1}} J_{SWLS}(\boldsymbol{\alpha},\mathbf{I}) = \sum_{i =1}^n \frac{-2 g_{i}(x_{c})}{I_{i}} \left(I_{i} -\alpha_{1} g_{i}(x_{c}) - \alpha_{2} \right),
\end{equation}

\begin{equation}
    \frac{\partial}{\partial\alpha_{2}} J_{SWLS}(\boldsymbol{\alpha},\mathbf{I}) = \sum_{i =1}^n \frac{-2}{I_{i}} \left(I_{i} -\alpha_{1} g_{i}(x_{c}) - \alpha_{2} \right).
\end{equation}

Taking derivatives of these expressions with respect to $\boldsymbol{\alpha}$ and $I_{i}$ for some $i \in \{1,\dots,n\}$ leads into the elements of $\nabla^{20} J_{SWLS}(\boldsymbol{\alpha},\mathbf{I})$ and $\nabla^{11} J_{SWLS}(\boldsymbol{\alpha},\mathbf{I})$, respectively. The former second-order derivative corresponds to
\begin{equation}
    \begin{split}
        \nabla^{20} J_{SWLS}(\boldsymbol{\alpha},\mathbf{I}) &= 2 \begin{bmatrix}
            \sum_{i =1}^n \frac{g_{i}^{2}(x_{c})}{I_{i}} & \sum_{i =1}^n \frac{g_{i}(x_{c})}{I_{i}} \\
            \sum_{i =1}^n \frac{g_{i}(x_{c})}{I_{i}} & \sum_{i =1}^n \frac{1}{I_{i}}
        \end{bmatrix}.
    \end{split}
    \label{eq:proof_SWLS_nabla20}
\end{equation}

From this last line, it stands out that $\nabla^{20} J_{SWLS}(\boldsymbol{\alpha}, \mathbf{I})$ and $\nabla^{20} J_{WLS}(\boldsymbol{\alpha}, \mathbf{I})$ (see Equation~(\ref{eq:WLS_tau_exp})) share the same structure, which is evident by emphasizing the fact that the difference between both cost functions falls solely to the evaluation of $w_{i} = \frac{1}{I_{i}}$ for $i \in \{1,\dots,n\}$.

However, what's more interesting is the fact that precisely this weight configuration conveys to (apparently) different structures after applying the $\nabla^{11}$ operator, as $w_{i}$ is also derivable with respect to $I_{i}$, not just its adjacent squared-error term. Then,
\begin{equation}
    \begin{split}
        \frac{\partial^{2}}{\partial\alpha_{1} \partial I_{i}} J_{SWLS}(\boldsymbol{\alpha},\mathbf{I}) &= \frac{\partial}{\partial I_{i}} \left( \sum_{j =1}^n \frac{-2 g_{j}(x_{c})}{I_{j}} \left(I_{j} -\alpha_{1} g_{j}(x_{c}) - \alpha_{2} \right) \right) \\
        &= \frac{-2 g_{i}(x_{c})}{I_{i}^{2}} \left( \alpha_{1} g_{i}(x_{c}) + \alpha_{2} \right),
    \end{split}
    \label{eq:proof_SWLS_nabla11_1i}
\end{equation}

\begin{equation}
    \begin{split}
        \frac{\partial^{2}}{\partial\alpha_{2} \partial I_{i}} J_{SWLS}(\boldsymbol{\alpha},\mathbf{I}) &= \frac{\partial}{\partial I_{i}} \left( \sum_{j =1}^n \frac{-2}{I_{j}} \left(I_{j} -\alpha_{1} g_{j}(x_{c}) - \alpha_{2} \right) \right) \\
        &= \frac{-2}{I_{i}^{2}} \left( \alpha_{1} g_{i}(x_{c}) + \alpha_{2} \right),
    \end{split}
    \label{eq:proof_SWLS_nabla11_2i}
\end{equation}

\begin{equation}
    \begin{split}
        \Rightarrow \nabla^{11} J_{SWLS}(\boldsymbol{\alpha},\mathbf{I}) &= \begin{bmatrix}
            \frac{\partial^{2}}{\partial\alpha_{1}\partial I_{1}} J_{SWLS}(\boldsymbol{\alpha},\mathbf{I}) & \cdots & \frac{\partial^{2}}{\partial\alpha_{1}\partial I_{n}} J_{SWLS}(\boldsymbol{\alpha},\mathbf{I}) \\
            \frac{\partial^{2}}{\partial\alpha_{2}\partial I_{1}} J_{SWLS}(\boldsymbol{\alpha},\mathbf{I}) & \cdots & \frac{\partial^{2}}{\partial\alpha_{2}\partial I_{n}} J_{SWLS}(\boldsymbol{\alpha},\mathbf{I})
        \end{bmatrix} \\
        &= \begin{bmatrix}
            -\frac{2 g_{1}(x_{c})}{I_{1}^{2}} (\alpha_{1} g_{1}(x_{c}) + \alpha_{2}) & \cdots & -\frac{2 g_{n}(x_{c})}{I_{n}^{2}} (\alpha_{1} g_{n}(x_{c}) + \alpha_{2}) \\
            -\frac{2}{I_{1}^{2}} (\alpha_{1} g_{1}(x_{c}) + \alpha_{2}) & \cdots & -\frac{2}{I_{n}^{2}} (\alpha_{1} g_{n}(x_{c}) + \alpha_{2})
        \end{bmatrix}.
    \end{split}
\end{equation}

Having the expressions for $\nabla^{11} J_{SWLS}(\boldsymbol{\alpha},\mathbf{I})$ and $\nabla^{20} J_{SWLS}(\boldsymbol{\alpha},\mathbf{I})$ at hand, the first-order derivative of the estimator, $\tau_{SWLS}'(\mathbf{I})$, can be straightforwarly computed just as depicted in Equation~(\ref{eq:pdim_tau_tilde1}).

The next step towards finding the bounds on the estimator's performance is to elaborate on the expressions for the third-order derivatives of $J_{SWLS}(\boldsymbol{\alpha},\mathbf{I})$. On this task, the simplest of these arrays corresponds to the application of the $\nabla^{30}$ operator since it can be readily verified (from Equation~(\ref{eq:proof_SWLS_nabla20})) that, $\forall j,k,l \in \{1,2\}$, and for any pair $(\boldsymbol{\alpha},\mathbf{I})$ on the parameter-observation space,

\begin{equation}
    \frac{\partial^{3}}{\partial\alpha_{j} \partial\alpha_{k} \partial\alpha_{l}} J_{SWLS}(\boldsymbol{\alpha},\mathbf{I}) = 0.
\end{equation}

Therefore, the full $2 \times 2 \times 2$ array $\nabla^{30} J_{SWLS}(\boldsymbol{\alpha},\mathbf{I})$ corresponds to a null-array, {\em i.e.},
\begin{equation}
    \begin{split}
        \nabla^{30} J_{SWLS}(\boldsymbol{\alpha},\mathbf{I}) = \mathbf{0}.
    \end{split}
\end{equation}

Recalling on Equations~(\ref{eq:proof_SWLS_nabla11_1i}) and~(\ref{eq:proof_SWLS_nabla11_2i}), we can develop on the applications of the $\nabla^{21}$ and $\nabla^{12}$ operators. For the former case, considering an arbitrary index $i \in \{1,\dots,n\}$ we can take derivatives of said equations with respect to $\boldsymbol{\alpha}$, which leads to:
\begin{equation}
    \begin{split}
        \frac{\partial^{3}}{\partial\alpha_{1}^{2} \partial I_{i}} J_{SWLS}(\boldsymbol{\alpha},\mathbf{I}) &= \frac{\partial}{\partial\alpha_{1}} \left( \frac{-2 g_{i}(x_{c})}{I_{i}^{2}} \left( \alpha_{1} g_{i}(x_{c}) + \alpha_{2} \right) \right) \\
        &= -\frac{2 g_{i}^{2}(x_{c})}{I_{i}^{2}},
    \end{split}
\end{equation}

\begin{equation}
    \begin{split}
        \frac{\partial}{\partial\alpha_{1} \partial\alpha_{2} \partial I_{i}} J_{SWLS}(\boldsymbol{\alpha},\mathbf{I}) &= \frac{\partial}{\partial\alpha_{2} \partial\alpha_{1} \partial I_{i}} J_{SWLS}(\boldsymbol{\alpha},\mathbf{I}) \\
        &= \frac{\partial}{\partial\alpha_{2}} \left( \frac{-2 g_{i}(x_{c})}{I_{i}^{2}} \left( \alpha_{1} g_{i}(x_{c}) + \alpha_{2} \right) \right) \\
        &= -\frac{2 g_{i}(x_{c})}{I_{i}^{2}},
    \end{split}
\end{equation}

\begin{equation}
    \begin{split}
        \frac{\partial^{3}}{\partial\alpha_{2}^{2} \partial I_{i}} J_{SWLS}(\boldsymbol{\alpha},\mathbf{I}) &= \frac{\partial}{\partial\alpha_{2}} \left( \frac{-2}{I_{i}^{2}} \left( \alpha_{1} g_{i}(x_{c}) + \alpha_{2} \right) \right) \\
        &= -\frac{2}{I_{i}^{2}}.
    \end{split}
\end{equation}

Combining these last three equations the full $2 \times 2 \times n$ array $\nabla^{21} J_{SWLS}(\boldsymbol{\alpha},\mathbf{I})$ can be summarized as
\begin{equation}
    \begin{split}
        \left[ \nabla^{21} J_{SWLS}(\boldsymbol{\alpha},\mathbf{I}) \right]_{(\cdot,\cdot,i)} = \begin{bmatrix}
            -\frac{2 g_{i}^{2}(x_{c})}{I_{i}^{2}} & -\frac{2 g_{i}(x_{c})}{I_{i}^{2}} \\
            -\frac{2 g_{i}^{2}(x_{c})}{I_{i}^{2}} & -\frac{2}{I_{i}^{2}}
        \end{bmatrix}.
    \end{split}
\end{equation}

On the other hand, taking derivatives of Equations~(\ref{eq:proof_SWLS_nabla11_1i}) and~(\ref{eq:proof_SWLS_nabla11_2i}) with respect to some $I_{h}, h \in \{1,\dots,n\}$, it follows that both respective expressions become case-dependent, in other words, both third-order derivatives take different values depending on whether $i=h$ or not. For that matter:

\begin{equation}
    \frac{\partial^{3}}{\partial\alpha_{1} \partial I_{i} \partial I_{h}} J_{SWLS}(\boldsymbol{\alpha},\mathbf{I}) = \begin{cases}
    \frac{4 g_{i}(x_{c})}{I_{i}^{3}} \left( \alpha_{1} g_{i}(x_{c}) + \alpha_{2} \right), &\quad h=i \\
    0, &\quad h \neq i \\
    \end{cases}
\end{equation}

\begin{equation}
    \frac{\partial^{3}}{\partial\alpha_{2} \partial I_{i} \partial I_{h}} J_{SWLS}(\boldsymbol{\alpha},\mathbf{I}) = \begin{cases}
    \frac{4}{I_{i}^{3}} \left( \alpha_{1} g_{i}(x_{c}) + \alpha_{2} \right), &\quad h=i \\
    0, &\quad h \neq i \\
    \end{cases}
\end{equation}

Consequently, the full $2 \times n \times n$ array $\nabla^{12} J_{SWLS}(\boldsymbol{\alpha},\mathbf{I})$ can be interpreted as two $n \times n$ stacked diagonal matrices, which can be expressed, respectively, as follows:

\begin{equation}
    \begin{split}
        \left[ \nabla^{12} J_{SWLS}(\boldsymbol{\alpha},\mathbf{I}) \right]_{(1,\cdot,\cdot)} = \begin{bmatrix}
            \frac{4 g_{1}(x_{c})}{I_{1}^{3}} (\alpha_{1} g_{1}(x_{c}) + \alpha_{2}) &  &  \\
             & \ddots &  \\
             &  & \frac{4 g_{n}(x_{c})}{I_{n}^{3}} (\alpha_{1} g_{n}(x_{c}) + \alpha_{2})
        \end{bmatrix},
    \end{split}
\end{equation}

\begin{equation}
    \begin{split}
        \left[ \nabla^{12} J_{SWLS}(\boldsymbol{\alpha},\mathbf{I}) \right]_{(2,\cdot,\cdot)} = \begin{bmatrix}
            \frac{4}{I_{1}^{3}} (\alpha_{1} g_{1}(x_{c}) + \alpha_{2}) &  &  \\
             & \ddots &  \\
             &  & \frac{4}{I_{n}^{3}} (\alpha_{1} g_{n}(x_{c}) + \alpha_{2})
        \end{bmatrix}.
    \end{split}
\end{equation}

Finally, we can draw upon Equations~(\ref{eq:pdim_Pih}),~(\ref{eq:pdim_Qih}) and~(\ref{eq:proof_pdim_dtau2}) to assemble the last building blocks to formulate the second-order derivatives of $\tau_{SWLS}(\mathbf{I})$ and to subsequently apply the bounds depicted in Section \ref{sec:impl_extension}.

\begin{equation}
    P_{SWLS}^{i,h}(\mathbf{I}) = \left[ \left[ \nabla^{21} J_{SWLS}(\tau_{SWLS}(\mathbf{I}), \mathbf{I}) \right]_{(j,\cdot,h)} \cdot \left[ \tau_{SWLS}'(\mathbf{I}) \right]_{(\cdot,i)} \right]_{j=1}^{2},
\end{equation}

\begin{equation}
    Q_{SWLS}^{i,h}(\mathbf{I}) = \left[ \left[ \nabla^{21} J_{SWLS}(\tau_{SWLS}(\mathbf{I}), \mathbf{I}) \right]_{(j,\cdot,i)} \cdot \left[ \tau_{SWLS}'(\mathbf{I}) \right]_{(\cdot,h)} + \left[ \nabla^{12} J_{SWLS}(\tau_{SWLS}(\mathbf{I}), \mathbf{I}) \right]_{(j,i,h)} \right]_{j=1}^{2},
\end{equation}

\begin{equation}
    \frac{\partial^{2}}{\partial I_{i} \partial I_{h}} \tau_{SWLS}(\mathbf{I}) = - \left[ \nabla^{20} J_{SWLS}(\tau_{ML}(\mathbf{I}), \mathbf{I}) \right]^{-1} \cdot \left[ P_{SWLS}^{i,h}(\mathbf{I}) + Q_{SWLS}^{i,h}(\mathbf{I}) \right].
\end{equation}

The expressions derived so far represent the cornerstone from which the performance bounds for the SWLS estimator are formulated and straightforwardly implemented, concluding the proof.

\section{Commentaries on the SNR obtained for different Scenarios and Apertures}
\label{app:SNR}

We introduce here a measurable form of signal-to-noise ratio (SNR) as in \citet{mendez2013} that captures to some extent the quality of the signal as the source is observed through the CCD detector under determined sky conditions. Based on the observational model of Section \ref{sec:preliminaries}, the sampled signal $S$ (in photo-e$^{-}$) can be defined as:

\begin{equation}
    S = \tilde{F} \cdot \int_{x_{l}}^{x_{u}} \phi(x - x_{c}, \sigma) dx,
    \label{eq:S_mendez}
\end{equation}

\noindent where $x_{l}$ and $x_{u}$ are suitably chosen (but arbitrary) apertures such that an appreciable fraction of the total flux is included. For the Gaussian case considered above, this expression can be written as:

\begin{equation}
    S = \tilde{F} \cdot P(u_{+}), \hspace{3mm} P(u) = \frac{2}{\sqrt{\pi}} \int_{0}^{u} e^{-v^{2}} dv,
    \label{eq:S_mendez_gauss}
\end{equation}

\noindent where $u_{+} = (x_{u} - x_{l}) / \sqrt{2}\sigma$ denotes the chosen aperture, which is symmetric with respect to the source's position $x_{c}$, that is $x_{u} - x_{c} = x_{c} - x_{l}$.

The total noise, $N$, is considered to have contributions from the detector's read-out noise, the sky noise and the noise from the source itself \citep{mendez2013}, all of them assumed to follow a Poisson distribution, such that:

\begin{equation}
    N = \sqrt{S + N_{pix}(G \Delta x f_{s} + RON^{2})},
    \label{eq:N_mendez}
\end{equation}

\noindent where $N_{pix}$ denotes the number of pixels under the same region in which the signal $S$ was sampled, {\em i.e.}, the pixels used to get an estimation $\hat{\tilde{F}}$. Finally, the SNR is given by $S/N$.

\begin{figure}[!h]
    \centering
\includegraphics[width=0.5\textwidth]{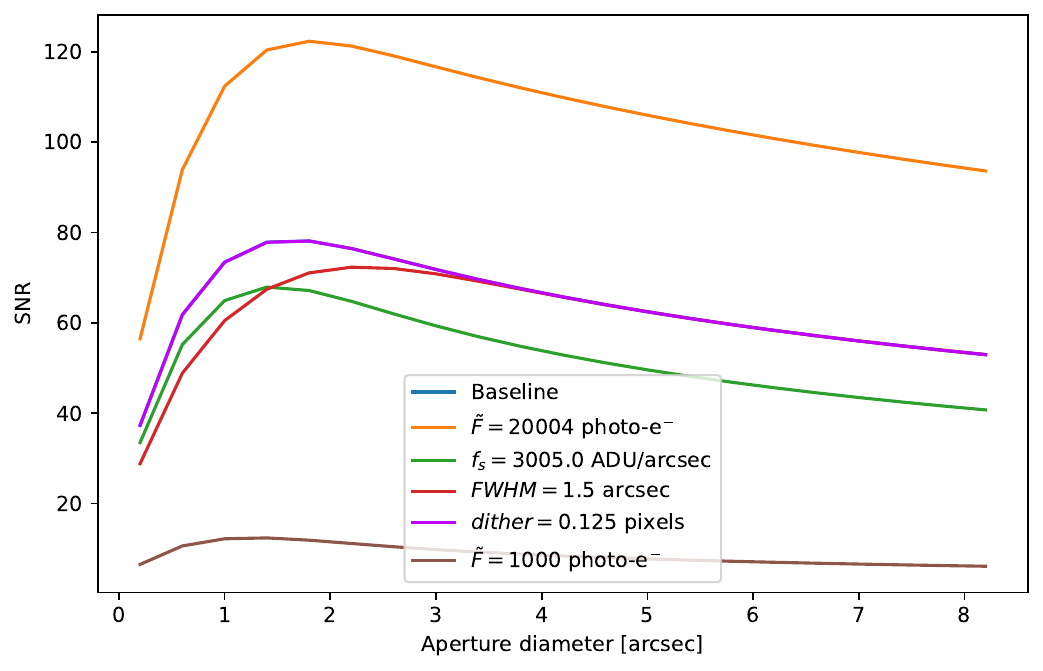}
    \caption{SNR for different observational scenarios, under an aperture pixel selection scheme.}
    \label{fig:SNR_other_cases}
\end{figure}

From these definitions, it becomes evident that SNR is a function not only of seeing and instrumental conditions, but also of the chosen aperture. For that reason, Figure~\ref{fig:SNR_other_cases} shows the SNR of the different simulated scenarios studied throughout this work and its variation with aperture\footnote{Note that curves corresponding to the baseline and the off-set scenario overlap with each other. Also, note that the aperture that provides the highest SNR changes with source magnitude and observing conditions.}.

\section{Proof: ML Estimator for the Joint Estimation Problem of Flux and Background}
\label{app:proof_ML_FB}

In order to apply the methodology proposed in Section \ref{sec:impl_extension}, we must show before that the ML functional has a unique optimal solution. To do so, we study its convexity. Let $\boldsymbol{\alpha}_{A} = (\alpha_{1}, \alpha_{2}), \boldsymbol{\alpha}_{B} = (\alpha_{1}', \alpha_{2}') \in \boldsymbol{\Theta}$ be two vectors on the parameter space and $t  \in [0,1]$. For some fixed observation vector $\mathbf{I}$, we have that
\begin{equation}
    \begin{split}
        J_{ML}(t\boldsymbol{\alpha}_{A} + (1-t)\boldsymbol{\alpha}_{B}, \mathbf{I}) &= \sum_{i =1}^n \left( t\alpha_{1} + (1-t)\alpha_{1}' \right) \cdot g_{i}(x_{c}) + \left( t\alpha_{2} + (1-t)\alpha_{2}' \right) \\
        & \hspace{13mm} - I_{i} \ln{\left(\left( t\alpha_{1} + (1-t)\alpha_{1}' \right) \cdot g_{i}(x_{c}) + \left( t\alpha_{2} + (1-t)\alpha_{2}' \right) \right)} \\
        &= \sum_{i =1}^n t\left( \alpha_{1} \cdot g_{i}(x_{c}) + \alpha_{2} \right) + (1-t)\left( \alpha_{1}' \cdot g_{i}(x_{c}) + \alpha_{2}' \right) \\
        & \hspace{13mm} - I_{i} \ln{\left( t\left( \alpha_{1} \cdot g_{i}(x_{c}) + \alpha_{2} \right) + (1-t)\left( \alpha_{1}' \cdot g_{i}(x_{c}) + \alpha_{2}' \right) \right)} \\
        &\leq \sum_{i =1}^n t\left( \alpha_{1} \cdot g_{i}(x_{c}) + \alpha_{2} \right) + (1-t)\left( \alpha_{1}' \cdot g_{i}(x_{c}) + \alpha_{2}' \right) \\
        &\hspace{13mm} - I_{i} \left[ t\ln{\left( \alpha_{1} \cdot g_{i}(x_{c}) + \alpha_{2} \right)} + (1-t)\ln{\left( \alpha_{1}' \cdot g_{i}(x_{c}) + \alpha_{2}' \right)}\right] \\
        &= \sum_{i =1}^n t\left( \alpha_{1} \cdot g_{i}(x_{c}) + \alpha_{2} - I_{i} \ln{(\alpha_{1} \cdot g_{i}(x_{c}) + \alpha_{2})} \right) \\
        &\hspace{13mm} + (1-t) \left( \alpha_{1}' \cdot g_{i}(x_{c}) + \alpha_{2}' - I_{i} \ln{(\alpha_{1}' \cdot g_{i}(x_{c}) + \alpha_{2}')} \right) \\
        &= t\sum_{i =1}^n \alpha_{1} \cdot g_{i}(x_{c}) + \alpha_{2} - I_{i} \ln{(\alpha_{1} \cdot g_{i}(x_{c}) + \alpha_{2})} \\
        &\hspace{4mm} + (1-t)\sum_{i =1}^n \alpha_{1}' \cdot g_{i}(x_{c}) + \alpha_{2}' - I_{i} \ln{(\alpha_{1}' \cdot g_{i}(x_{c}) + \alpha_{2}')} \\
        &= t J_{ML}(\boldsymbol{\alpha}_{A},\mathbf{I}) + (1-t)J_{ML}(\boldsymbol{\alpha}_{B},\mathbf{I}) 
    \end{split}
    \label{proof:ML_convexity}
\end{equation}\qed

\noindent where the inequality appears as a consequence of the concavity of the natural logarithm. 


Since $J_{ML}(\boldsymbol{\alpha},\mathbf{I})$ is convex on $\boldsymbol{\Theta}$ when $\mathbf{I}$ is fixed, we can proceed with the formulation of the bounds.

As a starting point, let's consider firstly the derivatives of $J_{ML}(\boldsymbol{\alpha},\mathbf{I})$ with respect to $\boldsymbol{\alpha}$, which define the first-order condition of Equation~(\ref{eq:implicit_foc}):
\begin{equation}
    \frac{\partial}{\partial\alpha_{1}} J_{ML}(\boldsymbol{\alpha},\mathbf{I}) = \sum_{i=1}^{n} g_{i}(x_{c}) - \frac{I_{i} g_{i}(x_{c})}{\alpha_{1} g_{i}(x_{c}) + \alpha_{2}},
\end{equation}

\begin{equation}
    \frac{\partial}{\partial\alpha_{2}} J_{ML}(\boldsymbol{\alpha},\mathbf{I}) = \sum_{i=1}^{n} 1 - \frac{I_{i}}{\alpha_{1} g_{i}(x_{c}) + \alpha_{2}}.
\end{equation}

Taking derivatives of these expressions with respect to $\boldsymbol{\alpha}$ and $I_{i}$ for some $i \in \{1,\dots,n\}$ results in the elements of $\nabla^{20} J_{ML}(\boldsymbol{\alpha},\mathbf{I})$ and $\nabla^{11} J_{ML}(\boldsymbol{\alpha},\mathbf{I})$, respectively. For the former matrix that is
\begin{equation}
    \begin{split}
        \nabla^{20} J_{ML}(\boldsymbol{\alpha},\mathbf{I}) &= \begin{bmatrix}
            \frac{\partial^{2}}{\partial\alpha_{1}^{2}} J_{ML}(\boldsymbol{\alpha},\mathbf{I}) & \frac{\partial^{2}}{\partial\alpha_{1}\partial\alpha_{2}} J_{ML}(\boldsymbol{\alpha},\mathbf{I}) \\
            \frac{\partial^{2}}{\partial\alpha_{2}\partial\alpha_{1}} J_{ML}(\boldsymbol{\alpha},\mathbf{I}) & \frac{\partial^{2}}{\partial\alpha_{2}^{2}} J_{ML}(\boldsymbol{\alpha},\mathbf{I})
        \end{bmatrix} \\
        &= \begin{bmatrix}
            \sum_{i=1}^{n} \frac{I_{i} g_{i}^{2}(x_{c})}{(\alpha_{1}g_{i}(x_{c}) + \alpha_{2})^{2}} & \sum_{i=1}^{n} \frac{I_{i} g_{i}(x_{c})}{(\alpha_{1}g_{i}(x_{c}) + \alpha_{2})^{2}} \\
            \sum_{i=1}^{n} \frac{I_{i} g_{i}(x_{c})}{(\alpha_{1}g_{i}(x_{c}) + \alpha_{2})^{2}} & \sum_{i=1}^{n} \frac{I_{i}}{(\alpha_{1}g_{i}(x_{c}) + \alpha_{2})^{2}}
        \end{bmatrix}.
    \end{split}
    \label{eq:proof_ML_nabla20}
\end{equation}

It may be interesting to note that $\nabla^{20} J_{ML}(\boldsymbol{\alpha},\mathbf{I})$ heavily resembles the structure of the Fisher matrix depicted in Equation~(\ref{eq:FB_fisher}), which is natural from the very definition of the Fisher Information Matrix and the ML estimator.

On the other hand, the elements of the latter matrix $\nabla^{11} J_{ML}(\boldsymbol{\alpha},\mathbf{I})$, take the following forms:

\begin{equation}
    \frac{\partial^{2}}{\partial\alpha_{1}\partial I_{i}} J_{ML}(\boldsymbol{\alpha},\mathbf{I}) = \frac{-g_{i}(x_{c})}{\alpha_{1} g_{i}(x_{c}) + \alpha_{2}},
\end{equation}

\begin{equation}
    \frac{\partial^{2}}{\partial\alpha_{2}\partial I_{i}} J_{ML}(\boldsymbol{\alpha},\mathbf{I}) = \frac{-1}{\alpha_{1} g_{i}(x_{c}) + \alpha_{2}},
\end{equation}

\noindent where $i\in\{1,\dots,n\}$. Therefore, applying the $\nabla^{11}$ operator on the ML cost function results in
\begin{equation}
    \begin{split}
        \nabla^{11} J_{ML}(\boldsymbol{\alpha},\mathbf{I}) &= \begin{bmatrix}
            \frac{\partial^{2}}{\partial\alpha_{1}\partial I_{1}} J_{ML}(\boldsymbol{\alpha},\mathbf{I}) & \cdots & \frac{\partial^{2}}{\partial\alpha_{1}\partial I_{n}} J_{ML}(\boldsymbol{\alpha},\mathbf{I}) \\
            \frac{\partial^{2}}{\partial\alpha_{2}\partial I_{1}} J_{ML}(\boldsymbol{\alpha},\mathbf{I}) & \cdots & \frac{\partial^{2}}{\partial\alpha_{2}\partial I_{n}} J_{ML}(\boldsymbol{\alpha},\mathbf{I})
        \end{bmatrix} \\
        &= \begin{bmatrix}
            \frac{-g_{1}(x_{c})}{\alpha_{1} g_{1}(x_{c}) + \alpha_{2}} & \cdots & \frac{-g_{n}(x_{c})}{\alpha_{1} g_{n}(x_{c}) + \alpha_{2}} \\
            \frac{-1}{\alpha_{1} g_{1}(x_{c}) + \alpha_{2}} & \cdots & \frac{-1}{\alpha_{1} g_{n}(x_{c}) + \alpha_{2}} 
        \end{bmatrix}.
    \end{split}
    \label{eq:proof_ML_nabla11}
\end{equation}

With both these matrices, the first-order derivative of $\tau_{ML}(\mathbf{I})$ is directly obtained from Equation~(\ref{eq:pdim_tau_tilde1}) as

\begin{equation}
    \tau_{ML}'(\mathbf{I}) = - \left[ \nabla^{20} J_{ML}(\tau_{ML}(\mathbf{I}), \mathbf{I}) \right]^{-1} \cdot \left[ \nabla^{11} J_{ML}(\tau_{ML}(\mathbf{I}), \mathbf{I}) \right].
    \label{eq:ML_tau_tilde1}
\end{equation}

In a similar fashion, the estimator's second-order derivative can be determined by means of evaluating Equations~(\ref{eq:pdim_Pih}),~(\ref{eq:pdim_Qih})~and~(\ref{eq:proof_pdim_dtau2}). In order to do so, it suffices to calculate the application of third-order operators $\nabla^{12}$, $\nabla^{21}$ and $\nabla^{30}$ on $J_{ML}(\boldsymbol{\alpha},\mathbf{I})$ and then evaluating the resulting expressions on $\boldsymbol{\alpha} = \tau_{ML}(\mathbf{I})$.

For that matter, let $i,h \in \{1,\dots,n\}$ be two indices, not necessarily distinct. Then:
\begin{equation}
    \begin{split}
        \left[ \nabla^{12} J_{ML}(\boldsymbol{\alpha},\mathbf{I}) \right]_{(1,h,i)} &= \frac{\partial}{\partial I_{h}} \left( \frac{\partial^{2}}{\partial\alpha_{1} \partial I_{i}} J_{ML}(\boldsymbol{\alpha}, \mathbf{I}) \right) \\
        &= \frac{\partial}{\partial I_{h}} \left( \frac{-g_{i}(x_{c})}{\alpha_{1} g_{i}(x_{c}) + \alpha_{2}} \right) \\
        &= 0,
    \end{split}
\end{equation}

\begin{equation}
    \begin{split}
        \left[ \nabla^{12} J_{ML}(\boldsymbol{\alpha},\mathbf{I}) \right]_{(2,h,i)} &= \frac{\partial}{\partial I_{h}} \left( \frac{\partial^{2}}{\partial\alpha_{2} \partial I_{i}} J_{ML}(\boldsymbol{\alpha}, \mathbf{I}) \right) \\
        &= \frac{\partial}{\partial I_{h}} \left( \frac{-1}{\alpha_{1} g_{i}(x_{c}) + \alpha_{2}} \right) \\
        &= 0.
    \end{split}
\end{equation}

Therefore,

\begin{equation}
    \nabla^{12} J_{ML}(\boldsymbol{\alpha},\mathbf{I}) = \mathbf{0},
\end{equation}

\noindent remarking once again the abuse of notation, where $\mathbf{0}$ denotes a null array.

On the other hand, the $i$-th ``slice" of $\nabla^{21} J_{ML}(\boldsymbol{\alpha},\mathbf{I})$ is readily calculated as
\begin{equation}
    \begin{split}
        \left[ \nabla^{21} J_{ML}(\boldsymbol{\alpha},\mathbf{I}) \right]_{(\cdot,\cdot,i)}  &= \begin{bmatrix}
            \frac{\partial}{\partial I_{i}} \left( \frac{\partial^{2}}{\partial\alpha_{1}^{2}} J_{ML}(\boldsymbol{\alpha},\mathbf{I}) \right)  & \frac{\partial}{\partial I_{i}} \left( \frac{\partial^{2}}{\partial\alpha_{1}\partial\alpha_{2}} J_{ML}(\boldsymbol{\alpha},\mathbf{I}) \right) \\
            \frac{\partial}{\partial I_{i}} \left( \frac{\partial^{2}}{\partial\alpha_{2}\partial\alpha_{1}} J_{ML}(\boldsymbol{\alpha},\mathbf{I}) \right) & \frac{\partial}{\partial I_{i}} \left( \frac{\partial^{2}}{\partial\alpha_{2}^{2}} J_{ML}(\boldsymbol{\alpha},\mathbf{I}) \right)
        \end{bmatrix} \\
        &= \begin{bmatrix}
            \frac{\partial}{\partial I_{i}} \left( \sum_{j=1}^{n} \frac{I_{j} g_{j}^{2}(x_{c})}{(\alpha_{1}g_{j}(x_{c}) + \alpha_{2})^{2}} \right) & \frac{\partial}{\partial I_{i}} \left( \sum_{j\in\mathcal{J}} \frac{I_{i} g_{i}(x_{c})}{(\alpha_{1}g_{i}(x_{c}) + \alpha_{2})^{2}} \right) \\
            \frac{\partial}{\partial I_{i}} \left( \sum_{j\in\mathcal{J}} \frac{I_{i} g_{i}(x_{c})}{(\alpha_{1}g_{i}(x_{c}) + \alpha_{2})^{2}} \right) & \frac{\partial}{\partial I_{i}} \left( \sum_{j\in\mathcal{J}} \frac{I_{i}}{(\alpha_{1}g_{i}(x_{c}) + \alpha_{2})^{2}} \right)
        \end{bmatrix} \\
        &= \begin{bmatrix}
            \frac{g_{i}^{2}(x_{c})}{(\alpha_{1}g_{i}(x_{c}) + \alpha_{2})^{2}} & \frac{g_{i}(x_{c})}{(\alpha_{1}g_{i}(x_{c}) + \alpha_{2})^{2}} \\
            \frac{g_{i}(x_{c})}{(\alpha_{1}g_{i}(x_{c}) + \alpha_{2})^{2}} & \frac{1}{(\alpha_{1}g_{i}(x_{c}) + \alpha_{2})^{2}}
        \end{bmatrix}.
    \end{split}
\end{equation}

Lastly, for $\nabla^{30} J_{ML}(\boldsymbol{\alpha},\mathbf{I})$ it follows that:
\begin{equation}
    \begin{split}
        \left[ \nabla^{30} J_{ML}(\boldsymbol{\alpha},\mathbf{I}) \right]_{(1,\cdot,\cdot)}  &= \begin{bmatrix}
            \frac{\partial}{\partial \alpha_{1}} \left( \frac{\partial^{2}}{\partial\alpha_{1}^{2}} J_{ML}(\boldsymbol{\alpha},\mathbf{I}) \right)  & \frac{\partial}{\partial \alpha_{1}} \left( \frac{\partial^{2}}{\partial\alpha_{1}\partial\alpha_{2}} J_{ML}(\boldsymbol{\alpha},\mathbf{I}) \right) \\
            \frac{\partial}{\partial \alpha_{1}} \left( \frac{\partial^{2}}{\partial\alpha_{2}\partial\alpha_{1}} J_{ML}(\boldsymbol{\alpha},\mathbf{I}) \right) & \frac{\partial}{\partial \alpha_{1}} \left( \frac{\partial^{2}}{\partial\alpha_{2}^{2}} J_{ML}(\boldsymbol{\alpha},\mathbf{I}) \right)
        \end{bmatrix} \\
        &= \begin{bmatrix}
            \frac{\partial}{\partial\alpha_{1}} \left( \sum_{i=1}^{n} \frac{I_{i} g_{i}^{2}(x_{c})}{(\alpha_{1}g_{i}(x_{c}) + \alpha_{2})^{2}} \right) & \frac{\partial}{\partial\alpha_{1}} \left( \sum_{i=1}^{n} \frac{I_{i} g_{i}(x_{c})}{(\alpha_{1}g_{i}(x_{c}) + \alpha_{2})^{2}} \right) \\
            \frac{\partial}{\partial\alpha_{1}} \left( \sum_{i=1}^{n} \frac{I_{i} g_{i}(x_{c})}{(\alpha_{1}g_{i}(x_{c}) + \alpha_{2})^{2}} \right) & \frac{\partial}{\partial\alpha_{1}} \left( \sum_{i=1}^{n} \frac{I_{i}}{(\alpha_{1}g_{i}(x_{c}) + \alpha_{2})^{2}} \right)
        \end{bmatrix} \\
        &= \begin{bmatrix}
            \sum_{i=1}^{n} \frac{-2I_{i} g_{i}^{3}(x_{c})}{(\alpha_{1}g_{i}(x_{c}) + \alpha_{2})^{3}} & \sum_{i=1}^{n} \frac{-2I_{i} g_{i}^{2}(x_{c})}{(\alpha_{1}g_{i}(x_{c}) + \alpha_{2})^{3}} \\
            \sum_{i=1}^{n} \frac{-2I_{i} g_{i}^{2}(x_{c})}{(\alpha_{1}g_{i}(x_{c}) + \alpha_{2})^{3}} & \sum_{i=1}^{n} \frac{-2I_{i} g_{i}(x_{c})}{(\alpha_{1}g_{i}(x_{c}) + \alpha_{2})^{3}}
        \end{bmatrix},
    \end{split}
\end{equation}

\begin{equation}
    \begin{split}
        \left[ \nabla^{30} J_{ML}(\boldsymbol{\alpha},\mathbf{I}) \right]_{(2,\cdot,\cdot)}  &= \begin{bmatrix}
            \frac{\partial}{\partial \alpha_{2}} \left( \frac{\partial^{2}}{\partial\alpha_{1}^{2}} J_{ML}(\boldsymbol{\alpha},\mathbf{I}) \right)  & \frac{\partial}{\partial \alpha_{2}} \left( \frac{\partial^{2}}{\partial\alpha_{1}\partial\alpha_{2}} J_{ML}(\boldsymbol{\alpha},\mathbf{I}) \right) \\
            \frac{\partial}{\partial \alpha_{2}} \left( \frac{\partial^{2}}{\partial\alpha_{2}\partial\alpha_{1}} J_{ML}(\boldsymbol{\alpha},\mathbf{I}) \right) & \frac{\partial}{\partial \alpha_{2}} \left( \frac{\partial^{2}}{\partial\alpha_{2}^{2}} J_{ML}(\boldsymbol{\alpha},\mathbf{I}) \right)
        \end{bmatrix} \\
        &= \begin{bmatrix}
            \frac{\partial}{\partial\alpha_{2}} \left( \sum_{i=1}^{n} \frac{I_{i} g_{i}^{2}(x_{c})}{(\alpha_{1}g_{i}(x_{c}) + \alpha_{2})^{2}} \right) & \frac{\partial}{\partial\alpha_{2}} \left( \sum_{i=1}^{n} \frac{I_{i} g_{i}(x_{c})}{(\alpha_{1}g_{i}(x_{c}) + \alpha_{2})^{2}} \right) \\
            \frac{\partial}{\partial\alpha_{2}} \left( \sum_{i=1}^{n} \frac{I_{i} g_{i}(x_{c})}{(\alpha_{1}g_{i}(x_{c}) + \alpha_{2})^{2}} \right) & \frac{\partial}{\partial\alpha_{2}} \left( \sum_{i=1}^{n} \frac{I_{i}}{(\alpha_{1}g_{i}(x_{c}) + \alpha_{2})^{2}} \right)
        \end{bmatrix} \\
        &= \begin{bmatrix}
            \sum_{i=1}^{n} \frac{-2I_{i} g_{i}^{2}(x_{c})}{(\alpha_{1}g_{i}(x_{c}) + \alpha_{2})^{3}} & \sum_{i=1}^{n} \frac{-2I_{i} g_{i}(x_{c})}{(\alpha_{1}g_{i}(x_{c}) + \alpha_{2})^{3}} \\
            \sum_{i=1}^{n} \frac{-2I_{i} g_{i}(x_{c})}{(\alpha_{1}g_{i}(x_{c}) + \alpha_{2})^{3}} & \sum_{i=1}^{n} \frac{-2I_{i}}{(\alpha_{1}g_{i}(x_{c}) + \alpha_{2})^{3}}
        \end{bmatrix}.
    \end{split}
\end{equation}

Having these three tensors at hand and considering the matrices previously calculated ({\em i.e.}, $\nabla^{20} J_{ML}(\tau_{ML}(\mathbf{I}),\mathbf{I})$ and $\tau_{ML}'(\mathbf{I})$), Equations~(\ref{eq:pdim_Pih}),~(\ref{eq:pdim_Qih})~and~(\ref{eq:proof_pdim_dtau2}) can be directly applied to, finally, get the second-order derivative $\tau_{ML}''(\mathbf{I})$. In other words:
\begin{equation}
    P_{ML}^{i,h}(\mathbf{I}) = \left[ \left[ \tau_{ML}'(\mathbf{I}) \right]_{(\cdot,i)}^{\intercal} \cdot \left[ \nabla^{30} J_{ML}(\tau_{ML}(\mathbf{I}), \mathbf{I}) \right]_{(j,\cdot,\cdot)} \cdot \left[ \tau_{ML}'(\mathbf{I}) \right]_{(\cdot,h)}  + \left[ \nabla^{21} J_{ML}(\tau_{ML}(\mathbf{I}), \mathbf{I}) \right]_{(j,\cdot,h)} \cdot \left[ \tau_{ML}'(\mathbf{I}) \right]_{(\cdot,i)} \right]_{j=1}^{2},
\end{equation}

\begin{equation}
    \begin{split}
        Q_{ML}^{i,h}(\mathbf{I}) &= \left[ \left[ \nabla^{21} J_{ML}(\tau_{ML}(\mathbf{I}), \mathbf{I}) \right]_{(j,\cdot,i)} \cdot \left[ \tau_{ML}'(\mathbf{I}) \right]_{(\cdot,h)} + \left[ \nabla^{12} J_{ML}(\tau_{ML}(\mathbf{I}), \mathbf{I}) \right]_{(j,i,h)} \right]_{j=1}^{2} \\
        &= \left[ \left[ \nabla^{21} J_{ML}(\tau_{ML}(\mathbf{I}), \mathbf{I}) \right]_{(j,\cdot,i)} \cdot \left[ \tau_{ML}'(\mathbf{I}) \right]_{(\cdot,h)} \right]_{j=1}^{2},
    \end{split}
\end{equation}

\begin{equation}
    \frac{\partial^{2}}{\partial I_{i} \partial I_{h}} \tau_{ML}(\mathbf{I}) = - \left[ \nabla^{20} J_{ML}(\tau_{ML}(\mathbf{I}), \mathbf{I}) \right]^{-1} \cdot \left[ P_{ML}^{i,h}(\mathbf{I}) + Q_{ML}^{i,h}(\mathbf{I}) \right].
\end{equation}

The expressions derived so far represent the cornerstone from which the performance bounds for the ML estimator are formulated and straightforwardly implemented, concluding the proof.

\end{document}